\documentclass{article}

\usepackage{arxiv}

\usepackage[utf8]{inputenc} 
\usepackage[T1]{fontenc}    
\usepackage{lmodern}
\usepackage{textgreek}
\usepackage{hyperref}       
\usepackage{url}            
\usepackage{booktabs}       
\usepackage{amsfonts}       
\usepackage{nicefrac}       
\usepackage{microtype}      
\usepackage{graphicx}
\usepackage[numbers,sort&compress]{natbib}
\usepackage{doi}
\usepackage{float}
\usepackage{subcaption}
\usepackage{multirow}
\usepackage{xcolor}
\usepackage{amsmath}
\usepackage{amssymb}
\usepackage{stmaryrd}

\title{Modeling CDRX and PDRX during hot forming of zircaloy-4}

\author{ \href{https://orcid.org/0000-0002-3382-4730}{\includegraphics[scale=0.06]{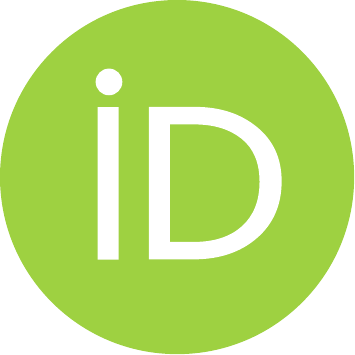}\hspace{1mm}Victor Grand}\thanks{Corresponding author.} \\
	Components Research Center\\
	Framatome\\
	73403, Ugine, France\\
	\texttt{victor.grand@framatome.com} \\
	\And
	\href{https://orcid.org/0000-0001-6804-1974}{\includegraphics[scale=0.06]{orcid.pdf}\hspace{1mm}Baptiste Flipon} \\
  	Mines Paris\\
	PSL University\\
 Centre for Material Forming (CEMEF), UMR CNRS\\
	06904, Sophia Antipolis, France \\
	\texttt{baptiste.flipon@minesparis.psl.eu} \\
 	\And
\href{https://orcid.org/0000-0000-0000-0000}{\includegraphics[scale=0.06]{orcid.pdf}\hspace{1mm}Alexis Gaillac} \\
	Components Research Center\\
	Framatome\\
	73403, Ugine, France\\
	\texttt{alexis.gaillac@framatome.com} \\
 	\And
	\href{https://orcid.org/0000-0002-6677-2850}{\includegraphics[scale=0.06]{orcid.pdf}\hspace{1mm}Marc Bernacki} \\
	Mines Paris\\
	PSL University\\
 Centre for Material Forming (CEMEF), UMR CNRS\\
	06904, Sophia Antipolis, France \\
	\texttt{marc.bernacki@minesparis.psl.eu} \\
}

\renewcommand{\headeright}{Preprint}
\renewcommand{\undertitle}{Preprint}
\renewcommand{\shorttitle}{CDRX and PDRX of Zy-4}

\hypersetup{
pdftitle={Modeling CDRX and PDRX during hot forming of zircaloy-4},
pdfsubject={q-ms.NC},
pdfauthor={Victor ~Grand, Baptiste ~Flipon, Alexis ~Gaillac, Marc ~Bernacki},
pdfkeywords={Continuous dynamic recrystallization, zirconium alloy, hot forming},
}

\begin{document}
\maketitle

\begin{abstract}
	A recently developed full field level-set model of continuous dynamic recrystallization is applied to simulate zircaloy-4 recrystallization during hot compression and subsequent heat treatment. The influence of strain rate, final strain and initial microstructure is investigated, by experimental and simulation tools. The recrystallization heterogeneity is quantified. This enables to confirm that quenched microstructures display a higher extent of heterogeneity. The simulation results replicate satisfactorily experimental observations. The simulation framework is especially able to capture such recrystallization heterogeneity induced by a different initial microstructure. Finally, the role of intragranular dislocation density heterogeneities over the preferential growth of recrystallized grains is pointed out thanks to additional simulations with different numerical formulations.
\end{abstract}


\section{Introduction}\label{sec:Introduction}

Nuclear energy constitutes an alternative to fossil fuels for fulfilling the increasing global energy demands. It presents a very low carbon footprint and its production can be controlled to match the needs. To ensure the highest quality and security requirements, the manufacturing processes of every nuclear component must be mastered. Fuel assemblies, mainly composed of zirconium alloy parts, are no exception to the rule. Thus, it is essential to continue improving the knowledge of microstructure evolution mechanisms taking place during hot forming of zirconium alloys. Recrystallization is responsible for the formation of new grains with a low dislocation density that can consume the deformed microstructure \cite{Rollett2017}. Previous works have concluded that under common hot forming conditions, zirconium alloys and in particular zircaloy-4 present typical features of continuous dynamic recrystallization (CDRX) \cite{Chauvy2006, Vanderesse2008_PhD, Gaudout2009, Ben_Ammar2012_PhD}. Chauvy \textit{et al.} proved that low angle grain boundaries (LAGB) form during hot deformation and progressively transform to high angle grain boundaries (HAGB) \cite{Chauvy2006}. The other studies confirmed that hot deformation of lamellar microstructures are characterized by the formation of fine grains surrounded by large colonies that are slightly deformed. If these studies focus upon the deformation of such complex lamellar microstructures, they do not address in details the influence of thermomechanical conditions and initial microstructure on recrystallization.

The few existing simulation results regarding zircaloy-4 recrystallization were obtained applying mean field models. Doing so, Dunlop \textit{et al.} were able to reproduce the static recrystallization kinetics for various heat treatment conditions. Gaudout \textit{et al.}, who applied the Gourdet-Montheillet CDRX model to zircaloy-4, predict the recrystallized fraction for different hot forming conditions \cite{Gaudout2009}. Since both strategies relied on a mean field model, they did not consider the presence of local microstructure heterogeneities which are very important for such materials.

The present article discusses in details the influence of thermomechanical conditions and initial microstructure upon recrystallization mechanisms, such as CDRX and post-dynamic recrystallization (PDRX), by coupling experimental and numerical investigations. To do so, samples with different initial microstructures are hot compressed and microstructures are characterized by EBSD. Full field simulations are performed using a level-set model \cite{Merriman1994} embedded within a finite element framework (FE-LS) \cite{Bernacki2008}. Experiment and simulation results are compared to evaluate the model abilities and limitations. Different underlying assumptions of the numerical model are assessed. Doing so, the impact of several microstructural features are highlighted.

\section{Materials and methods}\label{sec:MaterialsAndMethods}

\subsection{Materials}\label{subsec:Materials}

Zircaloy-4 samples presenting three different typical microstructures are selected. One presents an equiaxed (\textbf{Eq}) grain topology, with an average equivalent circle diameter ($\overline{ECD}$) equal to $5.3 ~ \mu m$. The others display a Widmanstätten microstructure inherited from the quench with a basket-weaved (\textbf{BW}) morphology for the second one and parallel plates (\textbf{PP}) for the last one.

\begin{figure}
    \centering
     \begin{subfigure}{0.49\textwidth}
        \centering
	    \includegraphics[width=0.95\linewidth]{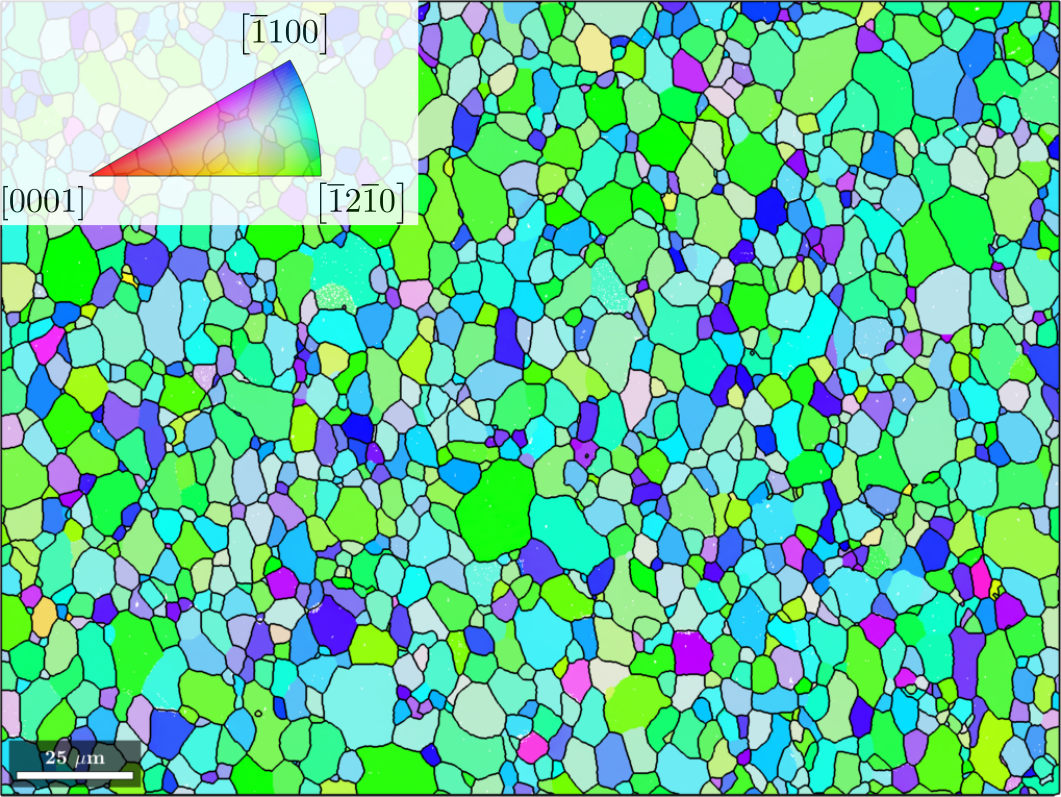}
        \caption{\label{subfig:InitialMicrostructureEq} \textbf{Eq} microstructure.}
    \end{subfigure}
    \begin{subfigure}{0.49\textwidth}
        \centering
        \includegraphics[width=0.95\linewidth]{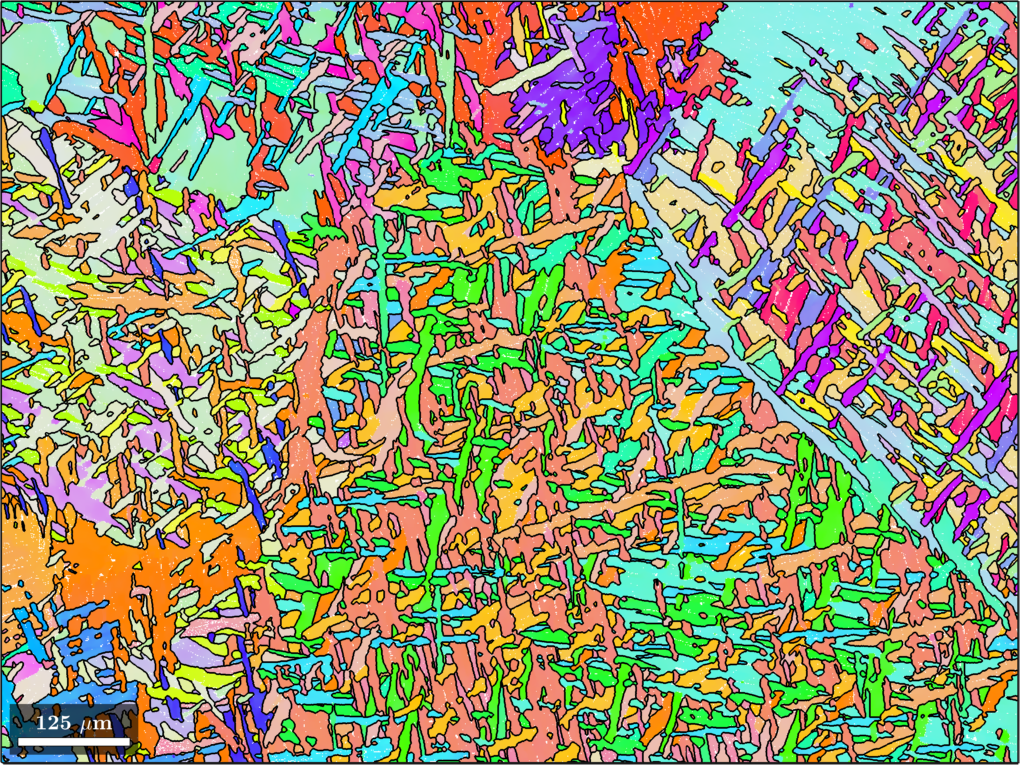}
        \caption{\label{subfig:InitialMicrostructureBW} \textbf{BW} microstructure.}
    \end{subfigure}\\
     \begin{subfigure}{0.49\textwidth}
        \centering
	    \includegraphics[width=0.95\linewidth]{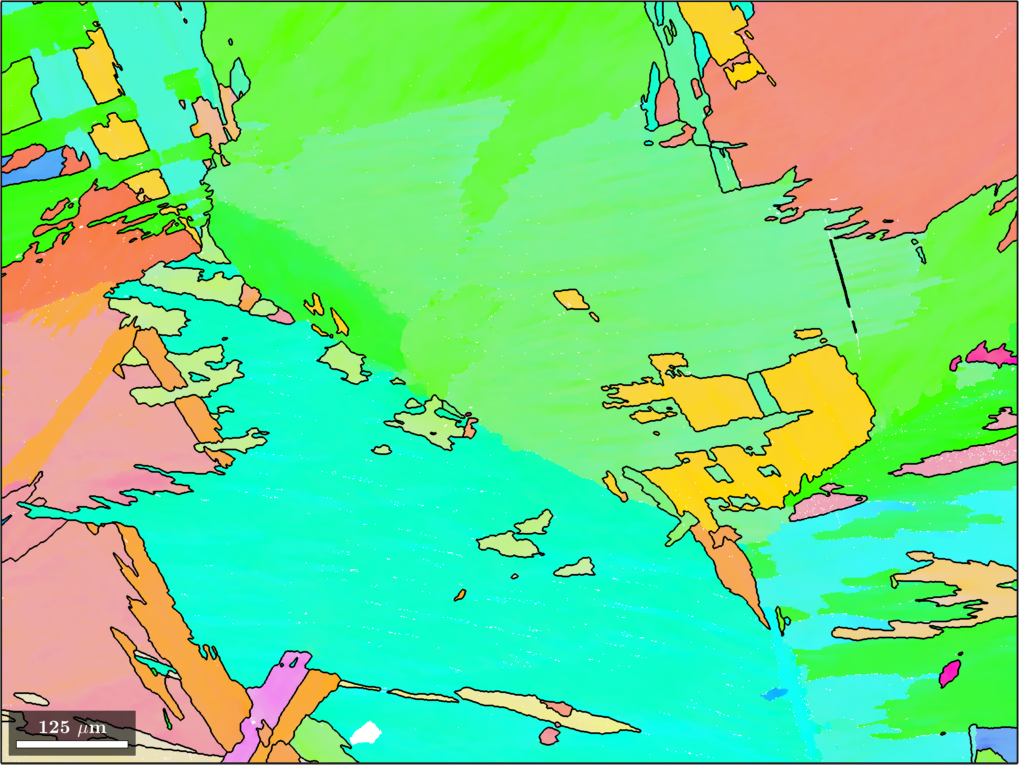}
        \caption{\label{subfig:InitialMicrostructurePP} \textbf{PP} microstructure.}
    \end{subfigure}
     \begin{subfigure}{0.49\textwidth}
        \centering
	    \includegraphics[width=0.95\linewidth]{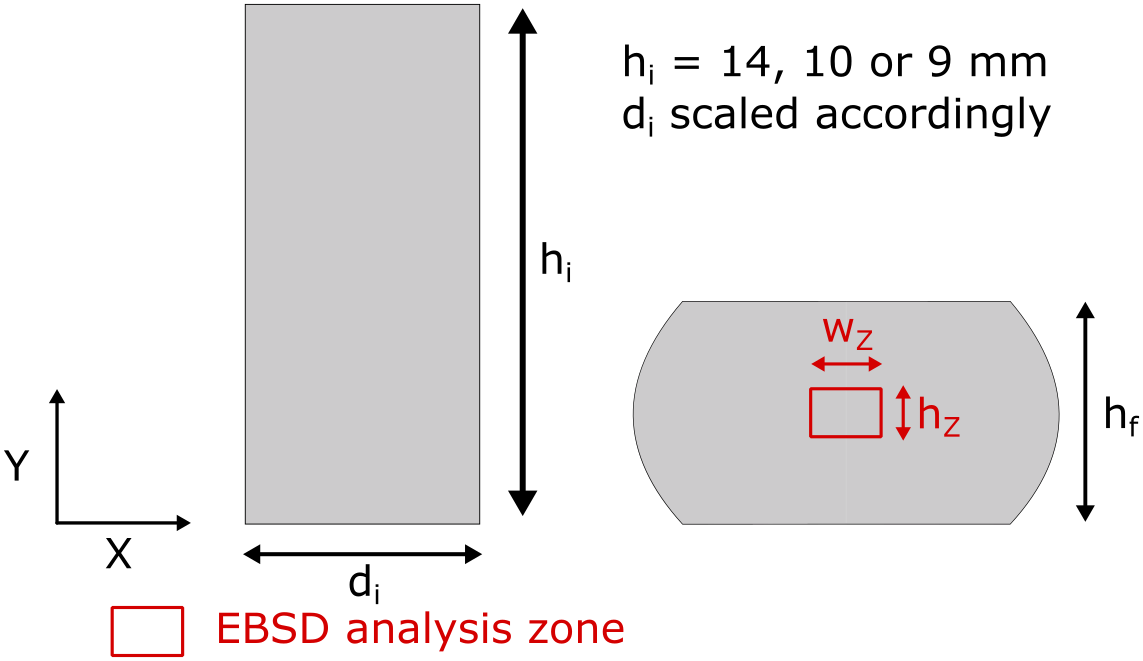}
        \caption{\label{subfig:schemeSample} Scheme sample.}
    \end{subfigure}
\caption{EBSD orientation maps representative of the three initial microstructures. Color codes are attributed thanks to an IPF Y color code, as illustrated in subfigure \ref{subfig:InitialMicrostructureEq}. The last subfigure presents sample geometry and EBSD observation zone.}
\label{fig:InitialMicrostructures}
\end{figure}

\begin{figure}
	\centering
    \begin{subfigure}[t]{0.40\textwidth}
        \centering
        \includegraphics[width=0.99\linewidth]{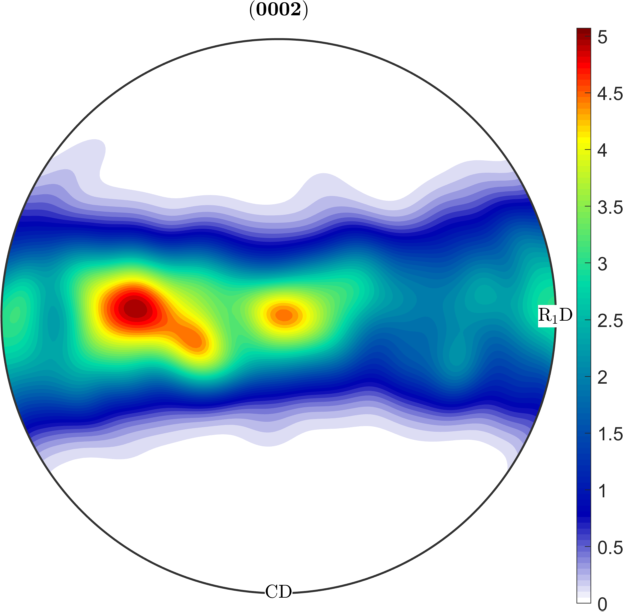}
        \caption{\label{fig:InitialEquiaxedPoleFigure_0002} $\lbrace 0002 \rbrace$ pole figure.}
    \end{subfigure}\hfill
    \begin{subfigure}[t]{0.40\textwidth}
        \centering
        \includegraphics[width=0.99\linewidth]{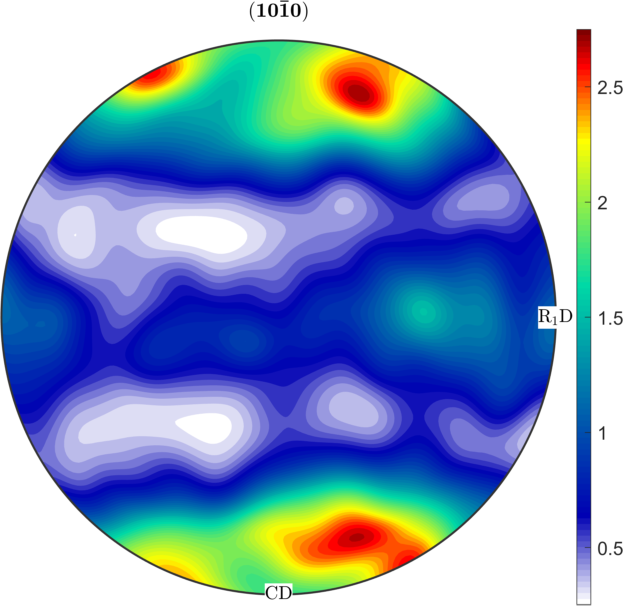}
        \caption{\label{fig:InitialEquiaxedPoleFigure_10-10} $\lbrace 10\overline{1}0 \rbrace$ pole figure.}
    \end{subfigure}\\
    \caption{\label{fig:InitialTextures} $\lbrace 0002 \rbrace$ and $\lbrace 10\overline{1}0 \rbrace$ pole figures from EBSD data.}
\end{figure}

The initial microstructures are characterized by EBSD. Orientation maps are provided in figure \ref{fig:InitialMicrostructures}. The pole figures for \textbf{Eq} microstructures are plotted in figure \ref{fig:InitialTextures}. Since the number of grains is very low for both \textbf{BW} and \textbf{PP} microstructures, the pole figures are not displayed since they are not representative of the sample global texture. Nevertheless, the orientation of the $\alpha$ grains nucleating during quench being determined by the variant selection rule, the texture of the \textbf{BW} and \textbf{PP} samples can be defined as pseudo isotropic. Figure \ref{fig:InitialTextures} shows that \textbf{Eq} sample is textured, with <c> axis being distributed within a radial plane.

\subsection{Methods}\label{subsec:Methods}

\subsubsection{Experimental methods}\label{subsubsec:ExperimentalMethods}

\paragraph{Thermomechanical testing}\label{par:ThermomechanicalTesting}

Hot compression tests, with or without subsequent heat treatment, are performed according to the conditions described in table \ref{tab:HotCompressionConditions}. They are made using a hydraulic testing machine (\textit{MTS Landmark 370-25}). A radiant furnace is used to heat the sample and dies. At the end of the experiment, the sample is quenched. The whole test is filmed in order to measure the delay between the end of the experiment and the effective quench. The average delay is equal to $1 ~s$. Samples and dies are coated with silicon nitride to limit friction. Before each test, the samples are held at temperature for $10$ minutes to ensure temperature homogeneity.

\begin{table}
\footnotesize	
    \begin{subtable}{0.95\textwidth}
        \centering
        \begin{tabular}{l c c c c c c c c c}\toprule
        $\varepsilon_{local}$ & \multicolumn{3}{c}{$\mathbf{0.01 ~ s^{-1}}$} & \multicolumn{3}{c}{$\mathbf{0.1 ~ s^{-1}}$} & \multicolumn{3}{c}{$\mathbf{1.0 ~ s^{-1}}$} \\
        \midrule
        $\mathbf{450 ^{\circ}C}$ & 0.45 & 0.75 & 1.0 & 0.45 & 0.75 & 1.0 & 0.45 & 0.75 & 1.0\\
        $\mathbf{550 ^{\circ}C}$ & 0.45 & 0.75 & 1.0 & 0.45 & 0.75 & 1.0 & 0.45 & 0.75 & 1.0\\
        $\mathbf{650 ^{\circ}C}$ & 0.45 & 0.75 & 1.0 & 0.45 & 0.75 & 1.0 & 0.45 & 0.75 & 1.0\\
        \bottomrule
        \end{tabular}
        \caption{\label{tab:TestMatrixHCR} Hot compression with equiaxed microstructures (\textbf{Eq}).}
    \end{subtable}
    \hfill
    \begin{subtable}{0.95\textwidth}
        \centering
        \begin{tabular}{l c c c c c c c c c}\toprule
        $\varepsilon_{local}$ & \multicolumn{3}{c}{$\mathbf{0.01 ~ s^{-1}}$} & \multicolumn{3}{c}{$\mathbf{0.1 ~ s^{-1}}$} & \multicolumn{3}{c}{$\mathbf{1.0 ~ s^{-1}}$} \\
        \midrule
        $\mathbf{450 ^{\circ}C}$ & 0.55 & 0.9 & 1.2 & \textcolor{gray}{\textbf{-}} & \textcolor{gray}{\textbf{-}} & 1.2 & \textcolor{gray}{\textbf{-}} & \textcolor{gray}{\textbf{-}} & \textcolor{gray}{\textbf{-}} \\
        $\mathbf{550 ^{\circ}C}$ & \textcolor{gray}{\textbf{-}} & \textcolor{gray}{\textbf{-}} & 1.2 & 0.55 & 0.9 & 1.2 & \textcolor{gray}{\textbf{-}} & \textcolor{gray}{\textbf{-}} & 1.2\\
        $\mathbf{650 ^{\circ}C}$ & \textcolor{gray}{\textbf{-}} & \textcolor{gray}{\textbf{-}} & \textcolor{gray}{\textbf{-}} & \textcolor{gray}{\textbf{-}} & \textcolor{gray}{\textbf{-}} & 1.2 & 0.55 & 0.9 & 1.2\\
        \bottomrule
        \end{tabular}
        \caption{\label{tab:TestMatrixHCVandHCP} Hot compression with lamellar microstructures (\textbf{BW} and \textbf{PP}).}
     \end{subtable}
     \begin{subtable}{0.95\textwidth}
     \centering
    \begin{tabular}{l l c c c c c c c c c c c c}\toprule
    \multicolumn{2}{c}{Holding time (s)} & \multicolumn{6}{c}{$\mathbf{0.1 ~ s^{-1}}$} & \multicolumn{6}{c}{$\mathbf{1.0 ~ s^{-1}}$} \\
    \midrule
    \multirow{2}{*}{$\mathbf{650 ^{\circ}C}$ } & \textbf{0.45} & \textcolor{gray}{\textbf{-}} & \textcolor{gray}{\textbf{-}} & \textcolor{gray}{\textbf{-}} & \textcolor{gray}{\textbf{-}} & \textcolor{gray}{\textbf{-}} & \textcolor{gray}{\textbf{-}} & 7 & 12 & 25 & 50 & 100 & 200 \\
    & \textbf{1.0} & 7 & 12 & 25 & 50 & 100 & 200 & 7 & 12 & 25 & 50 & 100 & 200 \\
    \bottomrule
    \end{tabular}
    \caption{\label{tab:TestMatrixtHCHR} Hot compression and holding at temperature with equiaxed microstructures (\textbf{Eq}).}
     \end{subtable}
     \begin{subtable}{0.95\textwidth}
     \centering
    \begin{tabular}{l l c c c c c c c c }\toprule
    \multicolumn{2}{c}{Holding time (s)} & \multicolumn{4}{c}{$\mathbf{0.1 ~ s^{-1}}$} & \multicolumn{4}{c}{$\mathbf{1.0 ~ s^{-1}}$} \\
    \midrule
    \multirow{2}{*}{$\mathbf{650 ^{\circ}C}$ } & \textbf{0.55} & \textcolor{gray}{\textbf{-}} & \textcolor{gray}{\textbf{-}} & \textcolor{gray}{\textbf{-}} & \textcolor{gray}{\textbf{-}} & 12 & 25 & 50 & 100 \\
    & \textbf{1.2} &  12 & 25 & 50 & 100 & 12 & 25 & 50 & 100\\
    \bottomrule
    \end{tabular}
    \caption{\label{tab:TestMatrixHCHVandHCHP} Hot compression and holding at temperature with lamellar microstructures (\textbf{BW} and \textbf{PP}).}
     \end{subtable}
     \caption{Description of the thermomechanical conditions of the hot compression campaigns.}
     \label{tab:HotCompressionConditions}
\end{table}

The strain levels displayed in table \ref{tab:HotCompressionConditions}, denoted $\varepsilon_{local}$, are computed from simulations performed using \textit{Forge NxT}\textsuperscript{\textregistered} software. To do so, friction coefficient is estimated by inverse analysis by measuring sample bulging. An average value for the friction coefficient is selected to compute $\varepsilon_{local}$. The values differ slightly between equiaxed and lamellar microstructures due to initial sample geometry differences. These differences being small, we estimate we can reasonably compare the results with different initial microstructures.

\paragraph{Characterization methods}\label{par:CharacterizationMethods}

After hot compression, the samples are characterized by electron back-scattered diffraction (EBSD) analysis. A \textit{Carl Zeiss SUPRA 40} field emission gun scanning electron microscope (FEG-SEM) equipped with a \textit{Bruker Quantax} system (EBSD \textit{e$^-$Flash$^{HR}$}) is used. Acceleration voltage is set to $20 ~ kV$ and step size to $100 ~ nm$. Each map is represented according to the sample reference frame with $Y$ being the compression direction and $X$ the radial one. The standard color code used for attributing colors to orientations is obtained from inverse pole figure color code in which $Y$ axis is projected in the standard triangle. The standard triangle corresponds to the crystal axis system reduced thanks to symmetry operations. These two conventions are presented in figure \ref{subfig:InitialMicrostructureEq}. 

Samples are cut such as the compression direction lies within the cutting plan. The observation zone is located at sample center. Its dimensions are: $130 \times 85 ~ \mu m^2$.

Post-processing of EBSD data is performed using MTEX toolbox \cite{Bachmann2011}. A half quadratic filter is applied before any post-processing operation to reduce the noise \cite{Hielscher2019}. If data are used as input for simulations, a filling operation is applied. Missing orientation data at non-indexed pixels are interpolated using the half quadratic filter.

Isolated indexed pixels are filtered out. To do so, a minimal number of 10 pixels per grain is considered. It corresponds to areas that are $300 ~ nm$ wide. Threshold angle values are set to $3 ^\circ$ for low angle grain boundaries ($\Delta \theta_{\textsc{lagb}} = 3 ^\circ$) and $15 ^\circ$ for high angle grain boundaries ($\Delta \theta_{\textsc{hagb}} = 15 ^\circ$). Geometrically necessary dislocation (GND) density is estimated using the method established by Pantleon \cite{Pantleon2008}. A grain is defined as recrystallized if its average GND density is lower than $10^{14} ~ m^{-2}$. GND density is preferred over other grain properties related to grain internal disorientation (grain orientation spread, GOS, or grain average kernel average misorientation, GAKAM, for instance). It presents the advantage of being directly transposable in simulations.

\subsubsection{Numerical tools}\label{subsubsec:NumericalTools}

A LS formulation, embedded within a FE framework is used to model CDRX and PDRX. LS functions ($\phi(x,t)$) track the position of interfaces with time \cite{Merriman1994, Bernacki2008, Bernacki2011}. They are initialized as signed Euclidean distance functions to grain boundaries (GB):

\begin{eqnarray}  
   	\begin{cases}
      \phi(x,t) = \pm ~ d\left( x, \Gamma(t) \right) , ~ x \in \Omega, \\
      \Gamma(t) = \lbrace x \in \Omega, ~ \phi(x,t) = 0 \rbrace,
    \end{cases}
\label{eq:InitLSfunction}
\end{eqnarray}

where $d$ is the signed Euclidean distance function and $\Omega$ the simulation domain. By convention, $\phi(x,t)$ is taken positive inside the grain and negative outside.

The movement of GB is predicted by solving the following transport equation:

\begin{eqnarray}
 \dfrac{\partial \phi}{\partial t} + \overrightarrow{v}\cdot \overrightarrow{\nabla\phi} = 0,
\label{eq:TransportEquation}       
\end{eqnarray}

with $\overrightarrow{v}$ being the velocity of interfaces. After each resolution, level-set functions are reinitialized to ensure LS functions remain signed distance functions \cite{Shakoor2015}. Voids and overlaps are corrected according to the method described by Merriman \textit{et al.} \cite{Merriman1994}.

GB velocity at mesoscopic scale can be defined as the sum of surface and volume energy effects. The term related to capillarity is computed naturally thanks to distance function properties:

\begin{eqnarray}
\overrightarrow{v}_c = -M \left( \overrightarrow{\nabla \gamma} \cdot \overrightarrow{n} - \gamma \kappa \right) \overrightarrow{n},
\end{eqnarray}

with $\overrightarrow{n} = - \overrightarrow{\nabla \phi}$, $\kappa = -\Delta \phi$ and $\gamma$ the GB energy. The way to extend $\gamma$ in the vicinity of GB interfaces to evaluate $\overrightarrow{\nabla \gamma}$ and to take into account this velocity through a convective-diffusive resolution of the transport LS equation (Eq.\ref{eq:TransportEquation}) are detailed in \cite{Fausty2020,Murgas2021}.

The other, related to differences of stored energy across GB, requires a dedicated treatment. Initially, Bernacki et al. \cite{Bernacki2008} proposed to weight the contribution of each grain stored energy by a function of the distance to the GB. This is done using equation \ref{eq:VelocityStoredEnergyBernacki}.

\begin{eqnarray}
\overrightarrow{v}_{e} \left( x, t \right) =  \sum_{i=1}^{N_d} \sum_{j=1}^{N_d} M_{ij} \chi_i(x,t)  f(\phi_i(x,t),l) \times \left[ E_j(x, t)-E_i(x, t) \right] \overrightarrow{n},
\label{eq:VelocityStoredEnergyBernacki}
\end{eqnarray}

with $\chi_i(x,t)$ the characteristic function of the $i^{th}$ LS function, i.e. $\chi_i(x,t) = 1$ where the LS function is positive and $\chi_i(x,t) = 0$ everywhere else. $f$ is a decreasing function being equal to 1 for $\phi_i = 0$ and 0 for $\phi_i = l$. To limit the number of LS functions in order to save computation time and memory, graph coloring and recoloring techniques have been developed to describe numerous non neighboring grains in each LS function \cite{Scholtes2015}. The set of all distance functions required to describe all the grains is now: $\Phi = \lbrace \phi_i, ~ i = 1, ~ ..., ~ N_d\rbrace$, with $N_d$ the number of distance functions, greatly smaller than $N_g$ the number of grains. This implementation implies the use of $N_d$ energy fields, that are initialized as: $E_i(x,0) = \chi_i(x,0) \times E(x,0) = \tau\rho_i(x,0)$, with $\tau = \frac{1}{2}\mu b^{2}$ the dislocation line energy, $b$ the norm of the Burgers vector and $\mu$ the shear modulus. These energy fields are tracked and updated to ensure they are consistent with the LS functions and the exact location of interfaces. 

Some additional details about this method are provided in several articles \cite{Bernacki2008, Maire2017} and an extension is proposed in the following \textbf{Stored energy gradients} paragraph. The implementations dedicated to CDRX are described extensively in ref. \cite{Grand2022}.

\paragraph{Low angle boundary characteristics}

CDRX is characterized by a high LAGB fraction. As a consequence, to model such phenomenon, it is essential to correctly describe the formation and properties of such interfaces. Formation of new LAGB is addressed using equations from the model developed by Gourdet and Montheillet \cite{Gourdet2003}. At each deformation increment $d\varepsilon$, the quantity of LAGB formed is described according to equation \ref{eq:SurfaceNewSubgrains}.

\begin{eqnarray}
    dS^+ = \dfrac{\alpha b K_2 \rho d\varepsilon}{\eta \theta_0},
    \label{eq:SurfaceNewSubgrains}       
\end{eqnarray}

where $dS^+$ refers to the length (respectively surface) of boundaries formed for 2D simulations (respectively 3D) and $\alpha = 1- \exp{\left( \dfrac{D}{D_0} \right) ^m}$ is a coefficient describing the fraction of dislocations recovered to form new subgrains. $D$ is the ECD, $D_0$ is a reference ECD and $m$ is a fixed coefficient. $K_2$ is the recovery parameter of the Yoshie-Laasraoui-Jonas equation \cite{Laasraoui1991}, $\rho$ the dislocation density,
$\eta$ is the number of sets of dislocations and $\theta_0$ the disorientation of newly formed subgrains.

The pre-existing LAGB experience a progressive disorientation prescribed by:

\begin{eqnarray}
    d\theta = \dfrac{b}{2 \eta} \left(1-\alpha\right) D K_2 \rho d\varepsilon.
    \label{eq:ProgressiveMisorientation}
\end{eqnarray}

The GB mobility is assumed isotropic. Its evolution with temperature is predicted using an Arrhenius relation, such as $M = M_0 \times \exp\left(\frac{-Q_m}{RT}\right)$, with $M_0$ the pre-exponential mobility factor and $Q_m$ the mobility apparent activation energy. The GB energy is described according to Read-Shockley equation \cite{Read1950}:

\begin{eqnarray}
\gamma  (\theta)  =
   	\begin{cases}
     \gamma_{max} \left( \dfrac{\theta}{\theta_{max}}\right)\left(1- \ln{\dfrac{\theta}{\theta_{max}}}\right) , ~ \theta < \theta_{max}, \\
      \gamma_{max}, ~ \theta \geq \theta_{max}.
    \end{cases}\label{eq:ReadSchockley}
\end{eqnarray}

$\gamma_{max}$ is the HAGB energy and $\theta_{max}$ the LAGB/HAGB disorientation threshold, set to $15^{\circ}$.

\paragraph{Second phase particle (SPP) pinning effect}

At each SPP position, remeshing operations are performed and elements removed. By imposing a null Neumann boundary conditions at mesh boundaries for our FE-LS resolution, the holes representing SPP are naturally exerting a tension on GB that are consistent with the Young-Herring equilibrium for incoherent particles \cite{Agnoli2014, Agnoli2012, Scholtes2015, Villaret2020}.

\paragraph{Hardening and recovery}

The evolution of dislocation density due to deformation is taken into account using Yoshie-Laasraoui-Jonas equation \cite{Laasraoui1991}:

\begin{eqnarray}
    d\rho = \left( K_1 - K_2 \rho  \right)d\varepsilon.
    \label{eq:YLJ}
\end{eqnarray}

$K_1$ is the hardening parameter. In the present study, $K_1$ varies from one grain to another. For each initial digital microstructure, it is set according to a distribution measured experimentally.

Evolution of $\left(K_1, ~ K_2\right)$ with 
thermomechanical conditions is described using equations \ref{eq:K1_functionThermomechanicalConditions} and \ref{eq:K2_functionThermomechanicalConditions} \cite{Montheillet2009}. The yield stress, $\sigma_0$, is described as a linear function of the Zener-Hollomon parameter logarithm (equation \ref{eq:sigma0_functionZenerHollomon}) \cite{Jonas1969}.

\begin{eqnarray}
K_1 = K_{1}^0 \dot{\varepsilon}^{m_h} \exp\left(\frac{m_h Q_h}{RT}\right), \label{eq:K1_functionThermomechanicalConditions}\\
K_2 = K_{2}^0 \dot{\varepsilon}^{-m_r} \exp\left(-\frac{m_r Q_r}{RT}\right),
\label{eq:K2_functionThermomechanicalConditions}\\
\sigma_0 = \sigma_{0}^s \times \ln \left( Z \right) + \sigma_{0}^i.
\label{eq:sigma0_functionZenerHollomon}
\end{eqnarray}

where $\sigma_{0}^s$ and $\sigma_{0}^i$ designate the slope and the intercept of the linear function, respectively.

\paragraph{Stored energy gradients}

\begin{figure}[h!]
\centering
\begin{subfigure}{0.49\textwidth}
  \centering
  \includegraphics[width=1.0\linewidth]{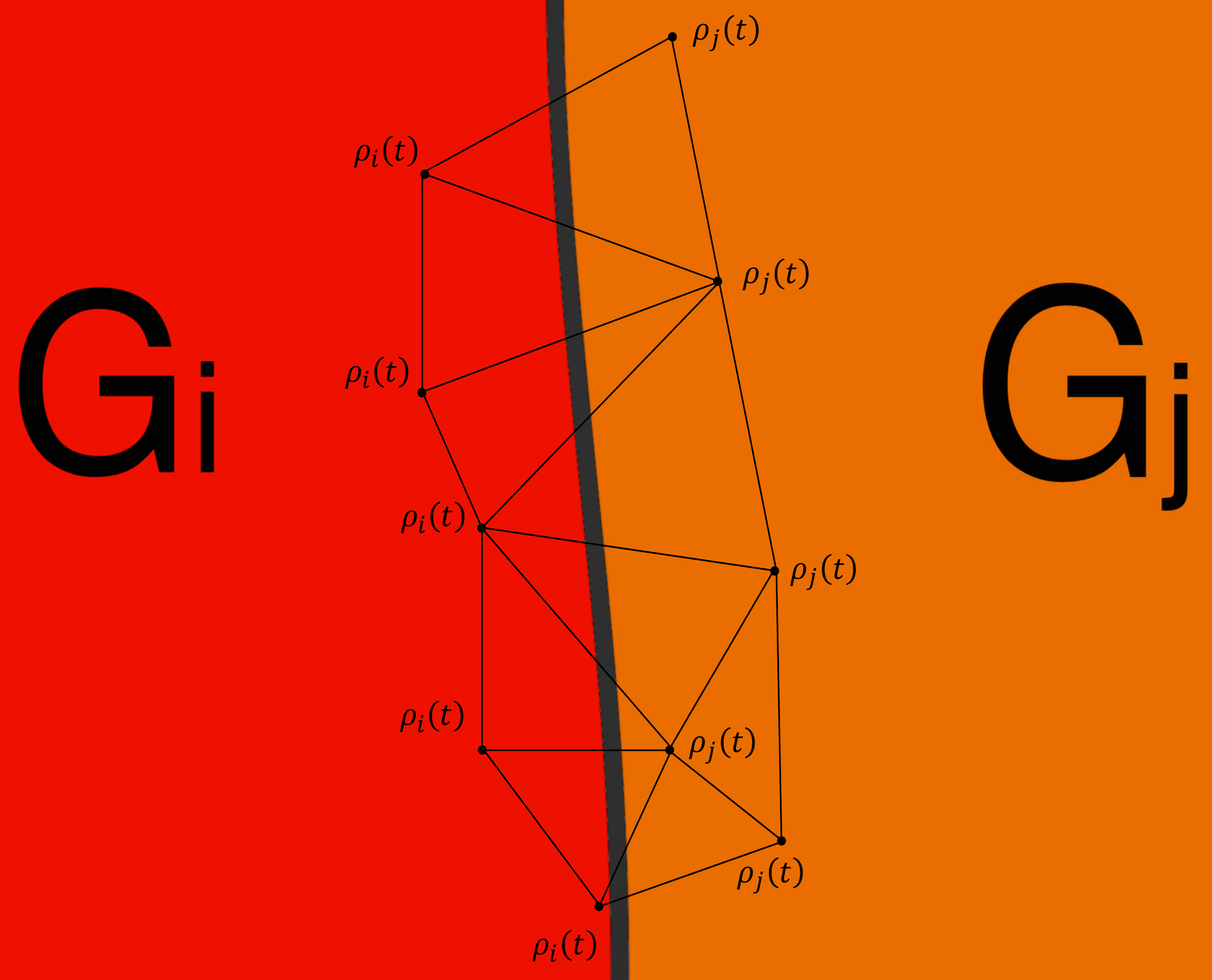}  
\end{subfigure}
\begin{subfigure}{0.49\textwidth}
  \centering
  \includegraphics[width=1.0\linewidth]{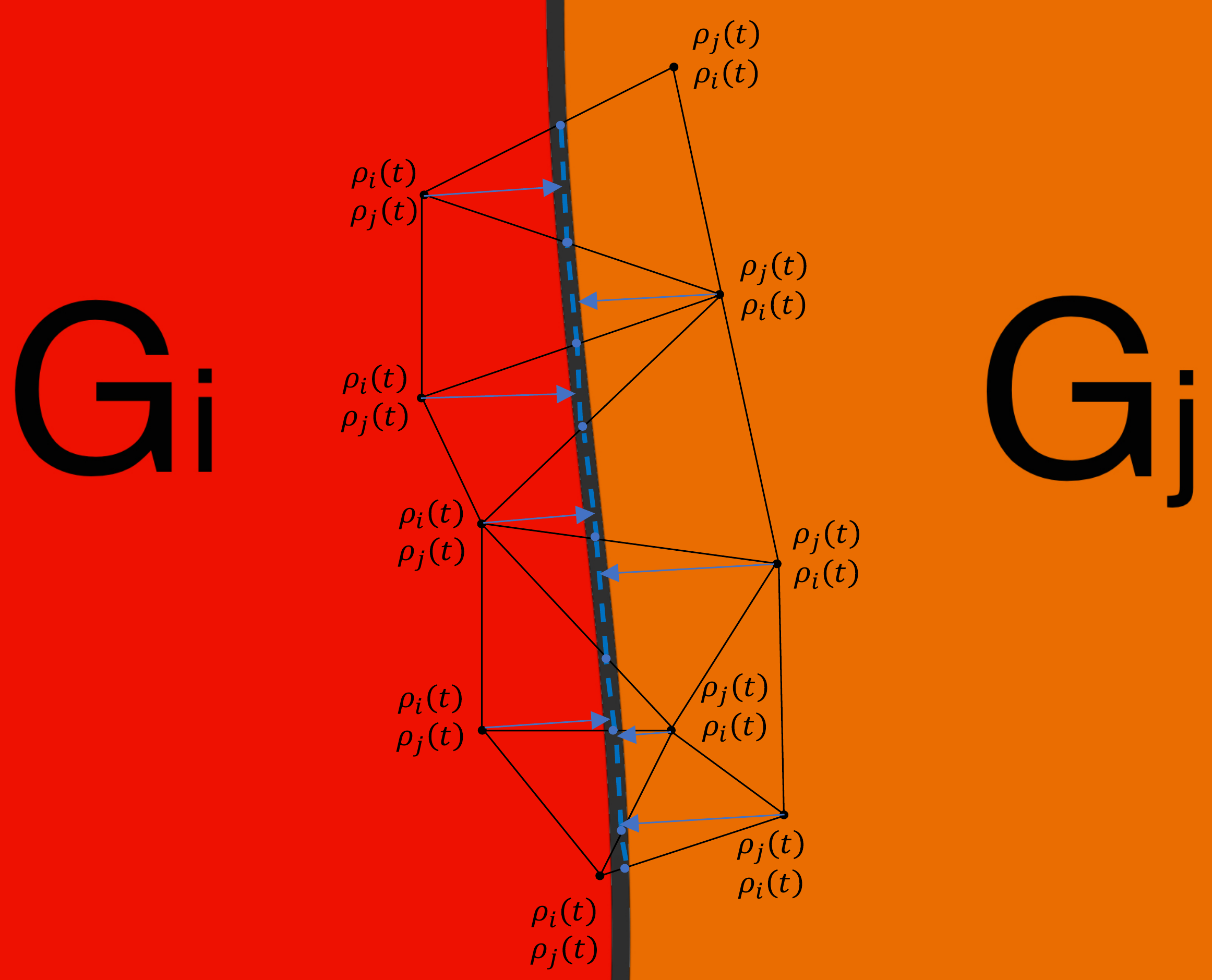} 
\end{subfigure}\\
\begin{subfigure}{0.49\textwidth}
  \centering
  \includegraphics[width=1.0\linewidth]{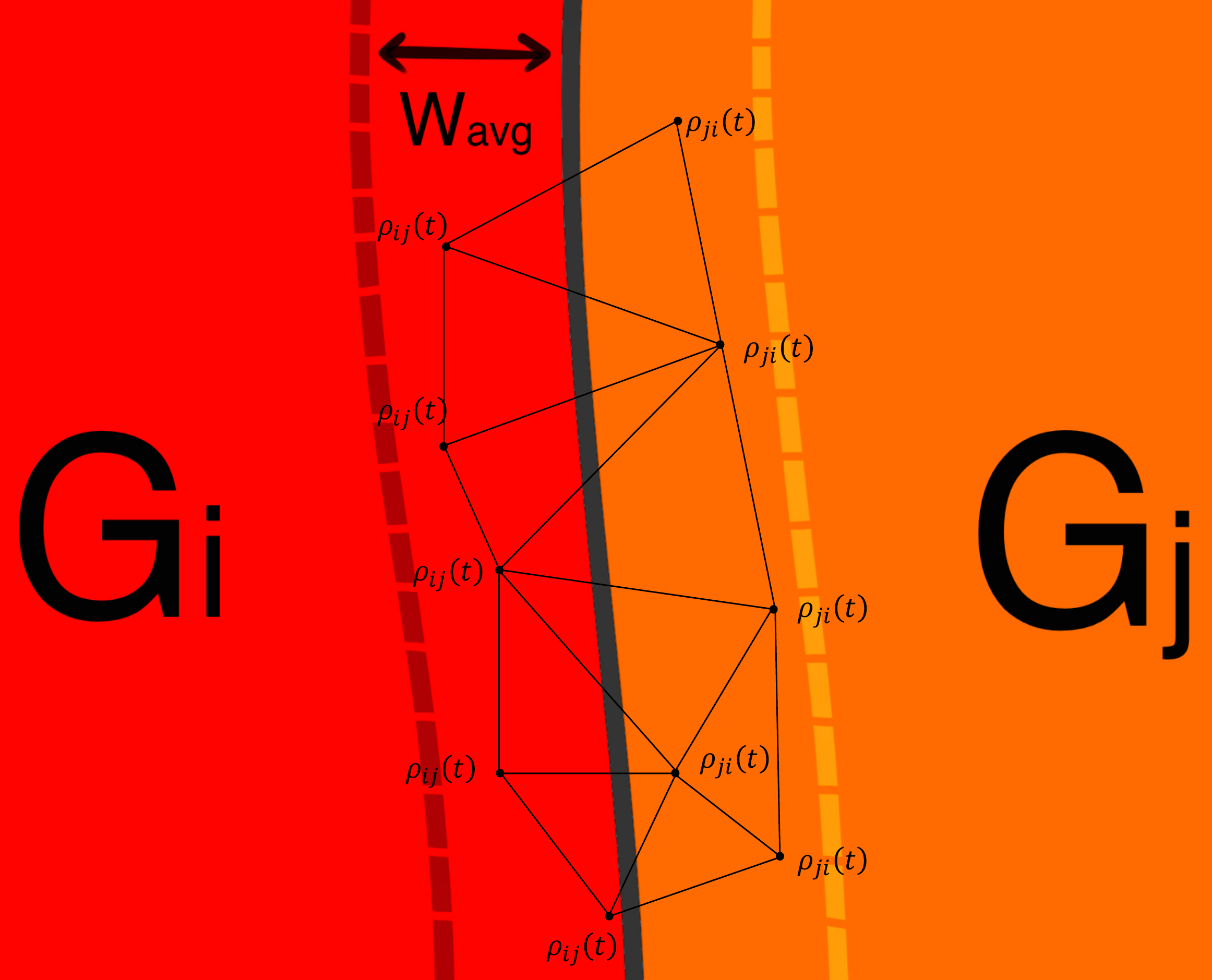}  
\end{subfigure}
\begin{subfigure}{0.49\textwidth}
  \centering
  \includegraphics[width=1.0\linewidth]{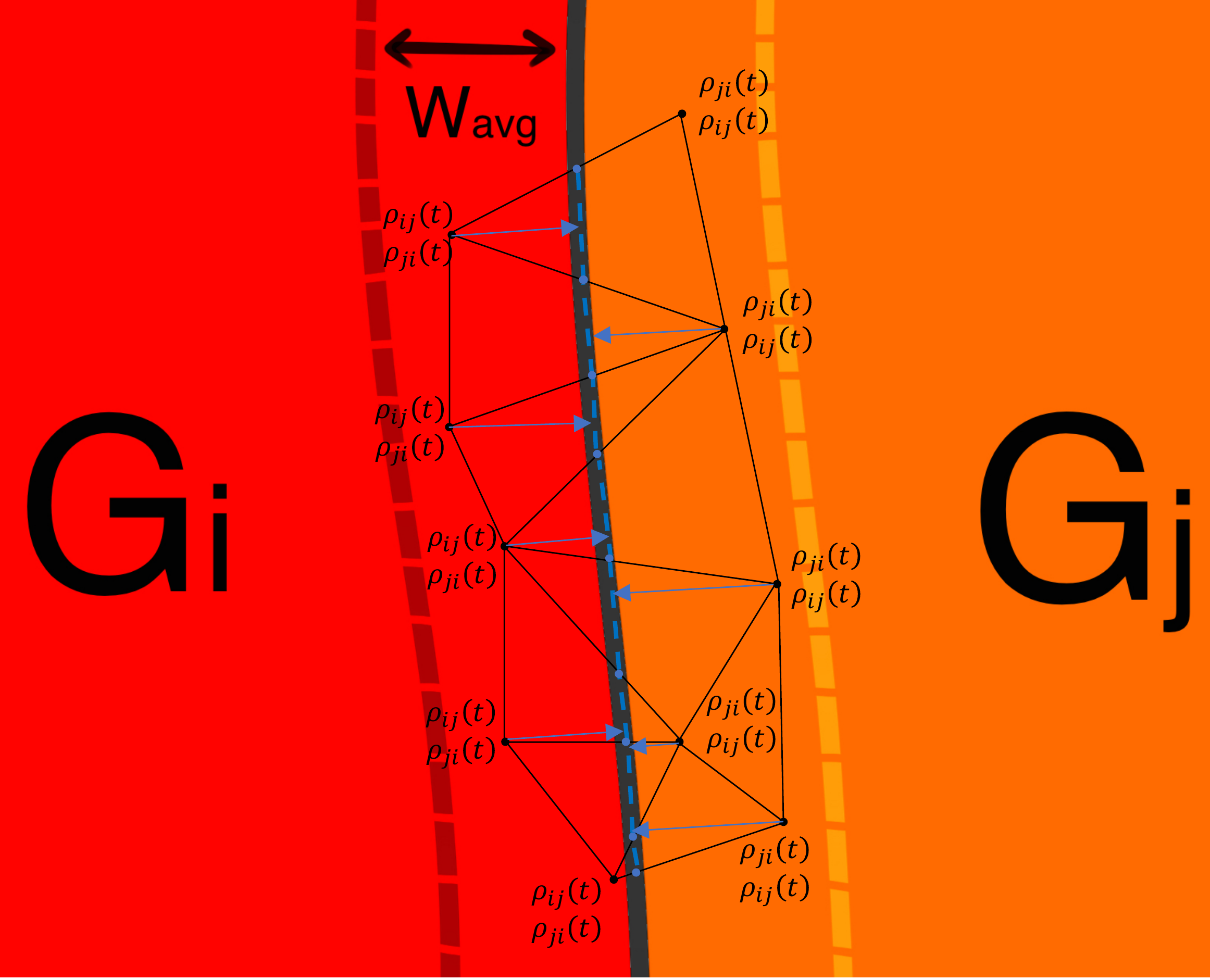}  
\end{subfigure}
\begin{subfigure}{0.49\textwidth}
  \centering
  \includegraphics[width=1.0\linewidth]{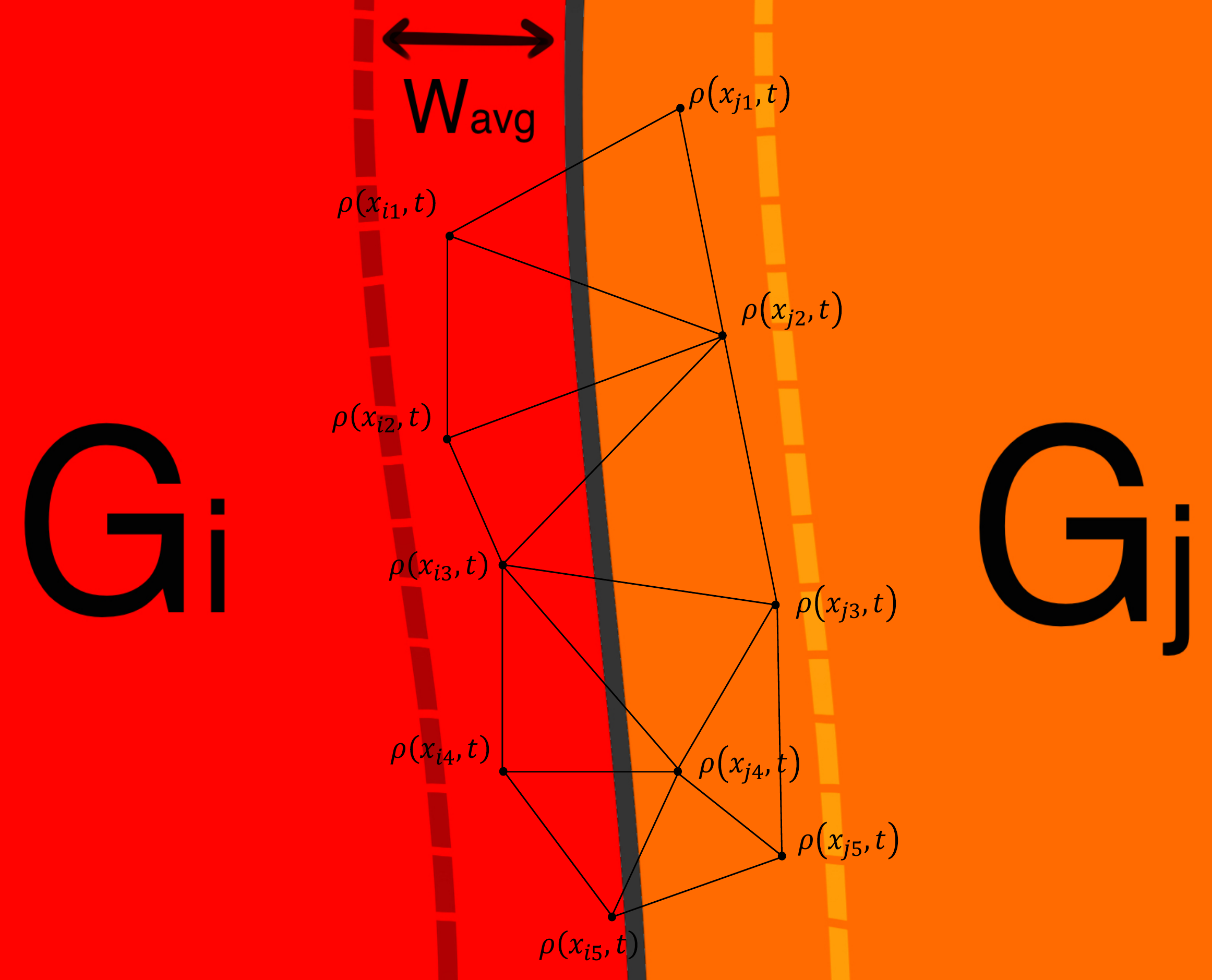}  
\end{subfigure}
\begin{subfigure}{0.49\textwidth}
  \centering
  \includegraphics[width=1.0\linewidth]{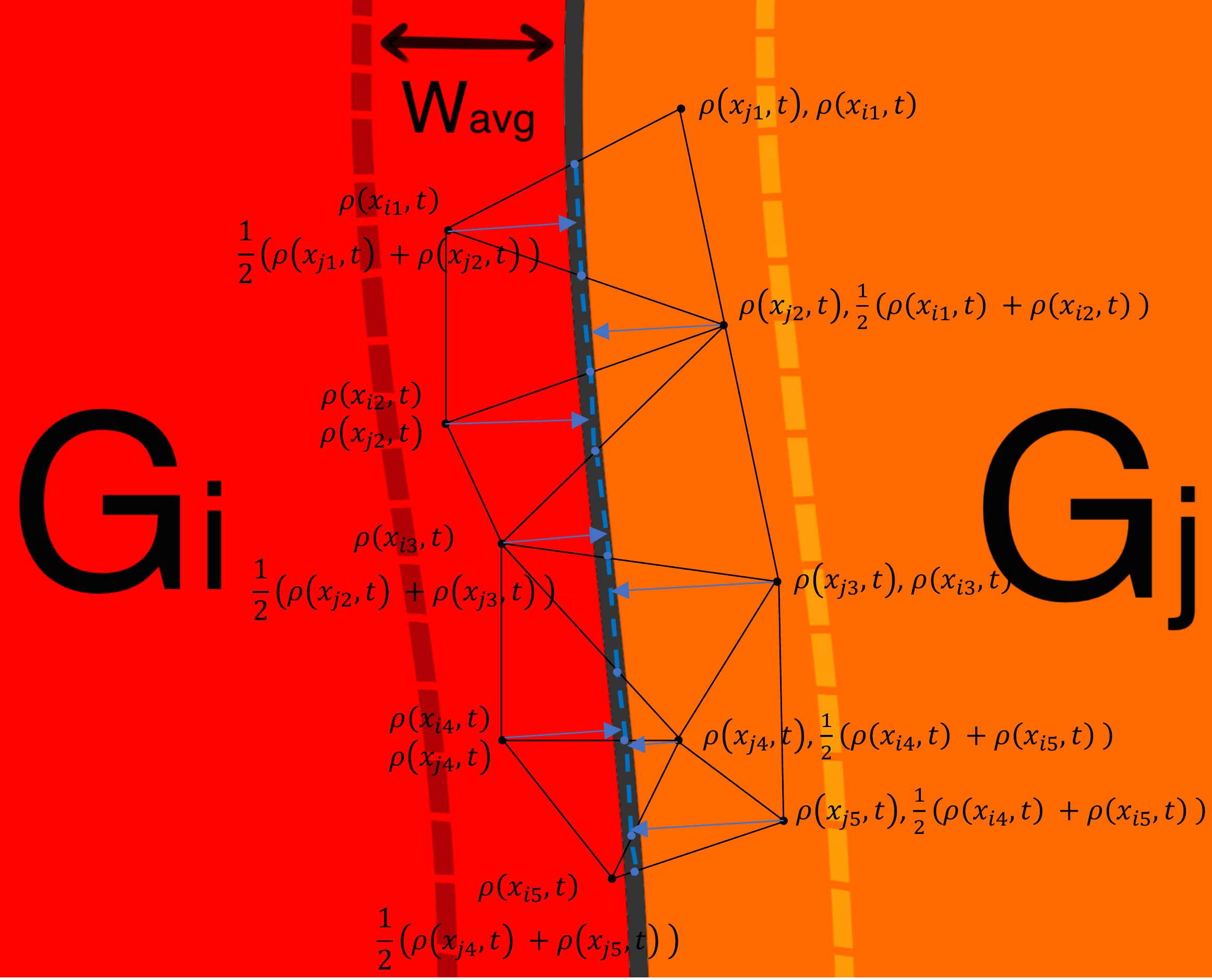}  
\end{subfigure}
\caption{Illustrating schemes concerning the evaluation of $E_j(x, t)-E_i(x,t)$ around the GB interfaces: (left side) available information per FE node before the reinitialization procedure, and (right side) after the reinitialization of $\phi_i$ and $\phi_j$.(top) Energy averaged per grain, (middle) energy averaged per interface and (bottom) not averaged.}
\label{fig:schemeenergy}
\end{figure}

Full field front-capturing recrystallization models such as the LS one or the multi phase-field approaches generally consider a unique and uniform value per grain for stored energy of deformation. However, in reality, dislocation density can vary significantly within a grain. Since the impact of localized and intense gradients over GB migration remains difficult to quantify, modeling it at mesoscopic scale is not straightforward. In context of LS method, Ilin \textit{et al.} proposed to consider a constant GB velocity per interface \cite{Ilin2018}. To do so, dislocation density is averaged at GB vicinity. The distance to the interface in which dislocation density is averaged ($w_{avg}$) is a parameter. Another modeling strategy is also considered here. This strategy consists in  computing locally, at each FE node, the energy difference. The different approaches are illustrated in Figure \ref{fig:schemeenergy}. The left side of this figure illustrates the initial configurations where for each grain, the energy is averaged per grain (top), per interface (middle) or not averaged (bottom). Few finite elements in black are described to illustrate the energy fields in the corresponding finite element nodes. The different notations used in these figures are described by the following equations:
$$
E_k\left(t\right) = \tau\rho_k\left(t\right)=\tau\int_{G_k}\rho\left(\bold{x},t\right)d\bold{x}, \quad 
$$
$$
E_{kl}\left(t\right) = \tau\rho_{kl}\left(t\right)=\tau\int_{G_{kl}}\rho\left(\bold{x},t\right)d\bold{x},
$$
with,
$$
G_{k}\left(\bold{x},t\right) = \lbrace \bold{x}\in\Omega, \phi_k\left(\bold{x},t\right)\geq 0\rbrace,\ G_{kl}\left(\bold{x},t\right) = \lbrace \bold{x}\in\Omega, min\left(\phi_k\left(\bold{x},t\right),\phi_l\left(\bold{x},t\right)+w_{avg}\right) \geq 0\rbrace
$$
At this stage, for the three strategies, the energy fields of each grain is perfectly known in the corresponding grain. However these fields must be extended outwards their respective LS functions to be able to compute the velocity field defined in Eq.\ref{eq:VelocityStoredEnergyBernacki}. The direct reinitialization algorithm introduced in \cite{Shakoor2015} computes, for each node, the distance to the nearest facet (segment in 2D, triangle in 3D) constituting the piece-wise linear 0-isovalue of the LS function as illustrated by the blue vectors in the right side of the Fig.\ref{fig:schemeenergy}. These facets which are not used in the FE resolution are also illustrated in blue (dotted segments) in the right side of the Fig.\ref{fig:schemeenergy}. This algorithm has been modified and now returns also all the associated features of the crossed element. Thus, when the $\phi_i$ function is reinitialized, the field describing the stored energy is extended in a fixed thickness in the neighboring grains. The right side of the Fig.\ref{fig:schemeenergy} illustrates, by using this procedure, all the information shared by the FE nodes around the interface when the energy is averaged per grain (top), per interface (middle) and not averaged (bottom). These information can then be used to evaluate, in each FE node, the terms $E_j(x, t)-E_i(x,t)$ of Eq.\ref{eq:VelocityStoredEnergyBernacki}.

To evaluate the effect of each implementation, five different approximation cases will be considered:
\begin{itemize}
\setlength\itemsep{-0.1em}
\item dislocation density averaged per grain,
\item dislocation density averaged per interface up to $100 ~ nm$ from the GB (\textit{Per interface;} $w_{avg}=100 ~ nm$),
\item dislocation density averaged per interface up to $200 ~ nm$ from the GB (\textit{Per interface;} $w_{avg}=200 ~ nm$).
\item Local dislocation density (\textit{Local}),
\item local dislocation density, with a pre-processing operation that consists in applying a 2D Gaussian filter to the dislocation density map. The standard deviation of the Gaussian filter is taken equal to $2 ~ px = 200 ~ nm$ (\textit{Local, Gaussian filtered}).
\end{itemize}

\section{Results}\label{sec:Results}

\subsection{Characterization of CDRX and PDRX}

\begin{figure}[h!]
\centering
\begin{subfigure}{0.49\textwidth}
  \centering
  \includegraphics[width=1.0\linewidth]{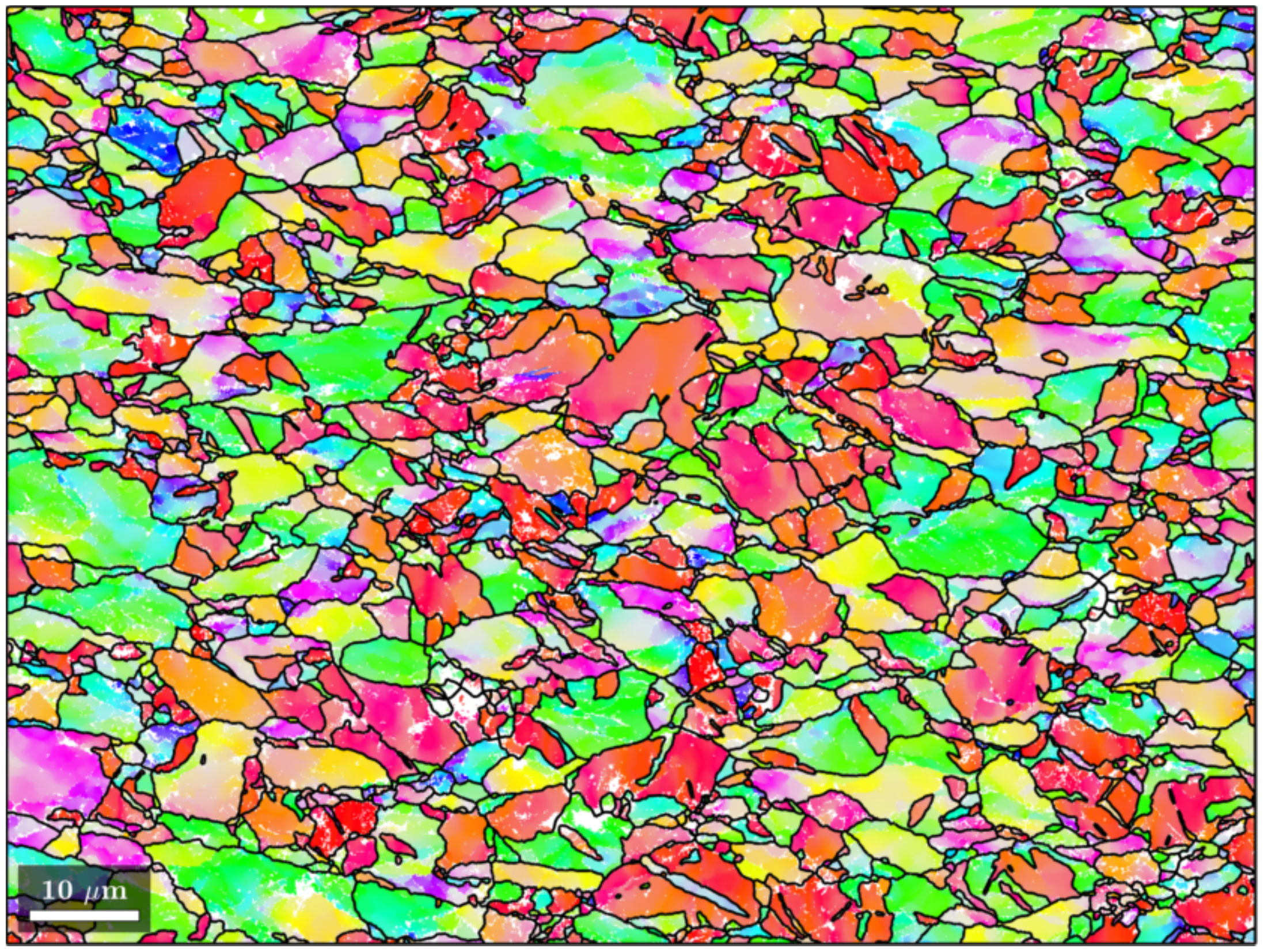}  
  \caption{\label{fig:HCR10_IPF_EBSD} $\varepsilon = 0.45$.}
\end{subfigure}
\begin{subfigure}{0.49\textwidth}
  \centering
  \includegraphics[width=1.0\linewidth]{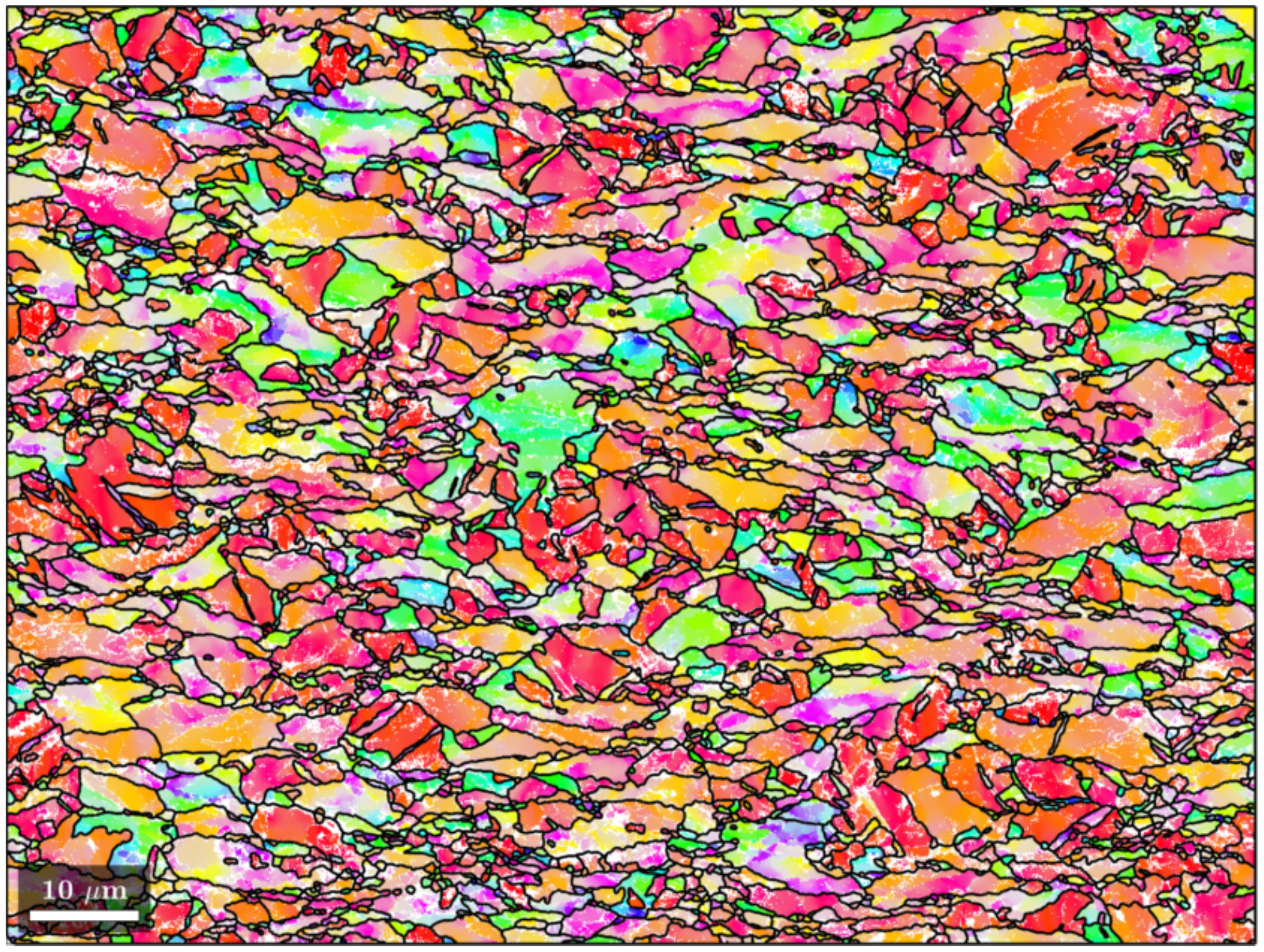} 
  \caption{\label{fig:HCHR11_IPF_EBSD} $\varepsilon = 0.75$.}
\end{subfigure}\\
\begin{subfigure}{0.49\textwidth}
  \centering
  \includegraphics[width=1.0\linewidth]{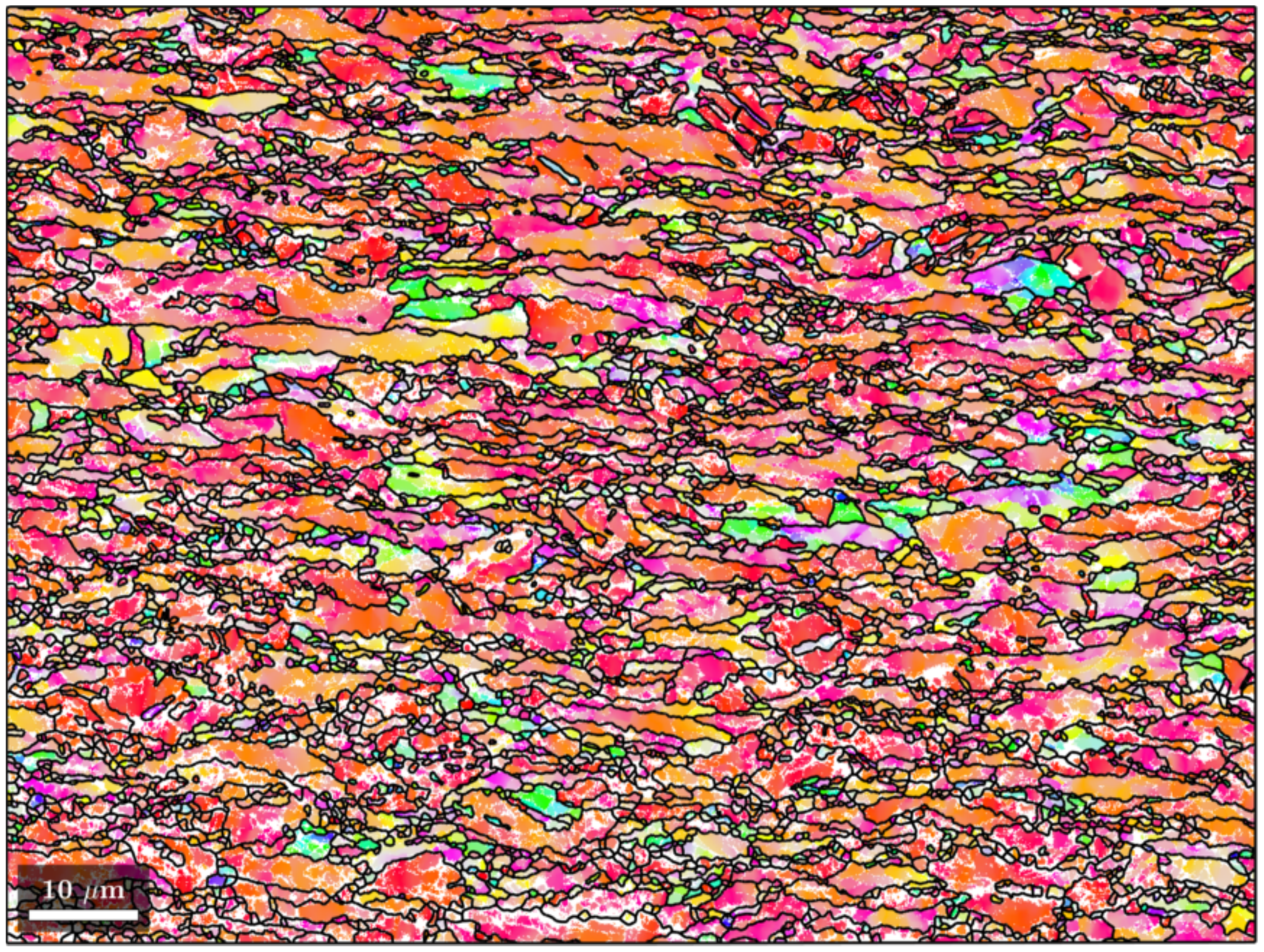}  
  \caption{\label{fig:HCHR12_IPF_EBSD} $\varepsilon = 1.0$.}
\end{subfigure}
\caption{EBSD orientation maps showing microstructure evolution during hot deformation ($T = 550^{\circ}C ; ~ \dot{\varepsilon} = 0.1 ~ s^{-1}$). \textbf{Eq} samples. IPF Y color code.}
\label{fig:EBSD_IPF_CDRX_InfluenceStrain}
\end{figure}

Orientation maps corresponding to samples hot deformed at three deformation levels are provided in figure \ref{fig:EBSD_IPF_CDRX_InfluenceStrain}. These figures confirm that to accommodate deformation, intragranular orientation gradients form progressively and lead to the development of subgrains. With deformation, grains become more and more elongated along horizontal direction and the number of small grains tend to increase notably. These small grains seem to be grouped in clusters.

\begin{figure}[h!]
\centering
\begin{subfigure}{0.49\textwidth}
  \centering
  \includegraphics[width=1.0\linewidth]{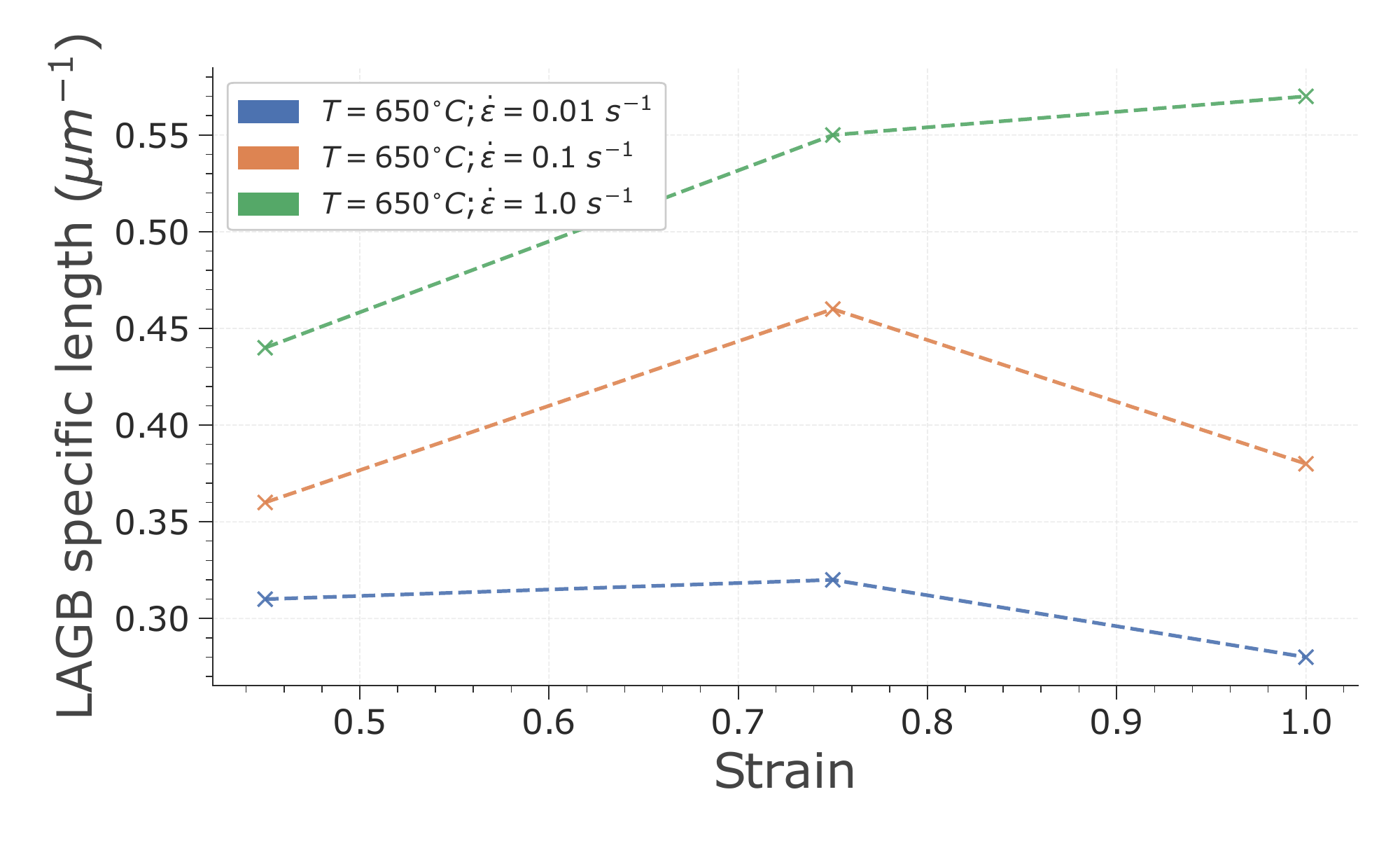} 
  \caption{\label{fig:LAGBlength_VS_Strain} LAGB specific length versus strain.}
\end{subfigure}
\begin{subfigure}{0.49\textwidth}
  \centering
  \includegraphics[width=1.0\linewidth]{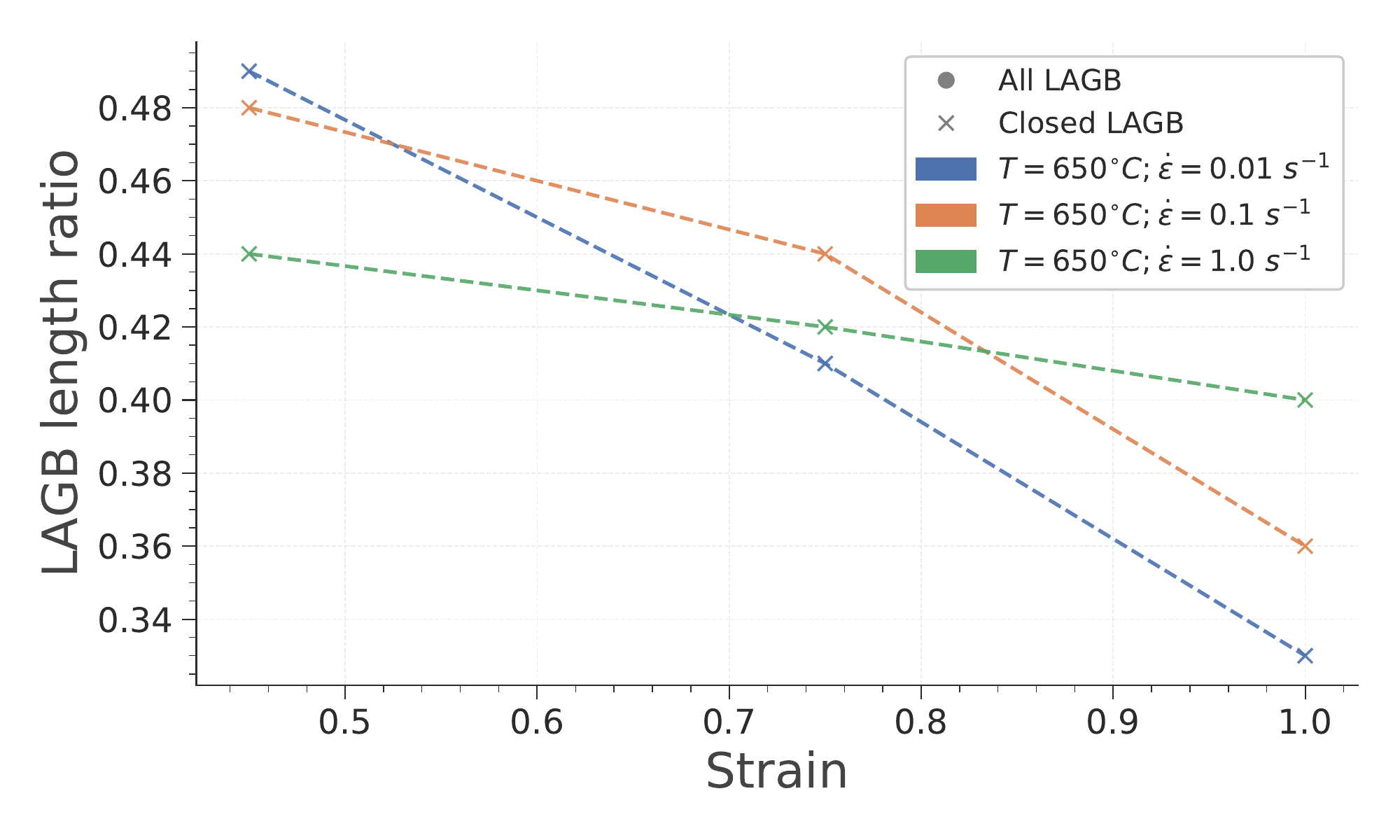} 
  \caption{\label{fig:LAGBlengthRatio_VS_Strain} LAGB length ratio versus strain.}
\end{subfigure}
\begin{subfigure}{0.49\textwidth}
  \centering
  \includegraphics[width=1.0\linewidth]{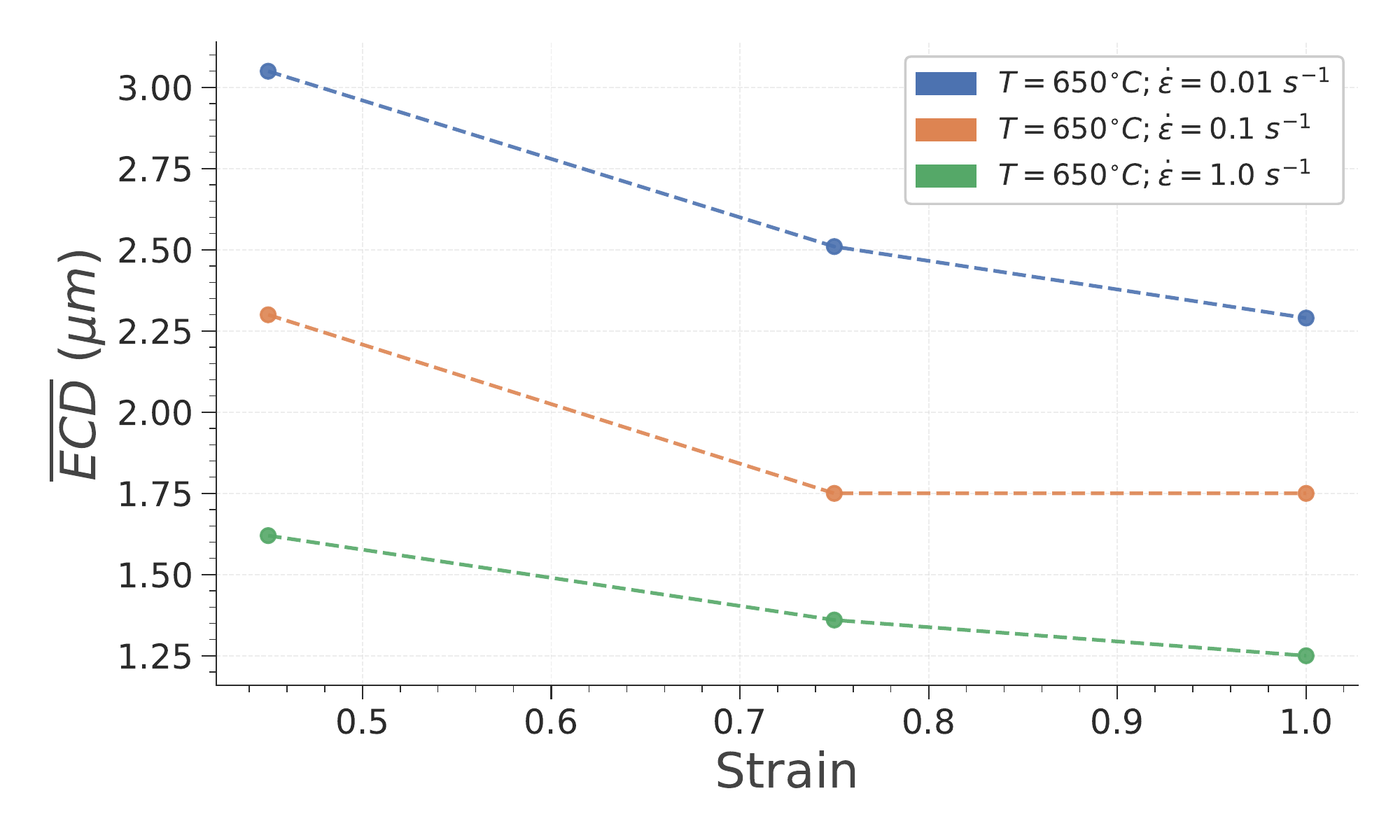}  
  \caption{\label{fig:ECD_VS_Strain} $\overline{ECD}$ versus strain.}
\end{subfigure}
\caption{Evolution of the microstructure topology during hot deformation at $T = 650^{\circ}C$. \textbf{Eq} microstructure.}
\label{fig:CDRX_EQ_GBparameters}
\end{figure}

To analyze quantitatively the impact of hot deformation upon microstructure topology, the evolution of LAGB specific length and of grain $\overline{ECD}$ are provided in figure \ref{fig:CDRX_EQ_GBparameters}. Figure \ref{fig:LAGBlengthRatio_VS_Strain} confirms the significant presence of LAGB. For all the conditions displayed here, at least $34 \%$ of the GB are LAGB. If the LAGB specific length generally increases with strain, it seems to stabilize at high strain levels. The LAGB length ratio decreases with strain, which indicates that the HAGB specific length increases at a faster rate. Finally, figure \ref{fig:ECD_VS_Strain} shows that the grain size decreases significantly with deformation.

\begin{figure}[h!]
	\centering
    \includegraphics[width=0.6\linewidth]{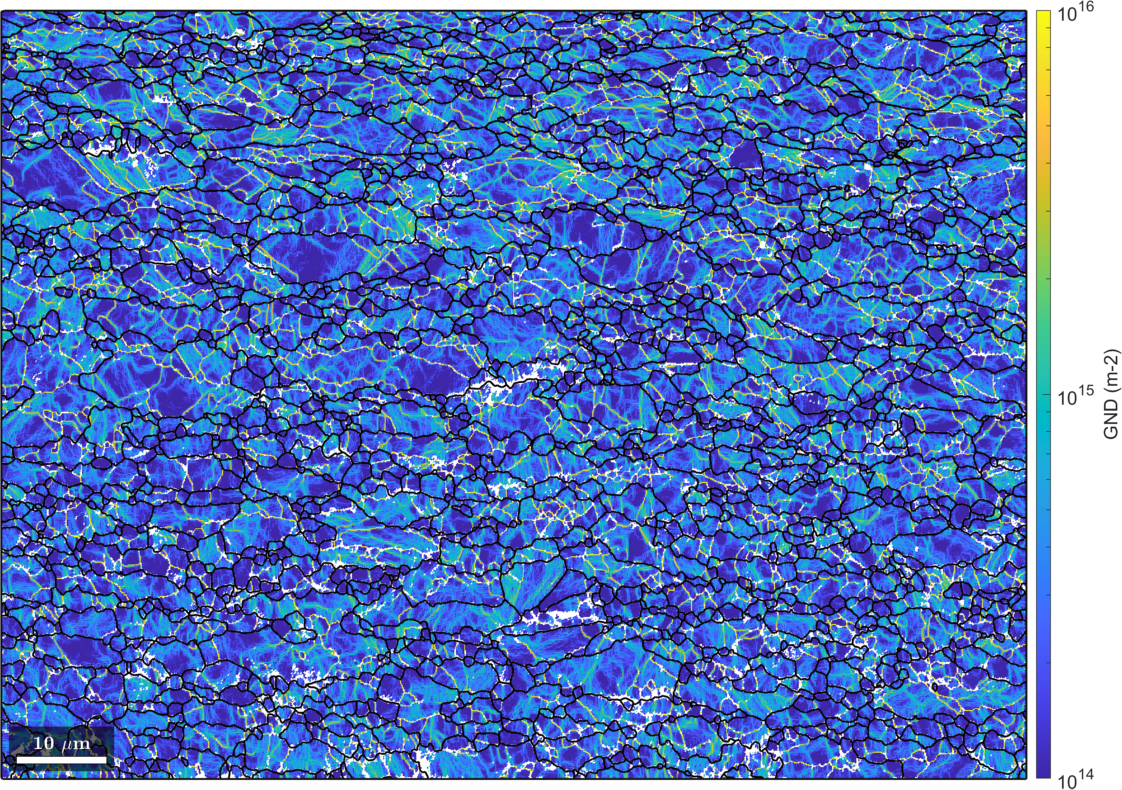}
    \caption{\label{fig:CDRX_GNDdensity} GND density map. \textbf{Eq} initial microstructure. Hot deformation conditions are $T=650^{\circ}C; ~ \dot{\varepsilon}=0.01 ~ s^{-1}; ~ \varepsilon=1.0$.}
\end{figure}

Figure \ref{fig:CDRX_GNDdensity} presents GND density map for a given hot deformation condition: $T=650^{\circ}C; ~ \dot{\varepsilon}=0.01 ~ s^{-1}; ~ \varepsilon=1.0$. This particular condition has been chosen since the high temperature, strain level and low strain rate favor CDRX. From this figure, one can observe that some of the smaller grains have a lower GND density. However, they do not stand out from the general grain population. Consequently, the recrystallized fraction remains low after hot deformation.

These results illustrate that CDRX operates by a progressive formation of LAGB. This phenomenon starts and prevails at low strains ($\varepsilon \leq 0.5$). Then, if sufficient stored energy of deformation is accumulated, LAGB progressively transform to HAGB. A large number of small grains is then observable and the LAGB length ratio decreases significantly. These small grains, however, do not present a very low GND density, as one could have expected. These microstructure changes are very progressive with deformation and distributed throughout the whole microstructure. It confirms that Zy-4 is experiencing CDRX over the range of conditions studied within the present work. 

\subsubsection{Effect of thermomechanical conditions over continuous and post-dynamic recrystallization}

\begin{figure}
\centering
\begin{subfigure}{0.49\textwidth}
  \centering
  \includegraphics[width=1.0\linewidth]{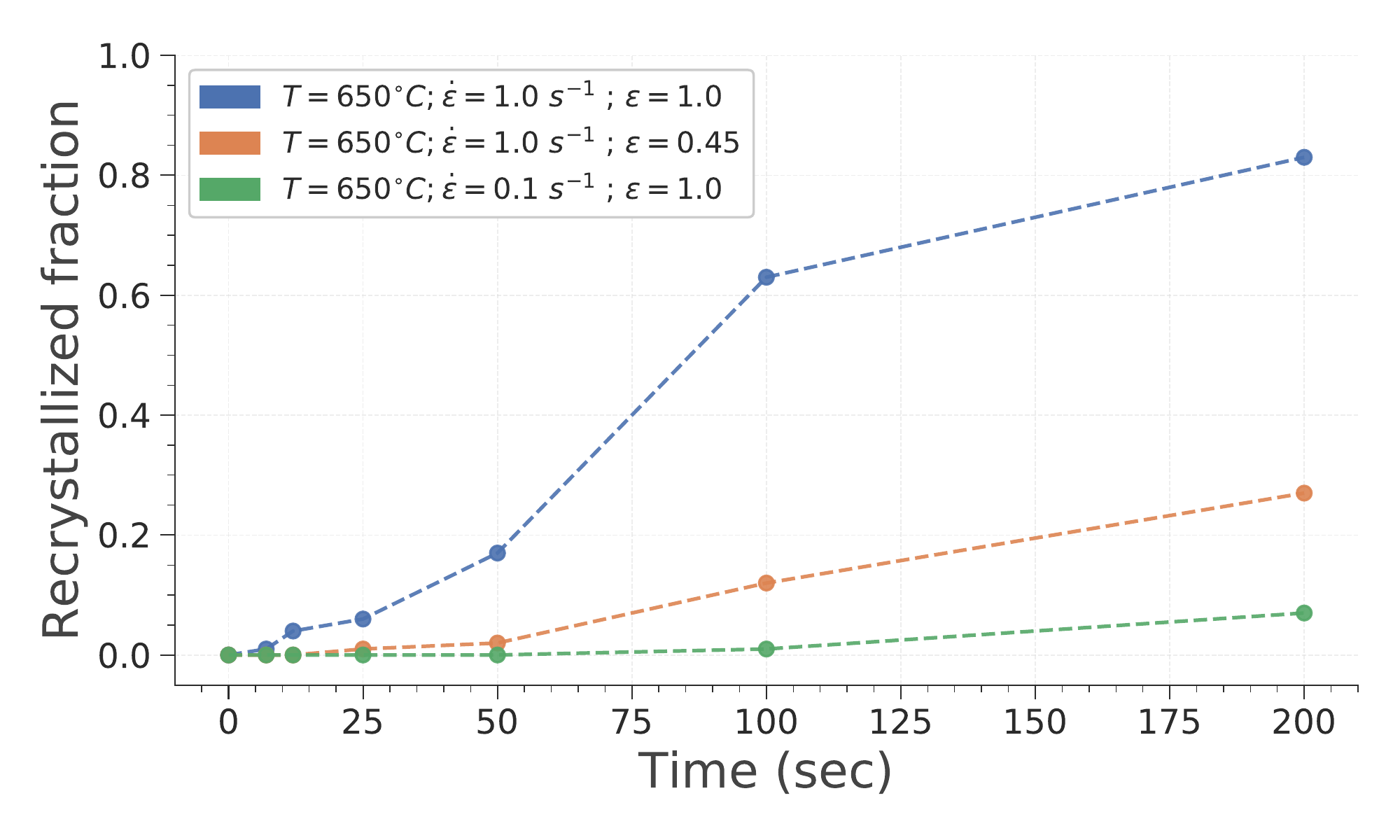}  
  \caption{\label{fig:RecrystallizedFractionPDRX_Eq} Recrystallized fraction versus time.}
\end{subfigure}
\begin{subfigure}{0.49\textwidth}
  \centering
  \includegraphics[width=1.0\linewidth]{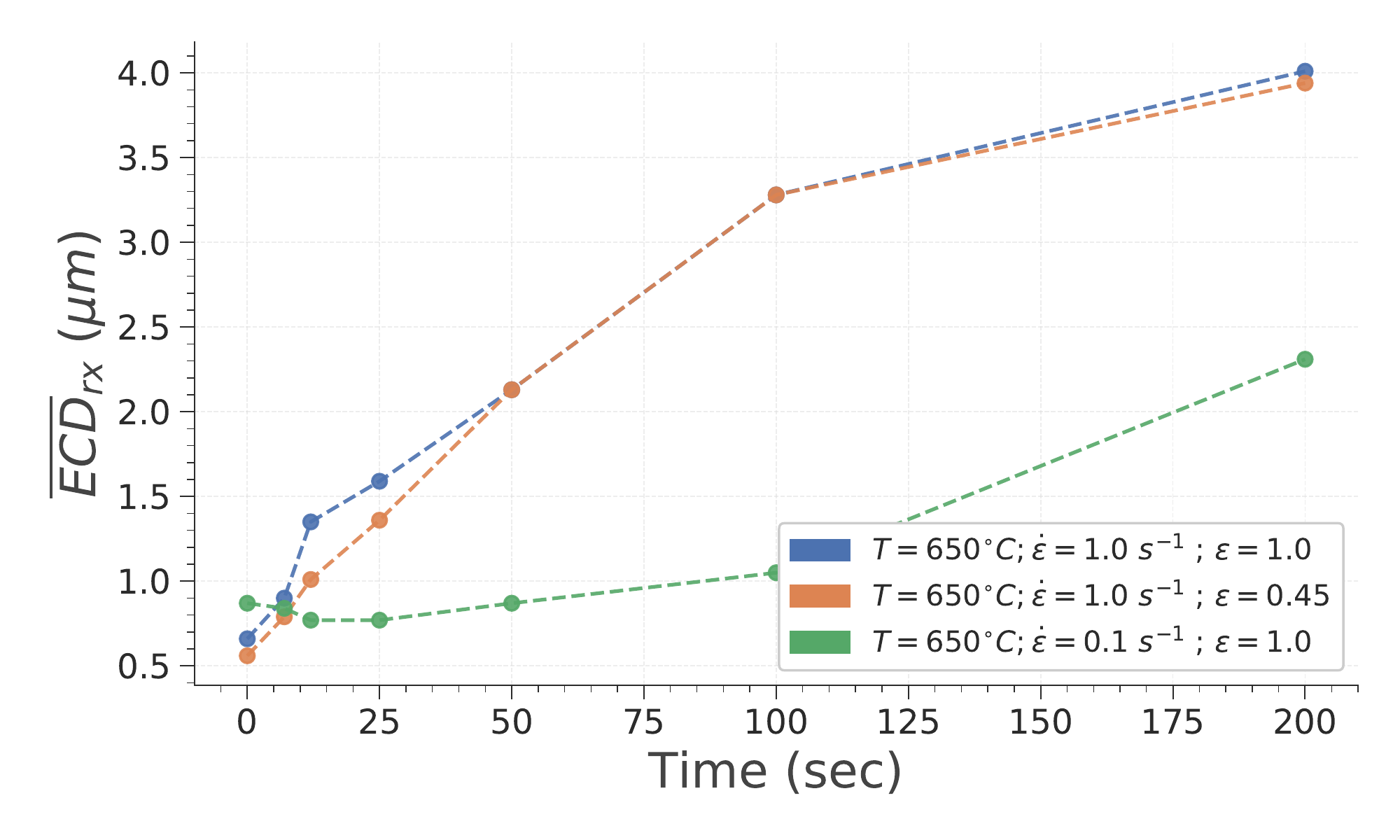} 
  \caption{\label{fig:MeanECDrxPDRX_Eq} $\overline{ECD}_{rx}$ versus time.}
\end{subfigure}
\caption{Evolution of properties related to recrystallization during holding at $T=650^{\circ}C$. \textbf{Eq} samples.} 
\label{fig:RX_properties_PDRX_Eq}
\end{figure}

$\overline{ECD}_{rx}$ and recrystallized fraction are plotted versus holding time, for different hot deformation conditions, in figure \ref{fig:RX_properties_PDRX_Eq}. It appears from figure \ref{fig:RecrystallizedFractionPDRX_Eq} that decreasing final strain or strain rate leads to a significant decrease of recrystallization kinetics during subsequent holding at temperature. Figure \ref{fig:MeanECDrxPDRX_Eq} shows that decreasing the strain level from $1.0$ to $0.45$ does not modify the recrystallized grain growth kinetics. On the other hand, decreasing the strain rate from $1.0 ~ s^{-1}$ to $0.1 ~ s^{-1}$ significantly lowers that kinetics.

\begin{figure}
\centering
\begin{subfigure}{0.49\textwidth}
  \centering
  \includegraphics[width=1.0\linewidth]{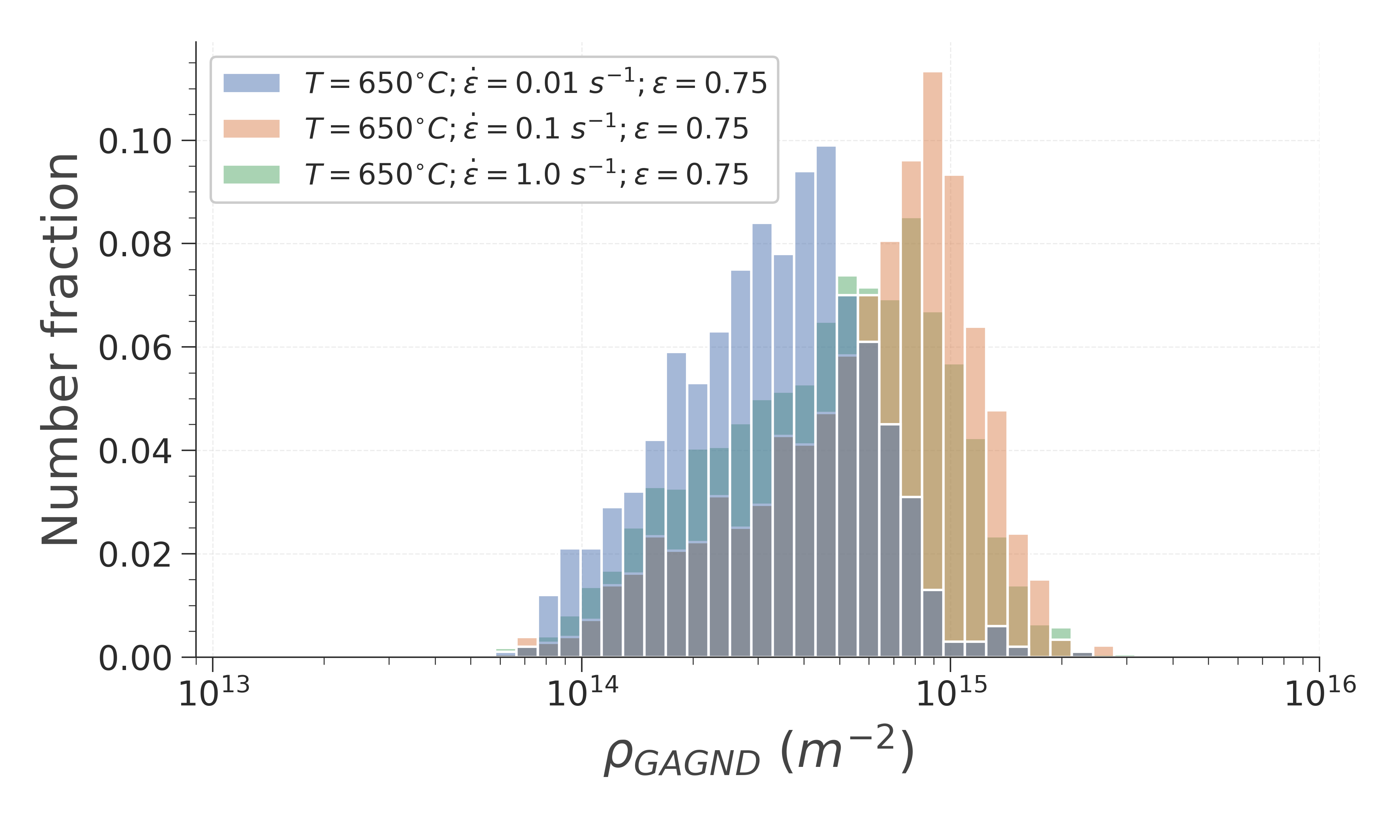}  
  \caption{\label{fig:Histo_CDRX_GNDdensity_weighted_Eps1_influenceStrainRate} $\varepsilon= 0.75$.}
\end{subfigure}
\begin{subfigure}{0.49\textwidth}
  \centering
  \includegraphics[width=1.0\linewidth]{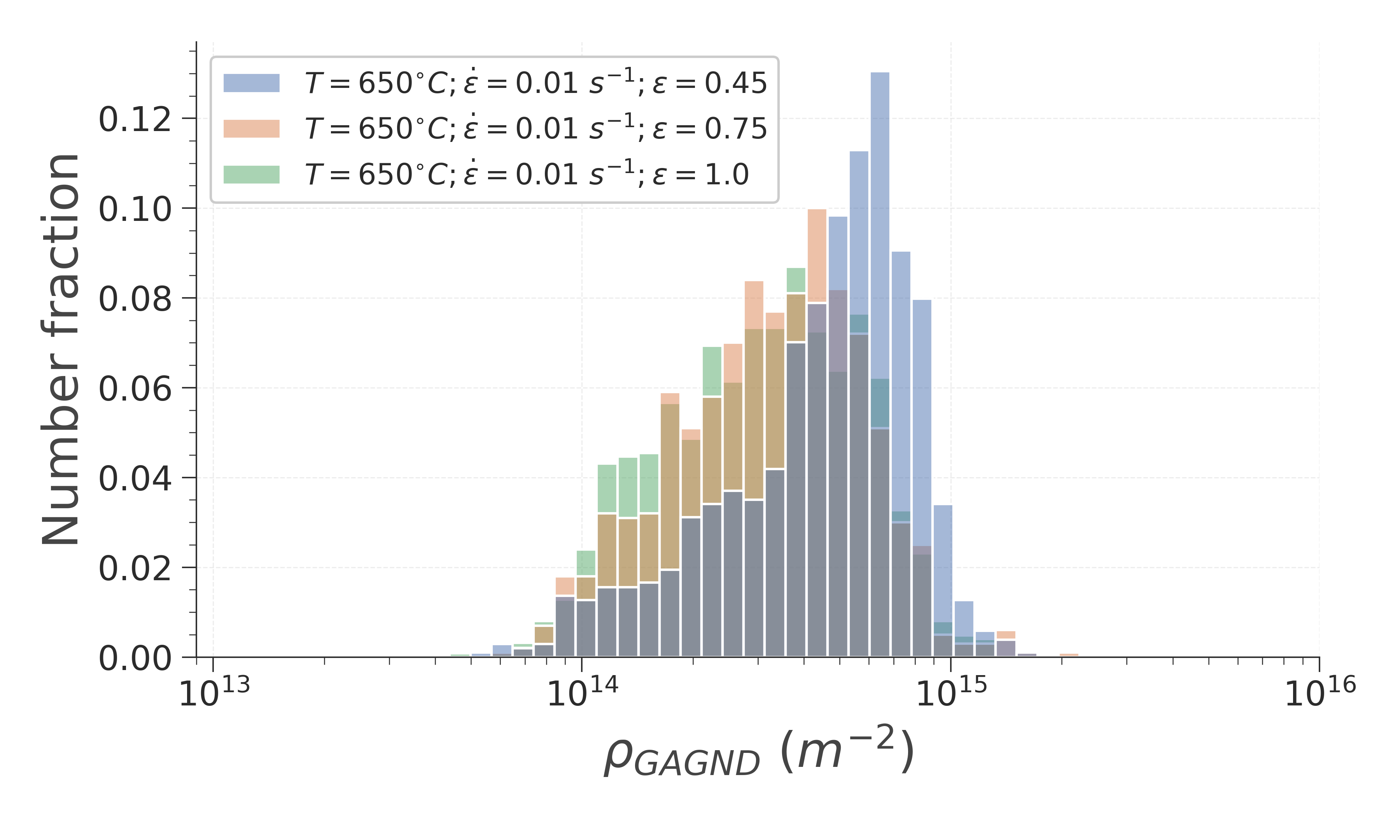} 
  \caption{\label{fig:Histo_CDRX_GNDdensity_weighted_Eps001_influenceFinalStrain} $\dot{\varepsilon}= 0.01 ~ s^{-1}$.}
\end{subfigure}
\caption{Grain average GND density number distribution.} 
\label{fig:Histo_CDRX_GNDdensity}
\end{figure}

To complete this observation, the grain average GND density distribution are plotted in figure \ref{fig:Histo_CDRX_GNDdensity}. One can observe from figure \ref{fig:Histo_CDRX_GNDdensity_weighted_Eps1_influenceStrainRate} that the grain average GND density distribution is shifted towards higher values with the strain rate. On the other hand, it appears that increasing the final strain tends to increase the number of grains with a low average dislocation density. 

These results corroborate the following hypothesis:
\begin{itemize}
    \item A strain rate increase leads to an increase of the stored energy of deformation and of the driving force for recrystallized grain growth. As a consequence, the average recrystallized grain size and the post-dynamic recrystallization kinetics both increase with strain rate.
    \item A strain increase results in a higher number of grains presenting a stored energy advantage. Therefore, the recrystallization kinetics increases with the deformation level but the average size of recrystallized grains does not.
\end{itemize}

\subsubsection{Effect of initial microstructure over continuous and post-dynamic recrystallization}

\begin{figure}[h!]
\centering
\begin{subfigure}{0.49\textwidth}
  \centering
  \includegraphics[width=1.0\linewidth]{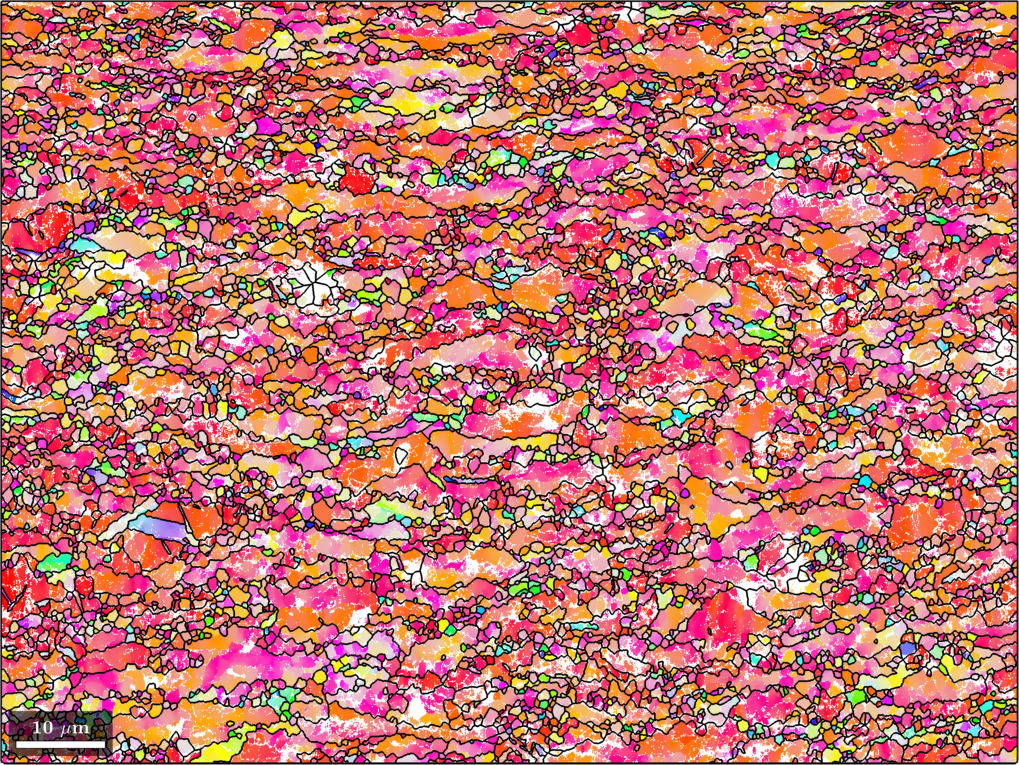}  
  \caption{\label{fig:HCR6_IPF} Equiaxed; $dt= 1.0 ~ s$.}
\end{subfigure}
\begin{subfigure}{0.49\textwidth}
  \centering
  \includegraphics[width=1.0\linewidth]{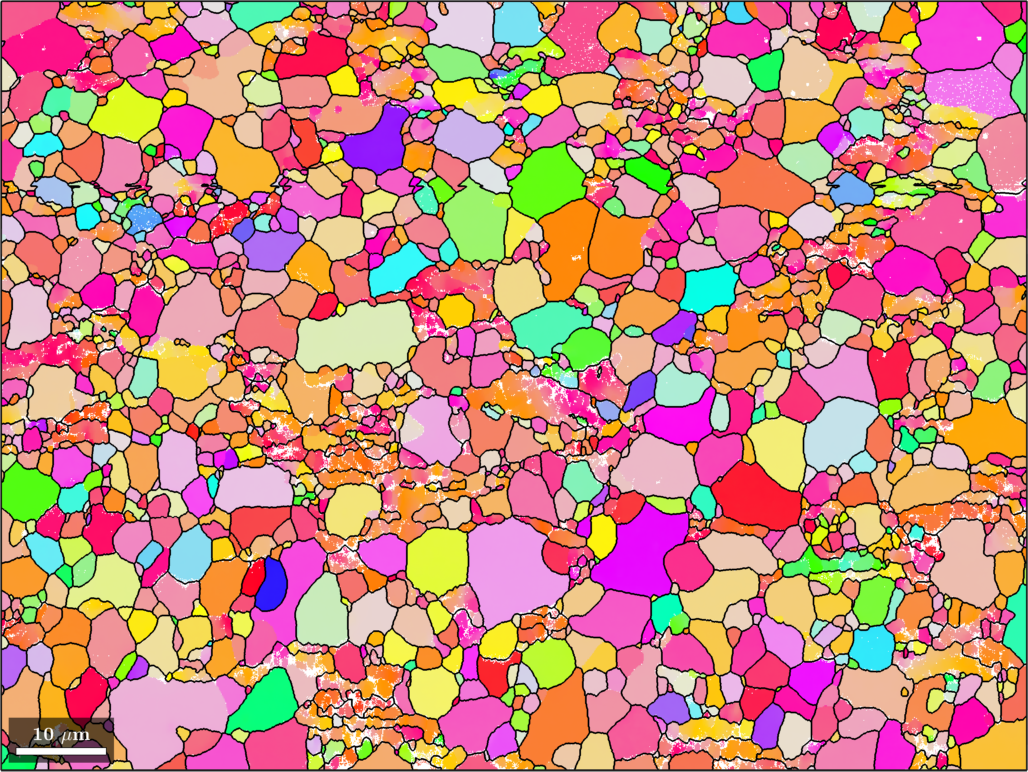} 
  \caption{\label{fig:HCHR14_IPF} Equiaxed; $dt= 103 ~ s$.}
\end{subfigure}\\
\begin{subfigure}{0.49\textwidth}
  \centering
  \includegraphics[width=1.0\linewidth]{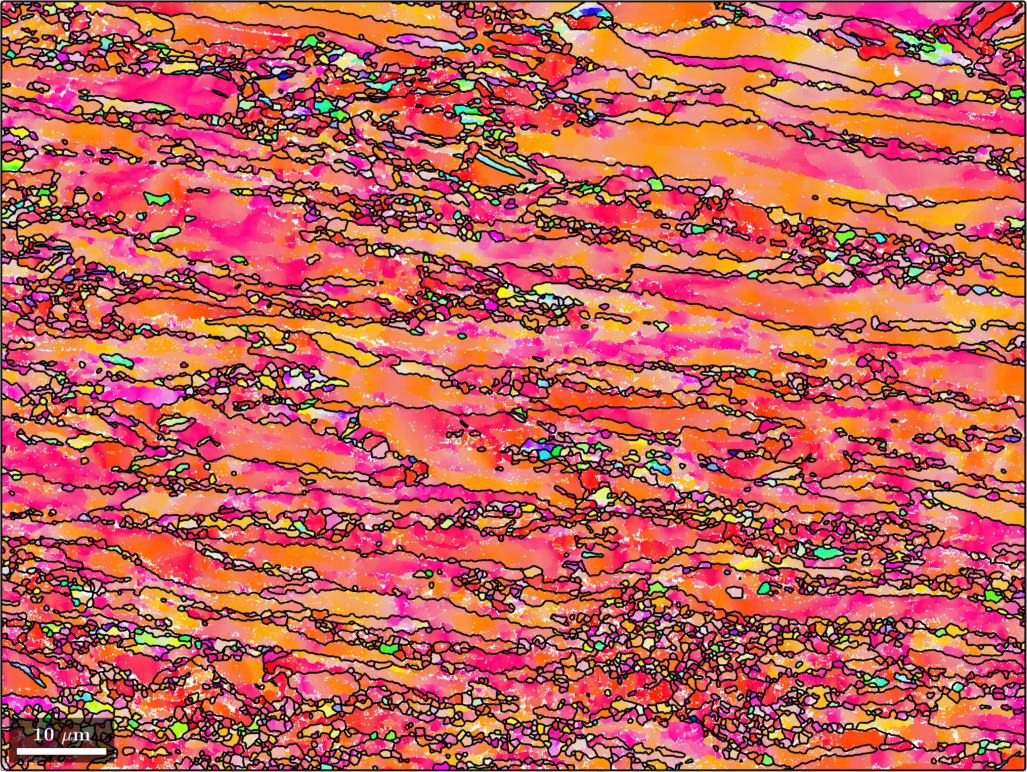}  
  \caption{\label{fig:HCV13_IPF} Basket-weaved; $dt= 1.1 ~ s$.}
\end{subfigure}
\begin{subfigure}{0.49\textwidth}
  \centering
  \includegraphics[width=1.0\linewidth]{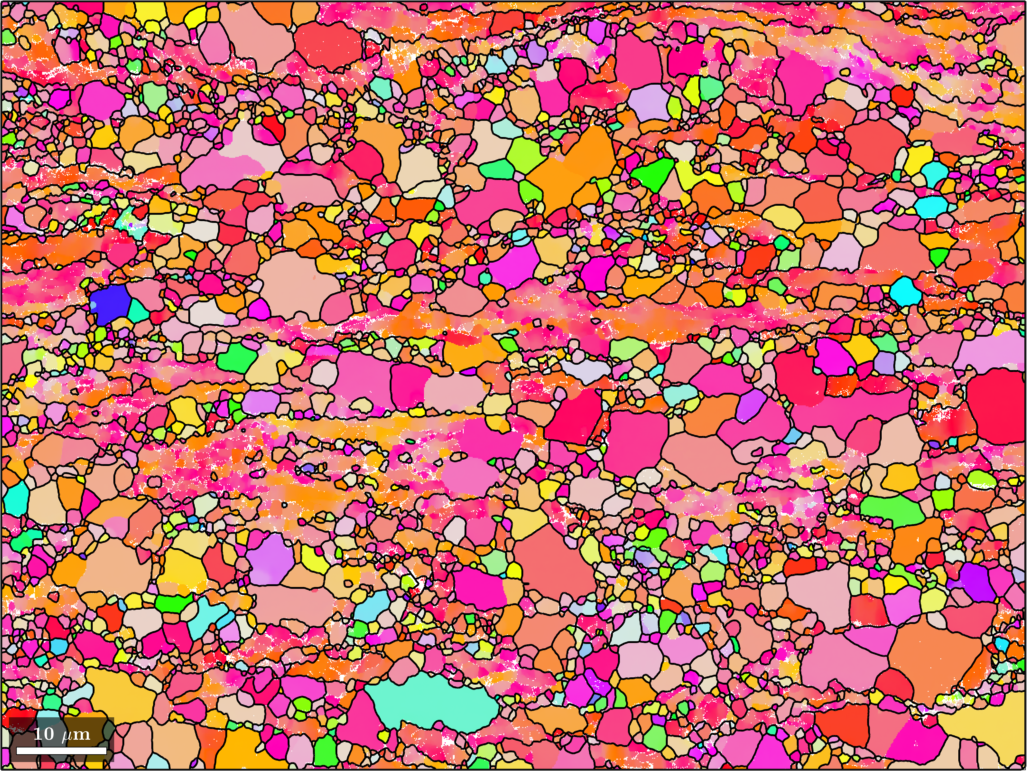} 
  \caption{\label{fig:HCHV8_IPF} Basket-weaved; $dt= 101 ~ s$.}
\end{subfigure}\\
\begin{subfigure}{0.49\textwidth}
  \centering
  \includegraphics[width=1.0\linewidth]{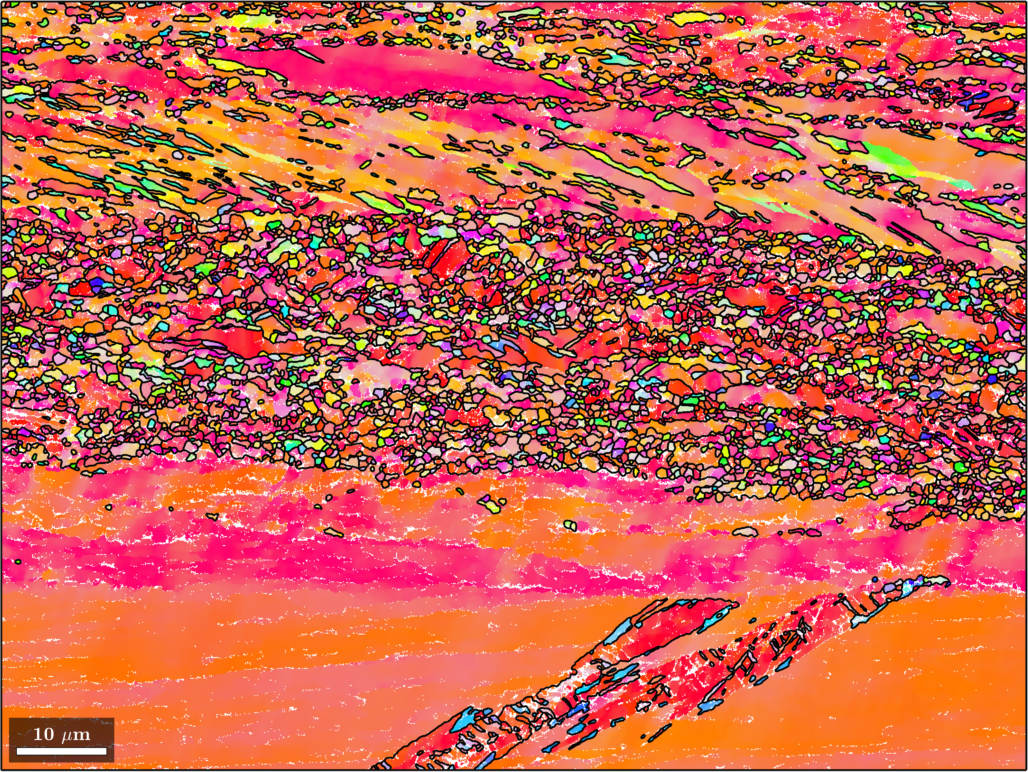}  
  \caption{\label{fig:HCP13_IPF} Parallel plates; $dt= 0.7 ~ s$.}
\end{subfigure}
\begin{subfigure}{0.49\textwidth}
  \centering
  \includegraphics[width=1.0\linewidth]{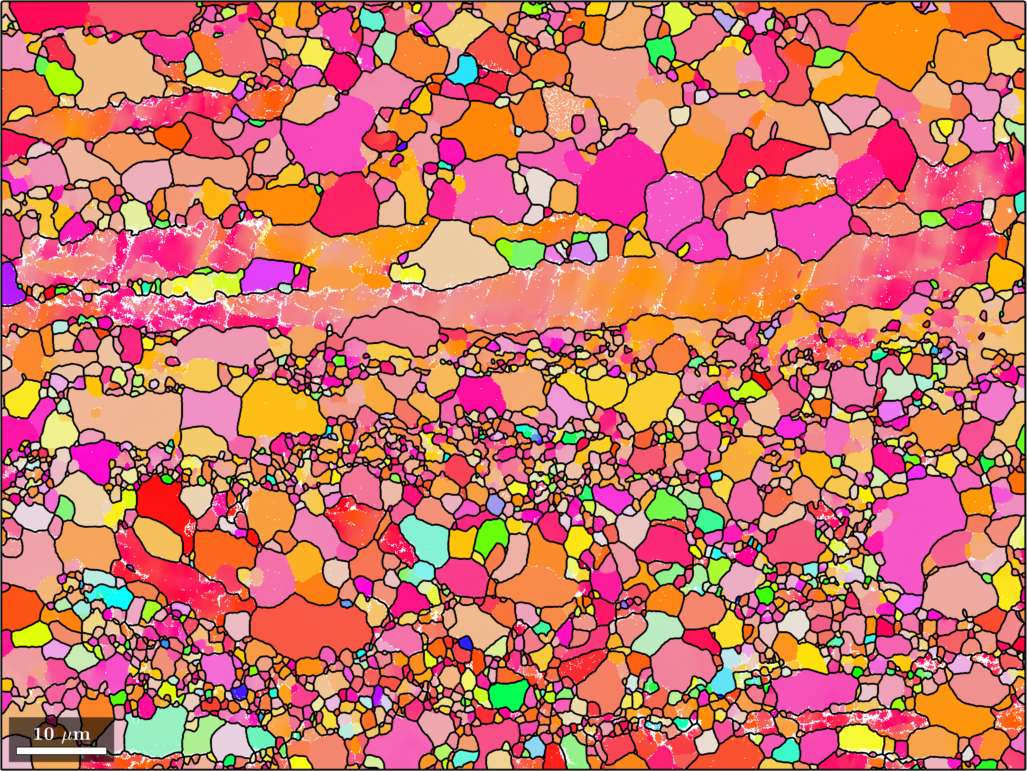} 
  \caption{\label{fig:HCHP8_IPF} Parallel plates; $dt= 100 ~ s$.}
\end{subfigure}\\
\caption{EBSD orientation maps of samples with different initial microstructure.} 
\label{fig:OrientationMaps_ImpactInitialMicrostructure}
\end{figure}

Six orientation maps are displayed in figure \ref{fig:OrientationMaps_ImpactInitialMicrostructure}. They correspond to samples after hot deformation and after $100 ~ s$ at $650^{\circ}C$. One can notice that right after deformation, microstructures present different degrees of heterogeneity at the observation scale. It seems that an initial parallel plate microstructure displays a higher heterogeneity extent than an initial basket-weaved one which presents more heterogeneities than an equiaxed one. This feature appears to be retained after subsequent heat treatment. This is confirmed by EBSD maps displaying larger observation zones and provided in appendix \ref{app:LargerOrientationMaps_ImpactInitialMicrostructure} (figure \ref{fig:LargerOrientationMaps_ImpactInitialMicrostructure}).

One hypothesis is proposed to explain the initial microstructure impact over recrystallization. It assumes that the heterogeneities observed during CDRX and PDRX are caused by deformation heterogeneities. These deformation heterogeneities being conditioned by initial texture, grain morphology and grain size. \textbf{Eq} samples present a strong texture and most of the grains initially have their <c> axis orthogonal to the compression direction. Since the spread of initial grain orientation is low, most of the grains deform rather similarly and the deformation incompatibilities are weak. In lamellar microstructures, on the other hand, there is no initial texture. The grain initial orientations are distributed in the orientation space following the variant selection rule. Therefore, grains are more susceptible to deform differently. This generates deformation incompatibilities close to grain boundaries. Grain boundary vicinities thus present a higher GND density and constitute preferential sites for formation and growth of recrystallized grains. Therefore, lamellar microstructures present a more heterogeneous recrystallization behavior.

Between \textbf{BW} and \textbf{PP} samples, the extent of heterogeneity is also rather different. In \textbf{BW} microstructures, most of the lamellae are in contact with lamellae presenting different orientations. Since the lamella thickness is low, the deformation incompatibilities affect almost all the microstructure. In \textbf{PP} microstructures, only lamellae located at ex-$\beta$ GB are in contact with lamellae of different orientations. Therefore, the areas around ex-$\beta$ GB are impacted by deformation incompatibilities whereas an important part of the microstructure does not experience such effect. Therefore, there are fewer preferential zones for formation and growth of recrystallized grains and the recrystallization is even more heterogeneous for \textbf{PP} microstructures.

\subsection{Modeling CDRX and PDRX}

The model parameters related to mechanical behavior are taken from literature. Grain boundary mobility parameters are fitted using experimental results from heat treatments performed on fully recrystallized samples and corresponding pure GG LS simulations as detailed in \cite{Alvarado2021} for nickel based superalloys. The parameters for subgrain formation and evolution are set based on the recommendations of Gourdet \cite{Gourdet1997_PhD}.

Hardening and recovery parameters are identified using hot compression test results. $\overline{K_1}$ and $K_2$ are identified by fitting the experimental macroscopic stress-strain curves from the hot compression tests thanks to Yoshie-Lasraaoui-Jonas and Taylor equations. Then, the $K_1$ distribution is set by measuring the experimental GAGND density distribution. This is based upon Yoshie-Lasraaoui-Jonas equation which predicts that: $\rho_{sat} = K_1/K_2$. Therefore, if all grains have reached their saturation GND density value, the $K_1$ distribution is equal to the $\rho_{sat}$ distribution multiplied by a factor $K_2$. Of course, such reasoning holds if all grains have reached their saturation GND density. Since no increase of average GND density is observed on all maps at different strain levels, such assumption can be considered. 

The values used for the present study are provided in table \ref{tab:ModelParameters} (appendix \ref{app:ModelParameters}).

Experimental data are immersed in the considered FE-LS strategy and an average dislocation field per grain is considered here. Six simulation cases with different initial microstructures and thermomechanical conditions are performed:
\begin{itemize}
    \item \textbf{Eq} initial microstructure, $T=650^{\circ}C; ~ \dot{\varepsilon}=1.0 ~ s^{-1}; ~ \varepsilon = 1.0; ~ dt=200 ~ s$,
    \item \textbf{Eq} initial microstructure, $T=650^{\circ}C; ~ \dot{\varepsilon}=0.1 ~ s^{-1}; ~ \varepsilon = 1.0; ~ dt=200 ~ s$,
    \item \textbf{Eq} initial microstructure, $T=650^{\circ}C; ~ \dot{\varepsilon}=1.0 ~ s^{-1}; ~ \varepsilon = 0.45; ~ dt=200 ~ s$,
    \item \textbf{BW} initial microstructure, $T=650^{\circ}C; ~ \dot{\varepsilon}=1.0 ~ s^{-1}; ~ \varepsilon = 1.2; ~ dt=200 ~ s$,
    \item \textbf{PP} initial microstructure, $T=650^{\circ}C; ~ \dot{\varepsilon}=1.0 ~ s^{-1}; ~ \varepsilon = 1.2; ~ dt=200 ~ s$.
\end{itemize}

Experimental and digital microstructures after $25$ and $100 ~ s$ are displayed in figure \ref{fig:PDRX_ExpVsSimu_GNDmaps} for the three different initial microstructures. Additional results are provided in appendix \ref{app:AdditionalSimulationResults}.

\begin{figure}[h!]
\centering
\begin{subfigure}{0.90\textwidth}
  \centering
  \includegraphics[width=1.0\linewidth]{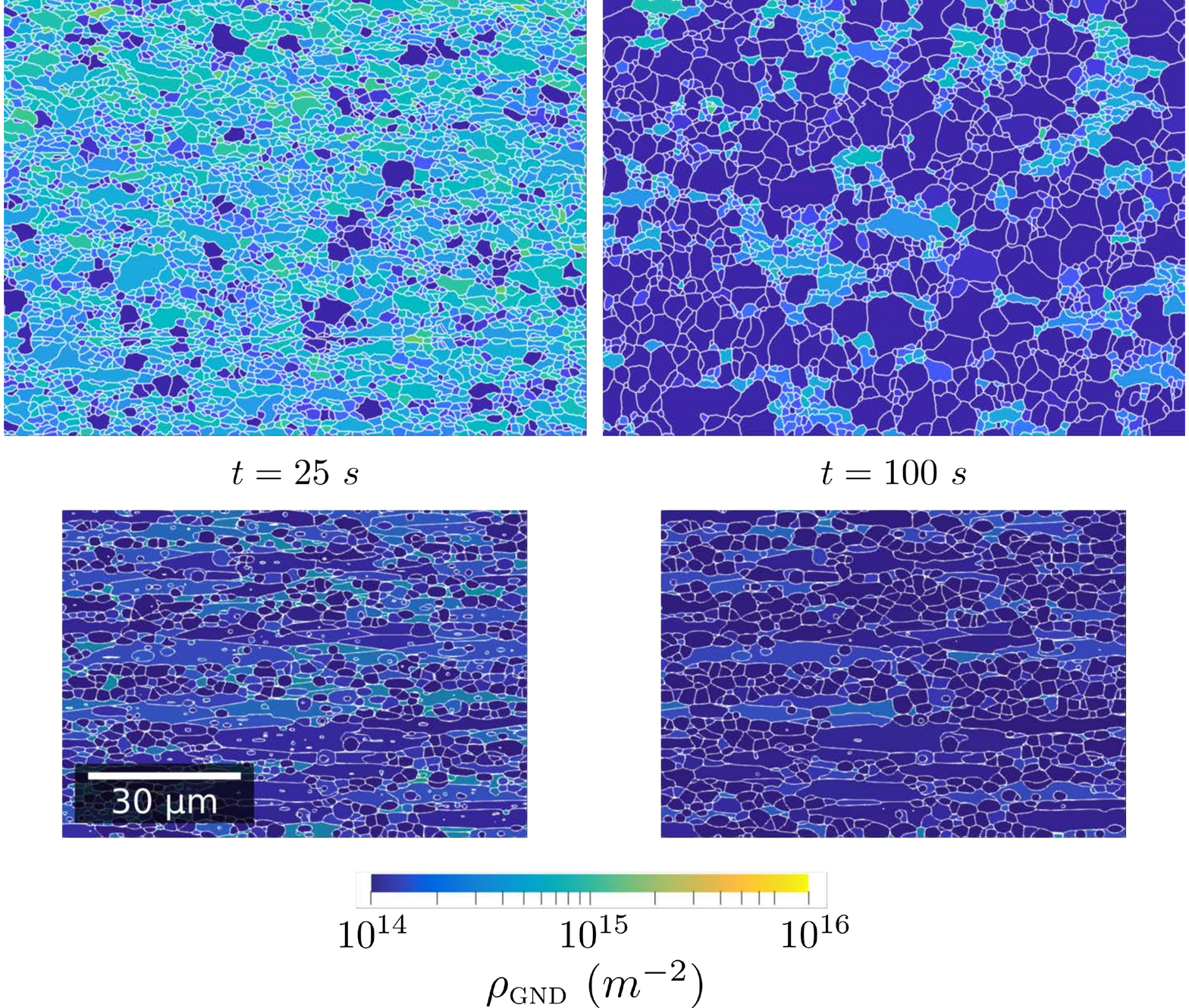}  \caption{\label{fig:CDRX_Eq_SR1_Eps135_ExpVsSimu} Equiaxed.}
\end{subfigure}\\
\begin{subfigure}{0.90\textwidth}
  \centering
  \includegraphics[width=1.0\linewidth]{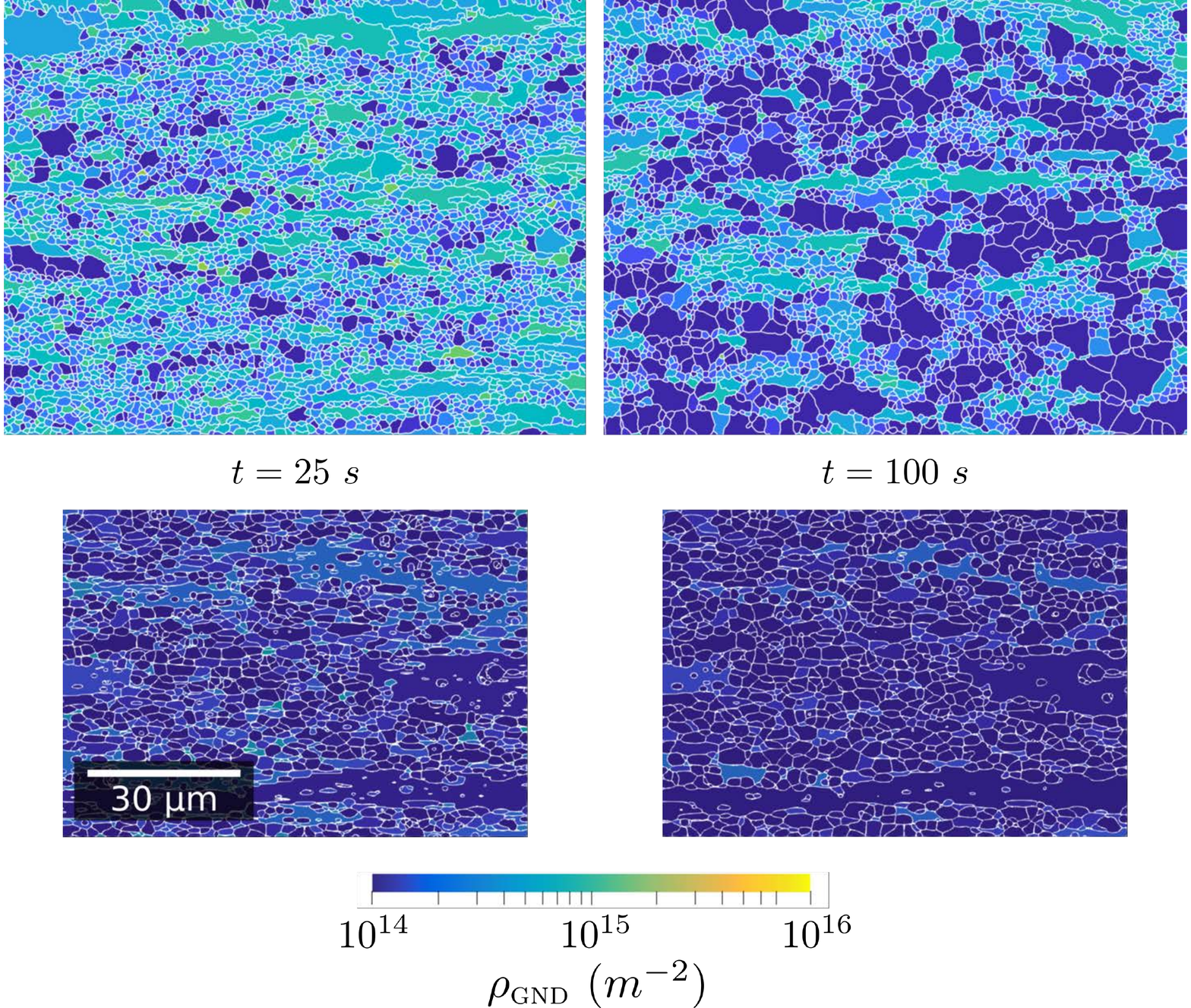} \caption{\label{fig:CDRX_PP_SR1_Eps135_ExpVsSimu} Basket-weaved.}
\end{subfigure}\\
\end{figure}
\begin{figure}\ContinuedFloat
\begin{subfigure}{0.90\textwidth}
  \centering
  \includegraphics[width=1.0\linewidth]{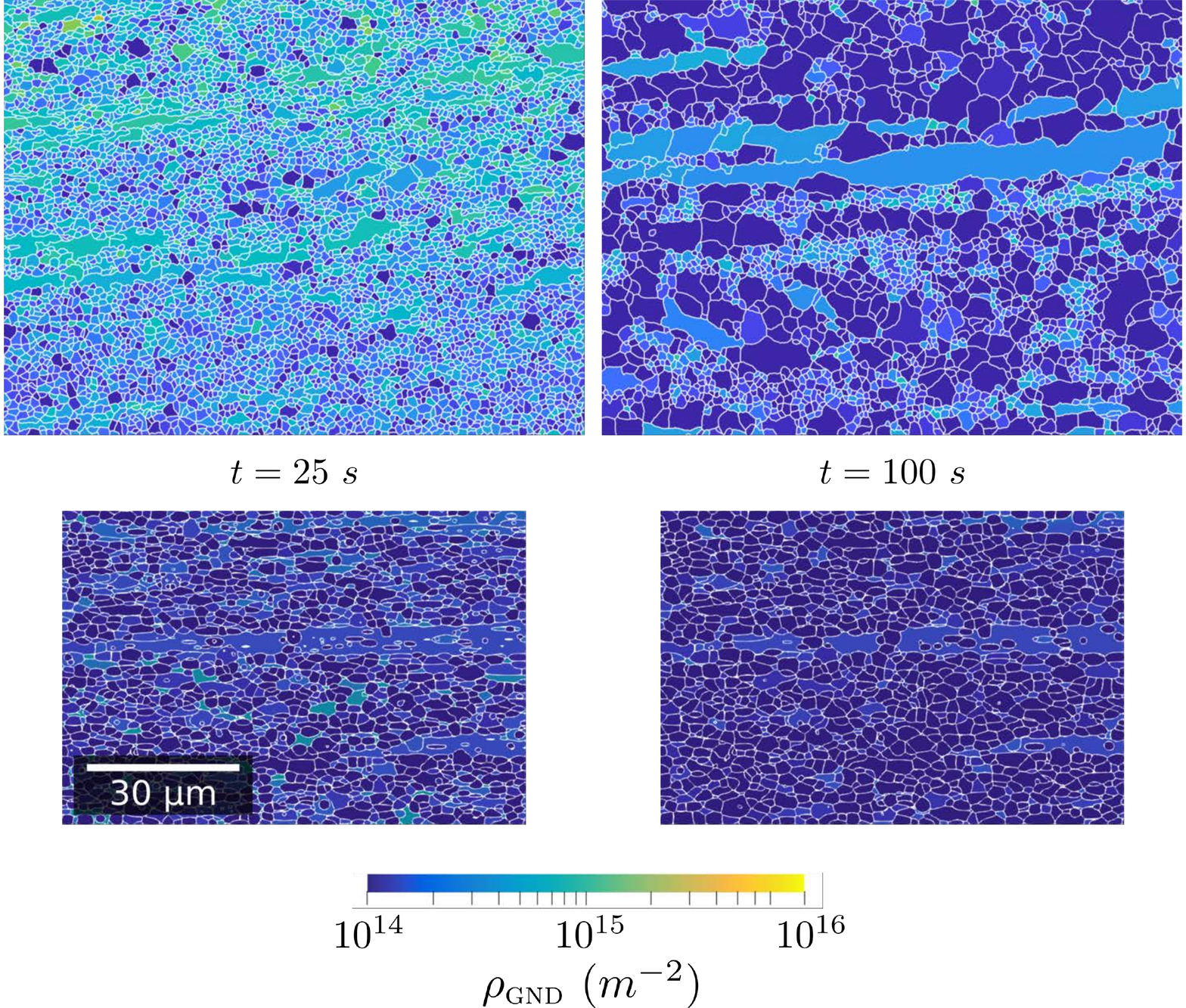}  \caption{\label{fig:CDRX_PP_SR1_Eps135_ExpVsSimu} Parallel plates.}
\end{subfigure}
\caption{Snapshots on experimental (top) and simulation (bottom) GND density maps for one (a) \textbf{Eq} case, (b) \textbf{BW} case, and (c) \textbf{PP} case. Hot deformation conditions are $T = 650^{\circ}C ; ~ \dot{\varepsilon} = 1.0 ~ s^{-1}; \varepsilon = 1.0-1.2$.}
\label{fig:PDRX_ExpVsSimu_GNDmaps}
\end{figure}

Several qualitative observations can be made from figure \ref{fig:PDRX_ExpVsSimu_GNDmaps}:
\begin{itemize}
    \item the qualitative correspondence between experimental and digital microstructures is satisfactory. The model ability to reproduce preferential zones where recrystallized grains form and grow is noteworthy. In addition, the model captures the different levels of heterogeneity observed depending on the initial microstructure.
    \item The model underestimates the GND density at low holding times,
    \item The number of small grains seems underestimated. Experimentally, a large number of small grains are kept up to $100 ~ s$ at $650^{\circ}C$. This is not observed in simulation where these small grains tend to shrink and disappear.
\end{itemize}

\begin{figure}[H]
\centering
\begin{subfigure}{0.49\textwidth}
  \centering
  \includegraphics[width=1.0\linewidth]{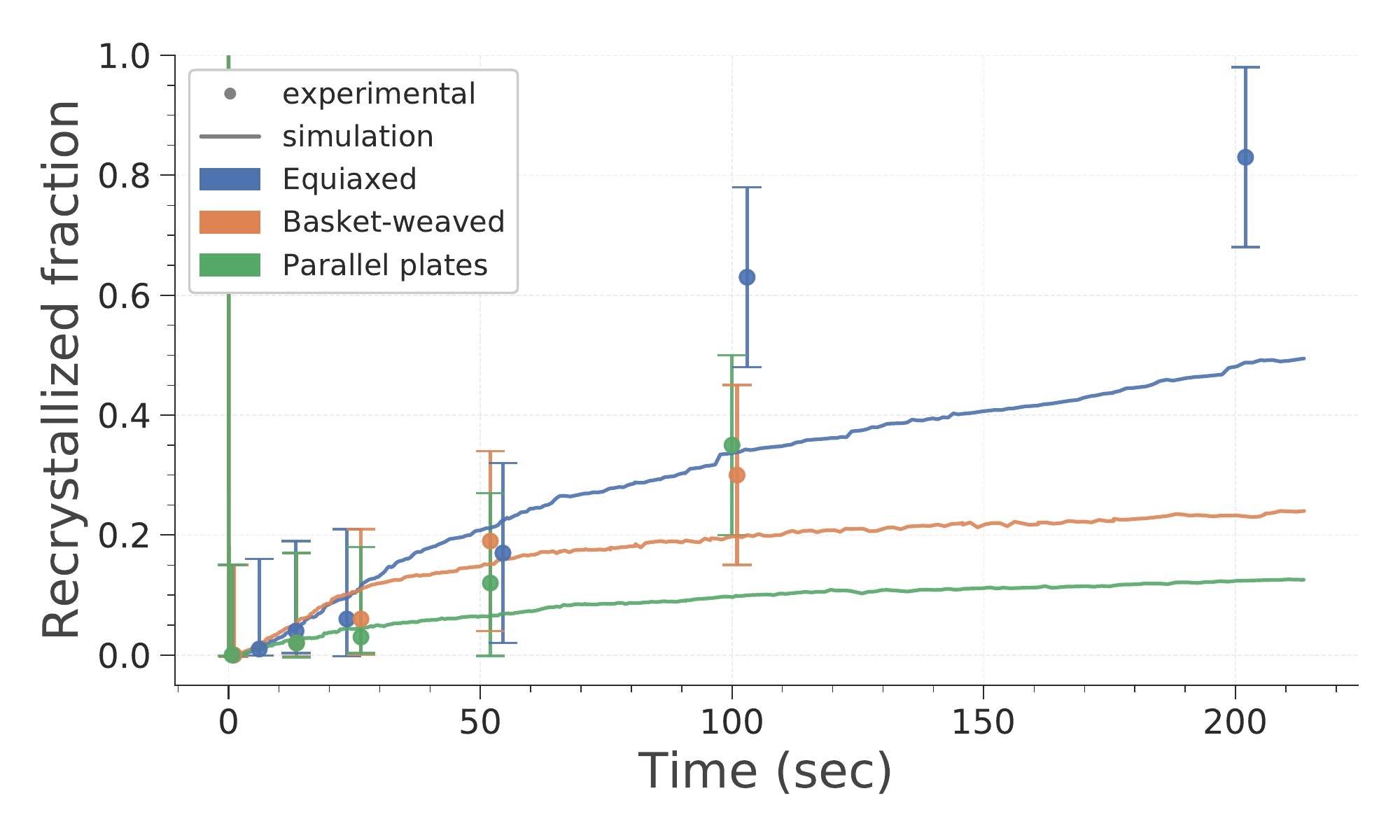}  \caption{\label{fig:CDRX_SR1_Eps135_ExpVsSimu_Rxfraction} Recrystallized fraction.}
\end{subfigure}
\begin{subfigure}{0.49\textwidth}
  \centering
  \includegraphics[width=1.0\linewidth]{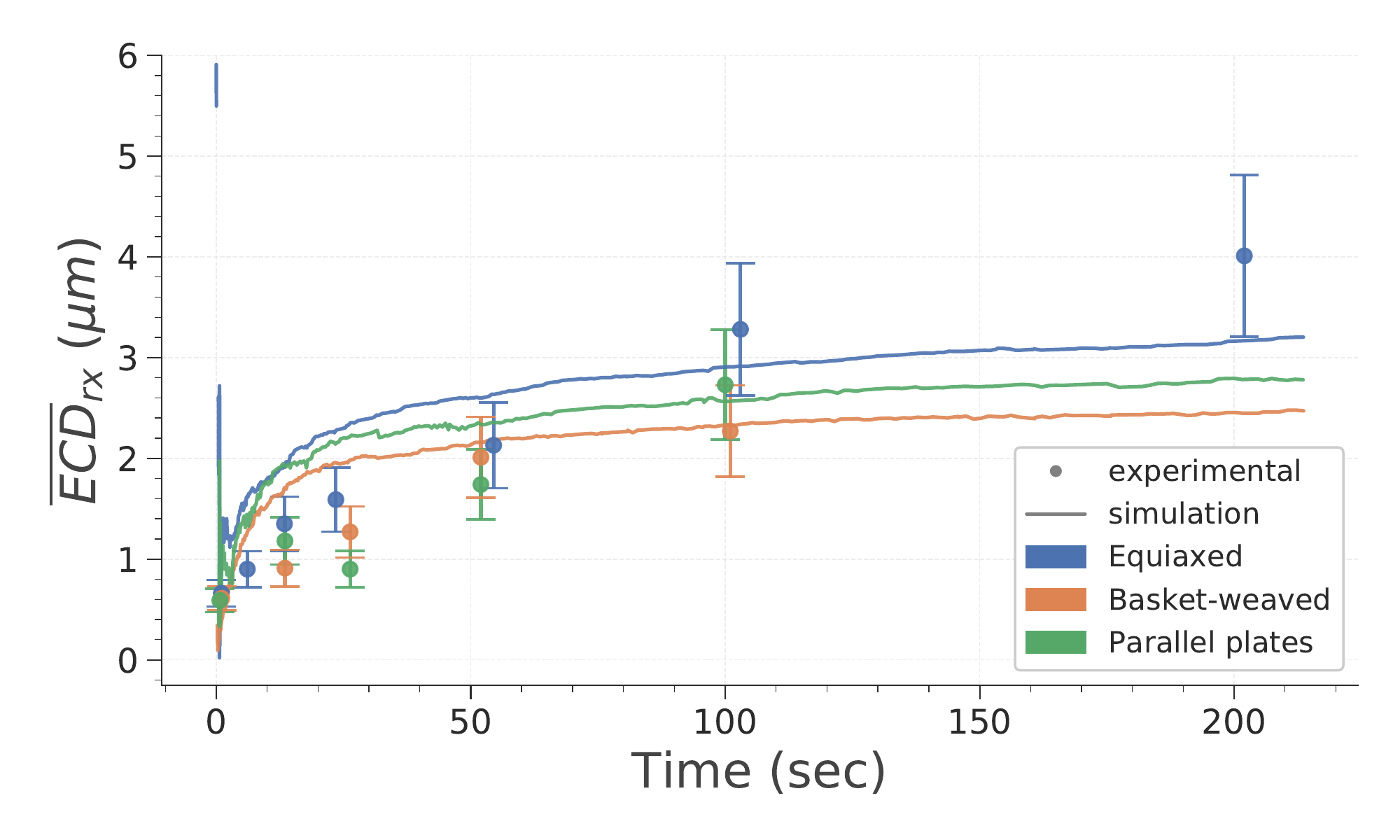} \caption{\label{fig:CDRX_SR1_Eps135_ExpVsSimu_ECDrx} Average recrystallized grain ECD.}
\end{subfigure}\\
\begin{subfigure}{0.49\textwidth}
  \centering
  \includegraphics[width=1.0\linewidth]{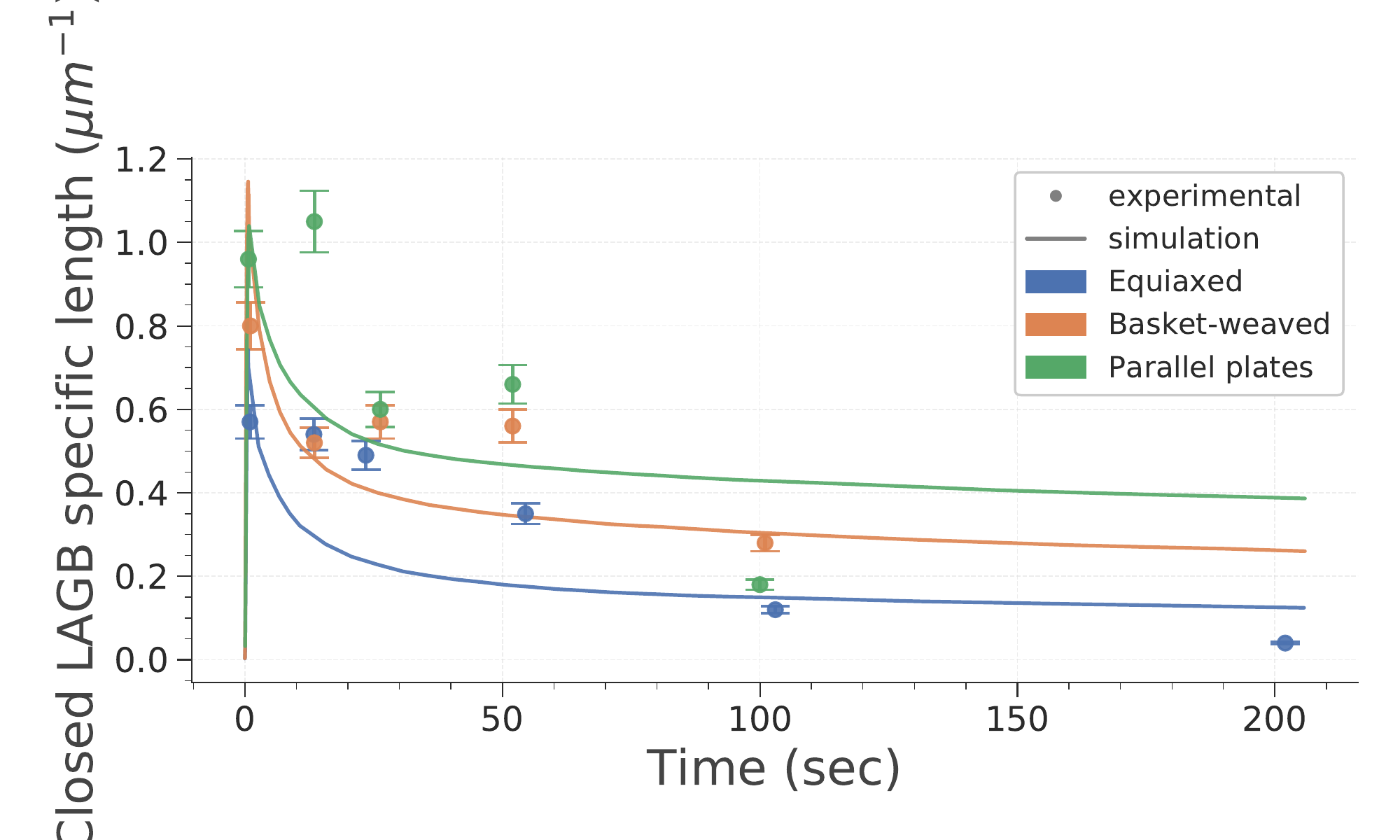}  \caption{\label{fig:CDRX_SR1_Eps135_ExpVsSimu_LAGBlength} LAGB specific length.}
\end{subfigure}
\caption{Evolution of recrystallized fraction, average recrystallized grain size and LAGB specific length with holding time. Three different initial microstructures are considered: \textbf{Eq}, \textbf{BW} and \textbf{PP}.  Hot deformation conditions are $T = 650^{\circ}C ; ~ \dot{\varepsilon} = 1.0 ~ s^{-1}; \varepsilon = 1.0-1.2$.}
\label{fig:CDRX_SR1_Eps135_InfluenceInitialMicrostructure}
\end{figure}

\begin{figure}[h!]
\centering
\begin{subfigure}{0.49\textwidth}
  \centering
  \includegraphics[width=1.0\linewidth]{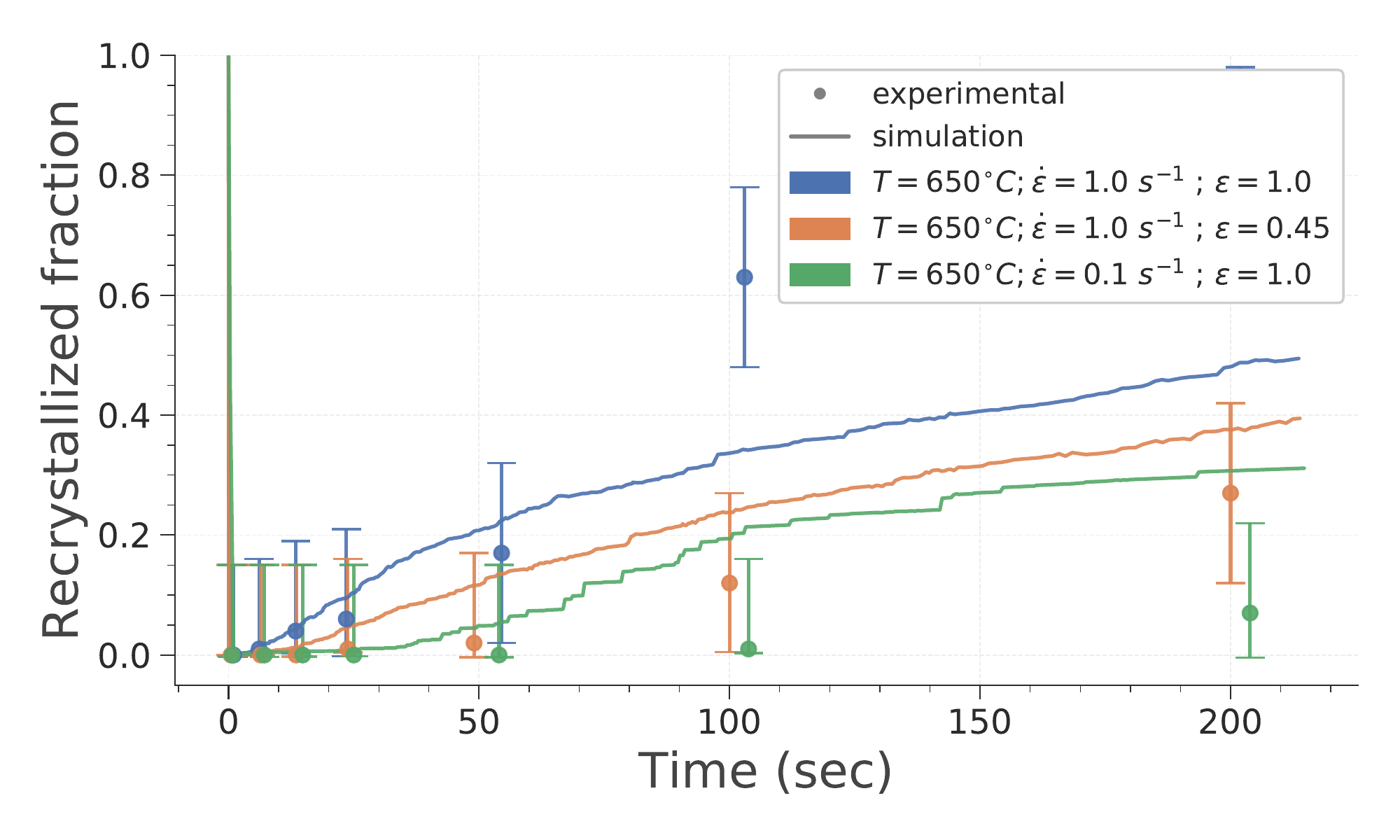}  \caption{\label{fig:CDRX_EQ_ExpVsSimu_Rxfraction} Recrystallized fraction.}
\end{subfigure}
\begin{subfigure}{0.49\textwidth}
  \centering
  \includegraphics[width=1.0\linewidth]{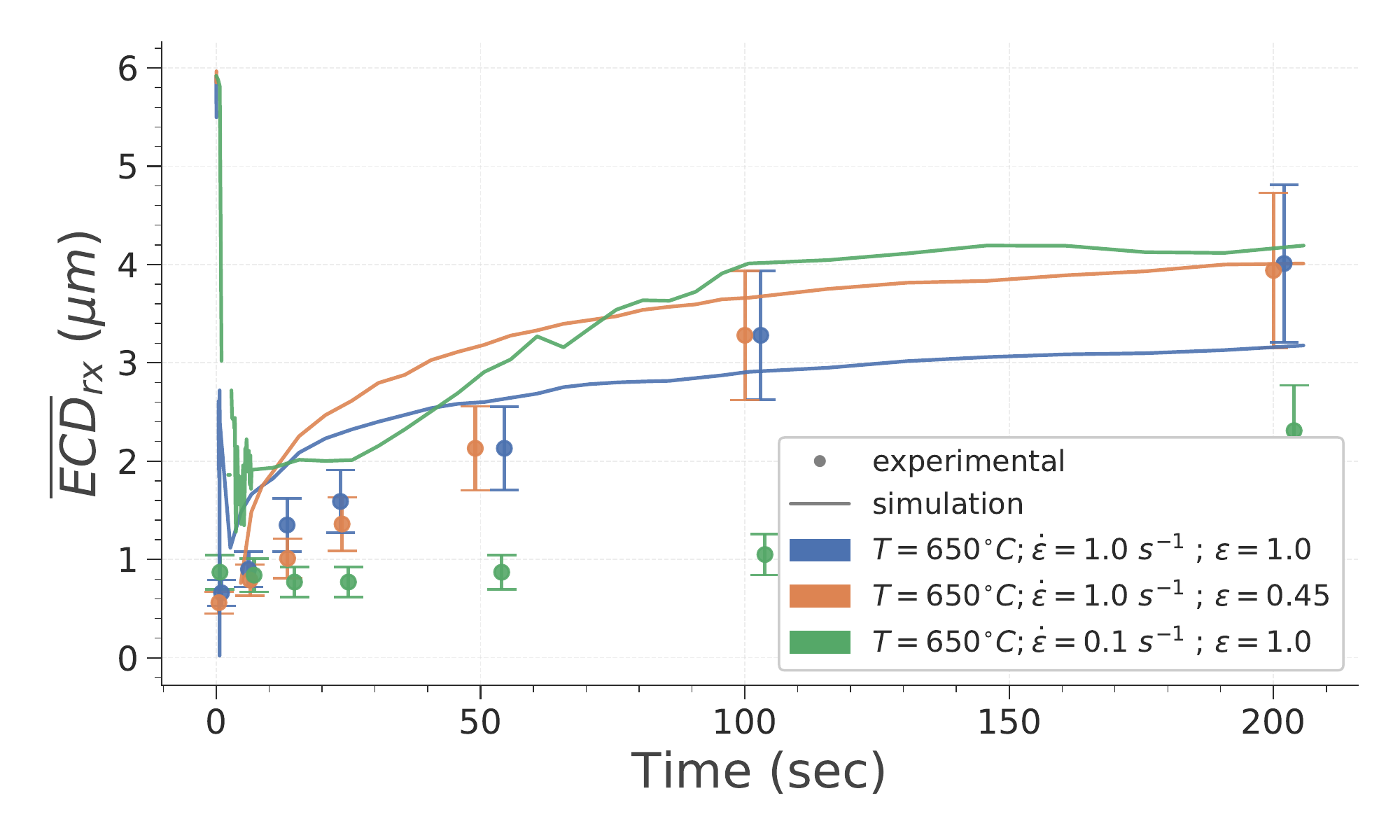} \caption{\label{fig:CDRX_SR1_Eps135_ExpVsSimu_ECDrx} Average recrystallized grain ECD.}
\end{subfigure}\\
\begin{subfigure}{0.49\textwidth}
  \centering
  \includegraphics[width=1.0\linewidth]{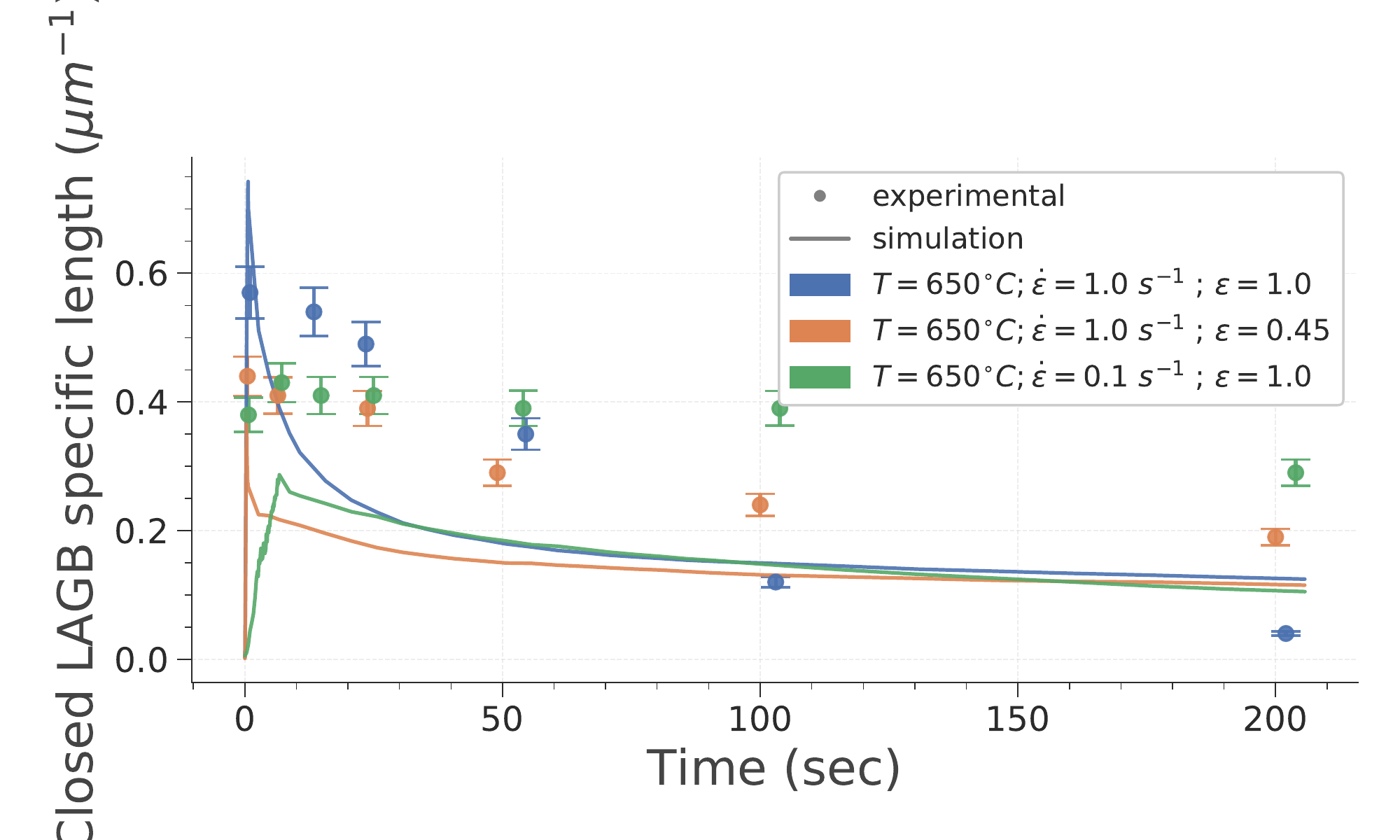}  \caption{\label{fig:CDRX_EQ_ExpVsSimu_LAGBlength} LAGB specific length.}
\end{subfigure}
\caption{Evolution of recrystallized fraction, average recrystallized grain size and LAGB specific length with holding time. \textbf{Eq} samples.}
\label{fig:CDRX_EQ_InfluenceThermomechanicalConditions}
\end{figure}

Recrystallized fraction, average recrystallized grain size and LAGB specific length are plotted versus holding time in figures \ref{fig:CDRX_SR1_Eps135_InfluenceInitialMicrostructure} and \ref{fig:CDRX_EQ_InfluenceThermomechanicalConditions}.

These results confirm that the model predictions for recrystallized fraction and average recrystallized grain size are satisfactory. It must be pointed out that the model presents some discrepancies. It significantly underestimates the recrystallized fraction and overestimates the average recrystallized grain size, for the following respective conditions: $T=650^{\circ}C; ~ \dot{\varepsilon}=1.0 ~ s^{-1}; ~ \varepsilon=1.0$ and $T=650^{\circ}C; ~ \dot{\varepsilon}=0.1 ~ s^{-1}; ~ \varepsilon=1.0$ (\textbf{Eq} microstructure). Finally, the model overestimates the decrease of LAGB specific length at the beginning of the PDRX regime, which is consistent with the observation regarding the shrinking of small grains and subgrains.

\begin{figure}[h!]
	\centering
    \includegraphics[width=0.6\linewidth]{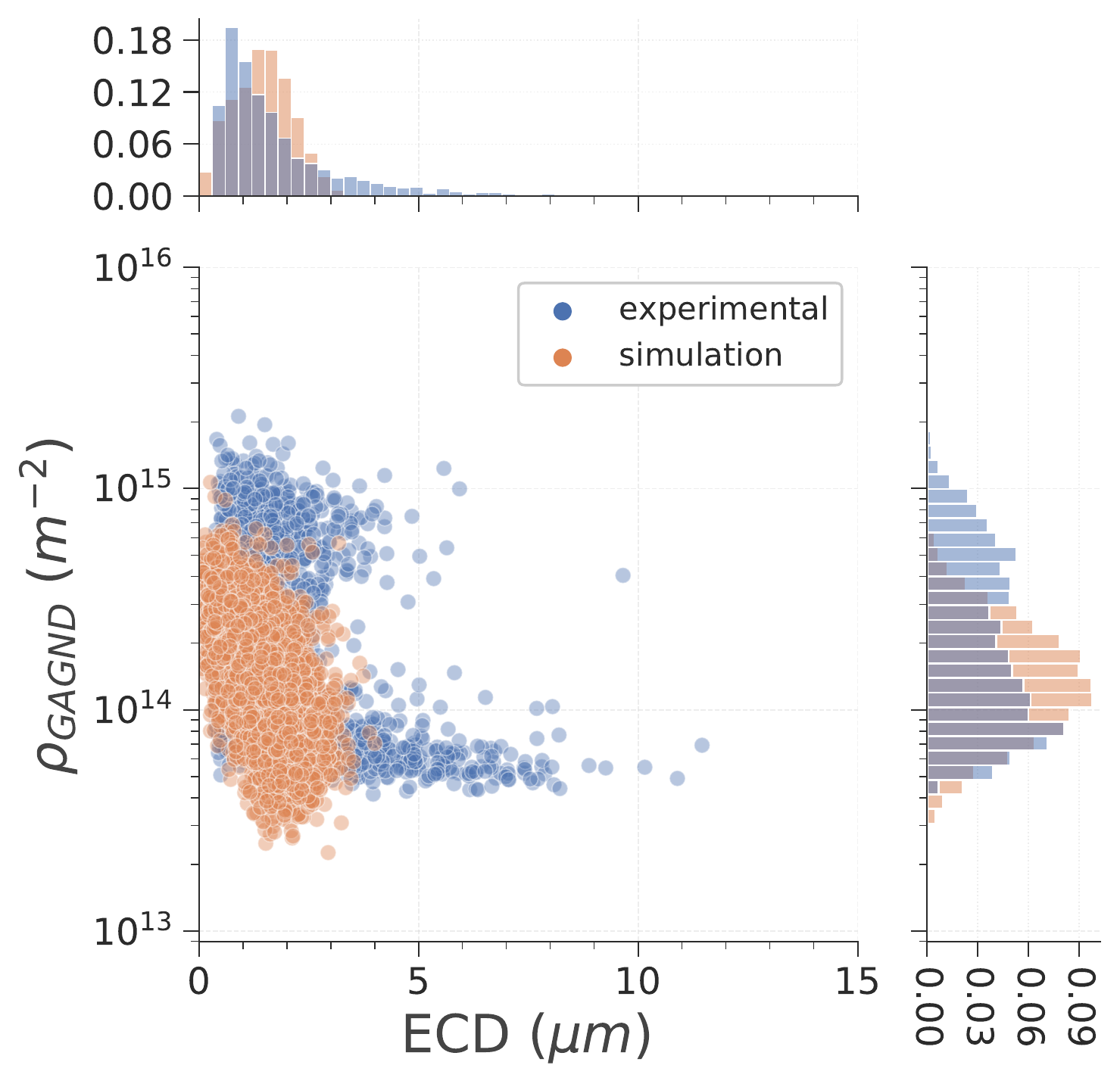}
    \caption{\label{fig:ScatterPlotStandardFormulation} Scatter plots displaying the grain ECD and grain average GND density distributions. Marginal number distributions. \textbf{Eq} initial microstructure. Hot deformation conditions are $T=650^{\circ}C; ~ \dot{\varepsilon}=1.0 ~ s^{-1}; ~ \varepsilon=1.0; ~ dt=100 ~ s$.}
\end{figure}

Figure \ref{fig:ScatterPlotStandardFormulation} provides new insights by displaying both grain ECD and grain average GND density distributions. From this plot, it appears clearly that simulations are significantly underpredicting the spread of both distributions. The simulations miss especially the growth of a limited number of subgrains and grains that present a low GND value and the persistence of large deformed grains.

Different modeling assumptions could potentially explain the differences between experimental and simulation results. The most probable are related to the anisotropy of GB properties and to the definition of GND density as a homogeneous variable within a subgrain or grain. Given that the formation of any particular texture component during PDRX is not observed, a dedicated study of the second hypothesis is preferred.

\subsubsection{Considering intragranular energy gradients in post-dynamic recrystallization simulations}

The deformation model considered within the present study does not allow to predict the formation of intragranular heterogeneities. Therefore, to evaluate the influence of such features, simulations of only the PDRX regime are proposed. In this case, initial microstructure after deformation is initialized using EBSD data. This enables to define an initial deformed microstructure presenting intragranular GND heterogeneities.

Simulations with the five configurations presented in section \ref{subsubsec:NumericalTools} are run for the following conditions: \textbf{BW} initial microstructure, $T=650^{\circ}C; ~ \dot{\varepsilon}=1.0 ~ s^{-1}; ~ \varepsilon = 1.2; ~ dt=200 ~ s$.

\begin{figure}[h!]
	\centering
    \includegraphics[width=0.6\linewidth]{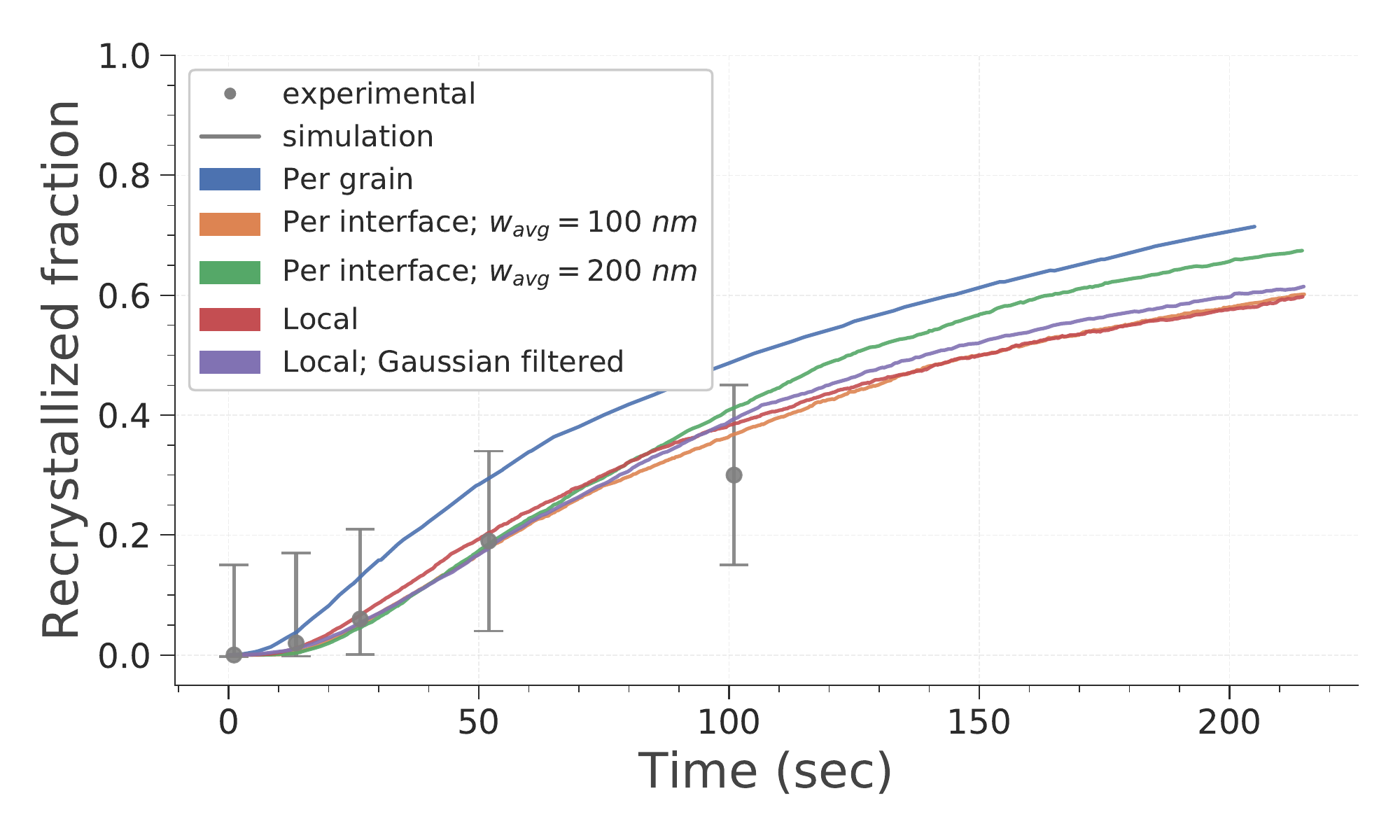}
    \caption{\label{fig:PDRXhetGND_mainFeatures} Evolution of recrystallized fraction with holding time, for the five different formulations. \textbf{BW} initial microstructure. Hot deformation conditions are $T=650^{\circ}C; ~ \dot{\varepsilon}=1.0 ~ s^{-1}; ~ \varepsilon=1.2$.}
\end{figure}

Figure \ref{fig:PDRXhetGND_mainFeatures} presents the recrystallized fraction evolution with holding time. Figure \ref{fig:PDRXhetGND_GNDmaps}, provided in appendix \ref{app:AdditionalSimulationResults} shows the five digital microstructures after a $100 ~ s$ heat treatment. Figure \ref{fig:PDRXhetGND_mainFeatures} points out that the four formulations that consider heterogeneous intragranular GND density predict a slower recrystallization kinetics which is in agreement with the experimental results. On the other hand, figure \ref{fig:PDRXhetGND_GNDmaps} confirm that all four formulations predict different local microstructure topology. As expected, considering local GND density difference leads to a significant increase of the interface tortuosity, especially at the beginning of the simulation. A zoom on a small zone displaying this aspect is presented in figure \ref{fig:TortuosityHetGND}. This could even lead to the formation of new grains. Such phenomenon present many similarities with the strain induced boundary migration (SIBM) mechanism.

\begin{figure}[H]
\centering
\begin{subfigure}{0.49\textwidth}
  \centering
  \includegraphics[width=1.0\linewidth]{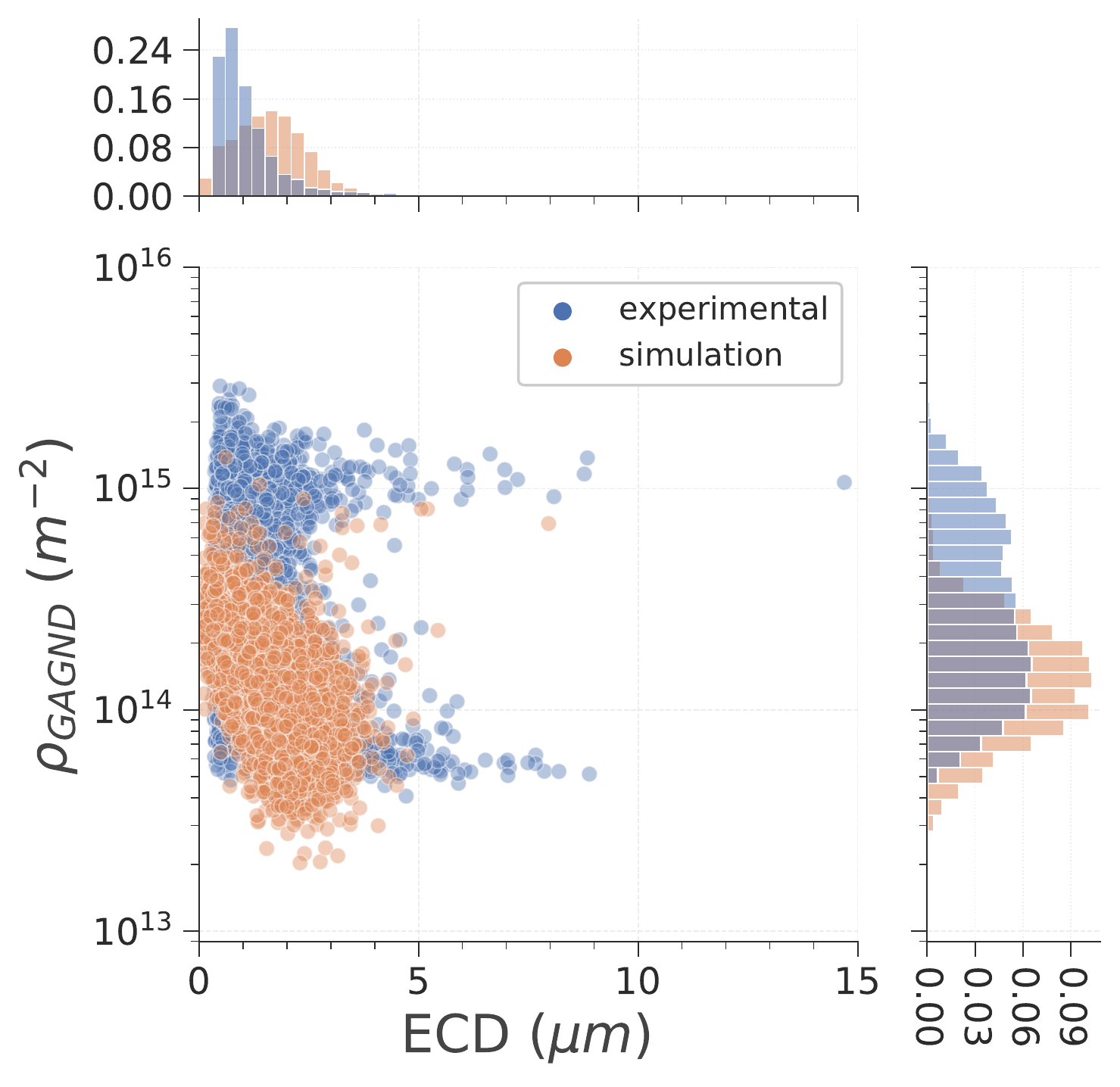}  \caption{\label{fig:PDRX_BW_CinetiqueRXParGrain} Per grain.}
\end{subfigure}
\begin{subfigure}{0.49\textwidth}
  \centering
  \includegraphics[width=1.0\linewidth]{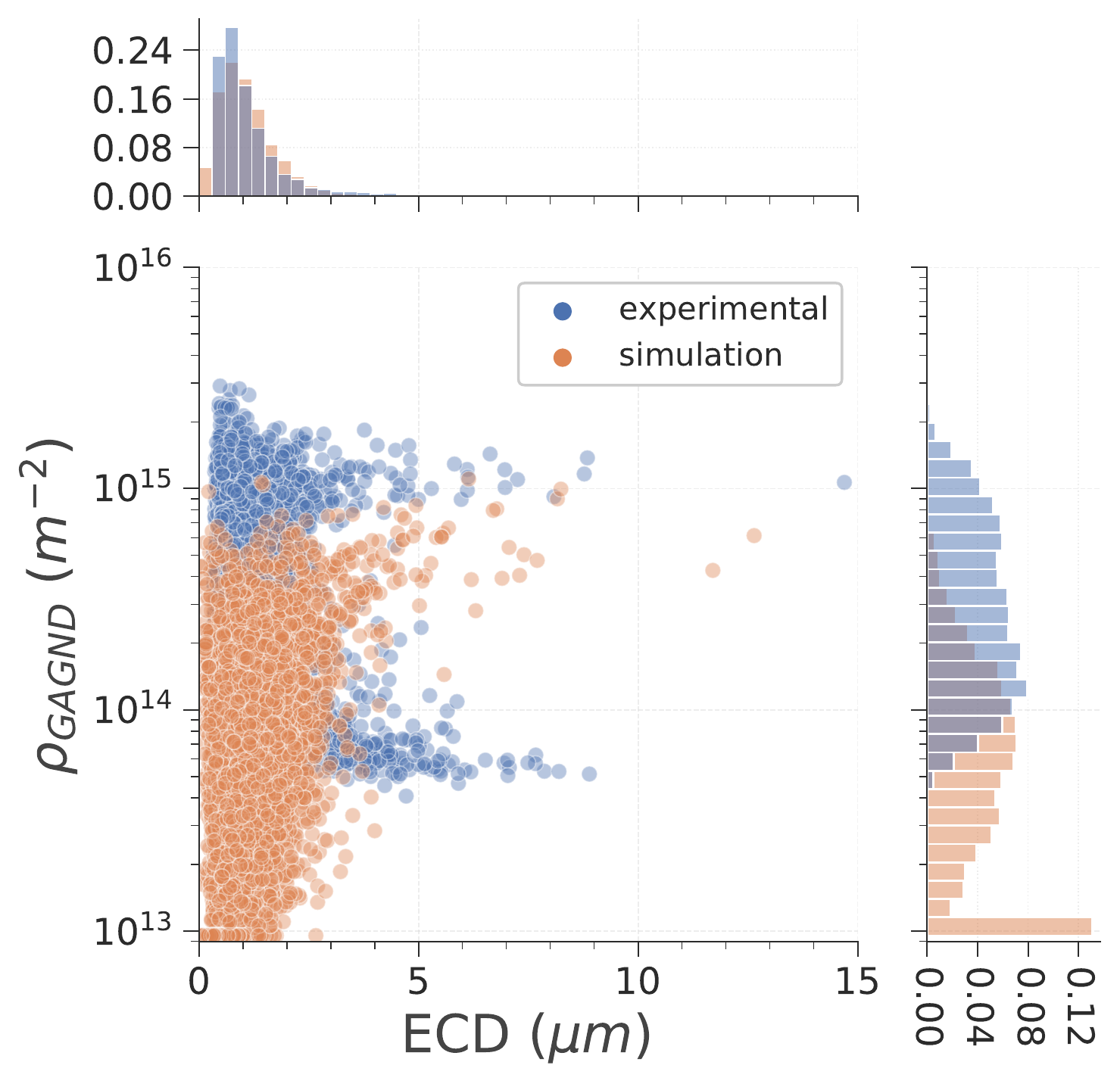} \caption{\label{fig:PDRX_BW_CinetiqueRXParInterface} Per interface.}
\end{subfigure}\\
\begin{subfigure}{0.49\textwidth}
  \centering
  \includegraphics[width=1.0\linewidth]{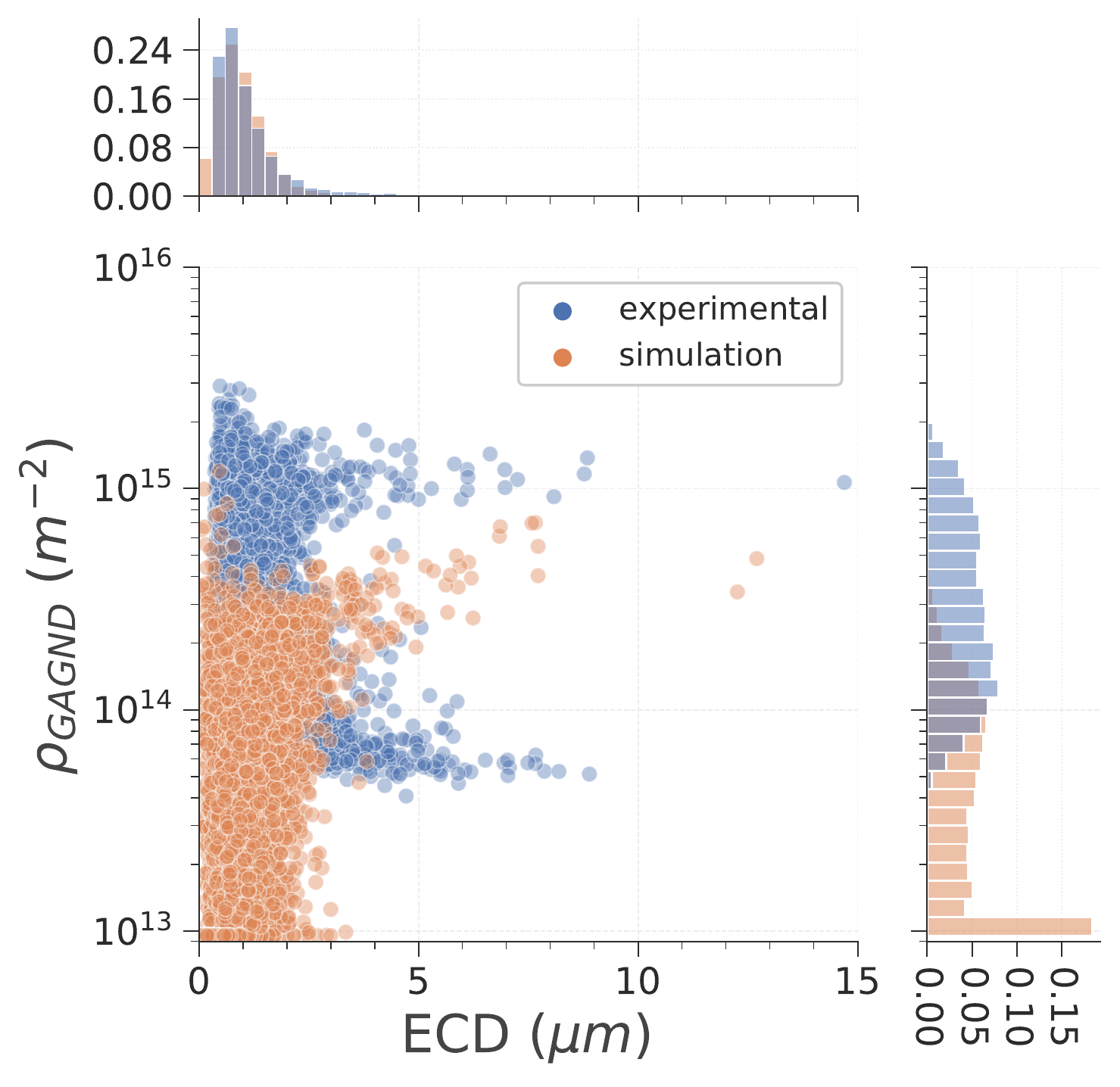}  \caption{\label{fig:PDRX_BW_CinetiqueRXFullHetero} Local.}
\end{subfigure}
\caption{Scatter plots displaying the grain ECD and grain average GND density distributions, for three simulations with different formulations for computation of driving pressure. Marginal number distributions. \textbf{BW} initial microstructure.  Hot deformation conditions are $T = 650^{\circ}C ; ~ \dot{\varepsilon} = 1.0 ~ s^{-1}; ~ \varepsilon = 1.0-1.2;~ dt = 100 ~ s$.}
\label{fig:PDRXhetGND_combinedScatterPlots}
\end{figure}

Figure \ref{fig:PDRXhetGND_combinedScatterPlots} displays experimental and simulation combined scatter plots after $100 ~ s$ for three of the five formulations.

These results highlight that \textit{Local} or \textit{Per interface} formulations predict significantly different grain size and grain average GND density distributions. These formulations better reproduce both GND density and grain size distributions. They enable to capture to a better extent the diversity of such microstructure. Large grains with a high GND density and small grains with a wide range of GND density are conserved up to longer holding times. 
Nevertheless, two major features are still not predicted with such numerical formulations. The number of large grains with a low dislocation density remain underestimated. This can be observed in figure \ref{fig:PDRXhetGND_GNDmaps}, provided in appendix. At the same time, the number of grains with a low average GND density is largely overestimated.

\section{Conclusion}\label{sec:Conclusion}

Zircaloy-4 dynamic and post-dynamic recrystallization have been studied using both experimental data and simulation tools. Samples presenting three initial microstructures (equiaxed, basket-weaved or parallel plate microstructures) have been hot deformed under various thermomechanical conditions. EBSD results have confirmed that:
\begin{itemize}
    \item zircaloy-4 recrystallization is characterized by a large number of LAGB, formed progressively and throughout the whole microstructure. Grains are progressively broken up and, with increasing strain, a large number of small grains and subgrains form.
    \item After hot deformation, the recrystallized fraction is almost systematically null. Indeed, one cannot distinguish a significant fraction of grains presenting a low internal energy.
    \item The increase of PDRX kinetics caused by a final strain or strain rate increase can be attributed to the higher number of grains with a low dislocation density or to the higher dislocation density, respectively.
    \item Initial microstructure impacts significantly recrystallization. It has been shown that the degree of heterogeneity is different for each of the three initial microstructures considered. One can classify the three typical microstructures in ascending order of heterogeneity: \textbf{PP} $>$ \textbf{BW} $>$ \textbf{Eq}.
\end{itemize}
A LS model recently extended has been applied to simulate CDRX and PDRX, for different thermomechanical conditions and initial microstructures. The results have shown that the model is able to capture with a very good agreement the initial microstructure influence over recrystallization. It is able to predict the persistence of large deformed grains upon subsequent heat treatment. Finally, additional simulations of PDRX considering intragranular heterogenous GND density fields have pointed out the impact of some simulation assumptions. The results confirmed that intragranular heterogeneities play a significant role in the formation of a wider grain size distribution. They also question about the adequate numerical formulation one should select to correctly take into account the effect of high stored energy gradients over GB migration during PDRX and the way to predict them during CDRX without using high numerical cost methodologies such as crystal plasticity FE formulations. 3D simulations and extension to other zirconium alloys are other perspectives of this work.

\clearpage

\appendix

\section{Supplementary EBSD orientation maps}\label{app:LargerOrientationMaps_ImpactInitialMicrostructure}

\begin{figure}[h!]
\centering
\begin{subfigure}{0.49\textwidth}
  \centering
  \includegraphics[width=1.0\linewidth]{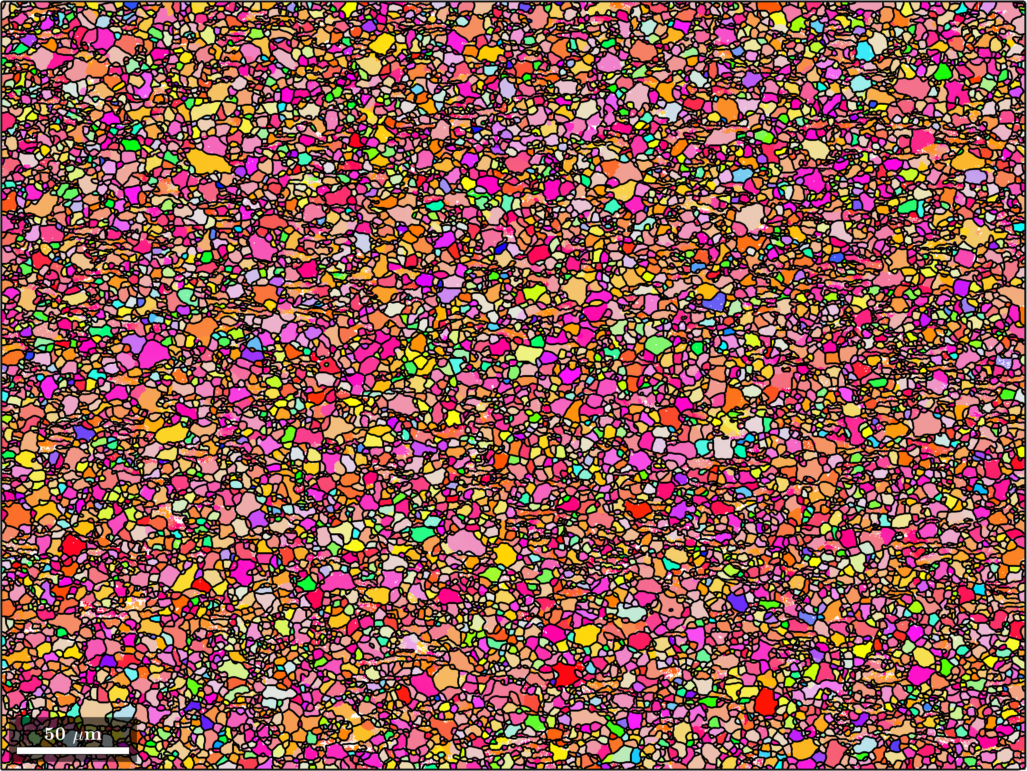}  
  \caption{\label{fig:HCR14_IPF_EBSD} Equiaxed.}
\end{subfigure}
\begin{subfigure}{0.49\textwidth}
  \centering
  \includegraphics[width=1.0\linewidth]{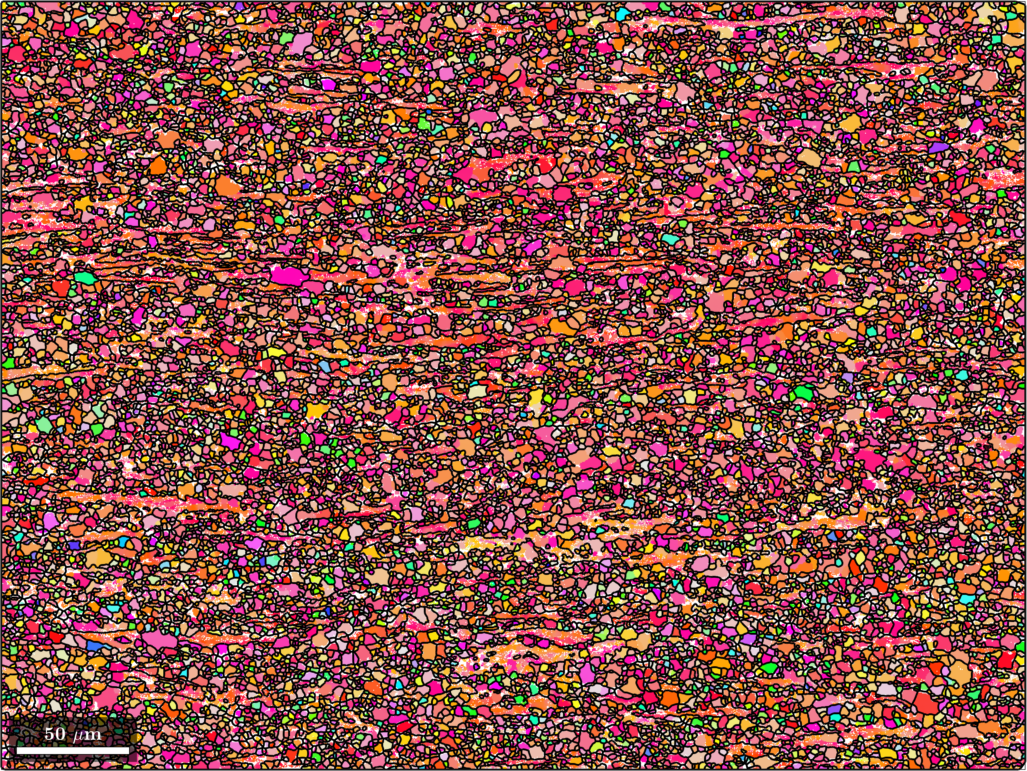} 
  \caption{\label{fig:HCHV8_IPF_EBSD} Basket-weaved.}
\end{subfigure}\\
\begin{subfigure}{0.49\textwidth}
  \centering
  \includegraphics[width=1.0\linewidth]{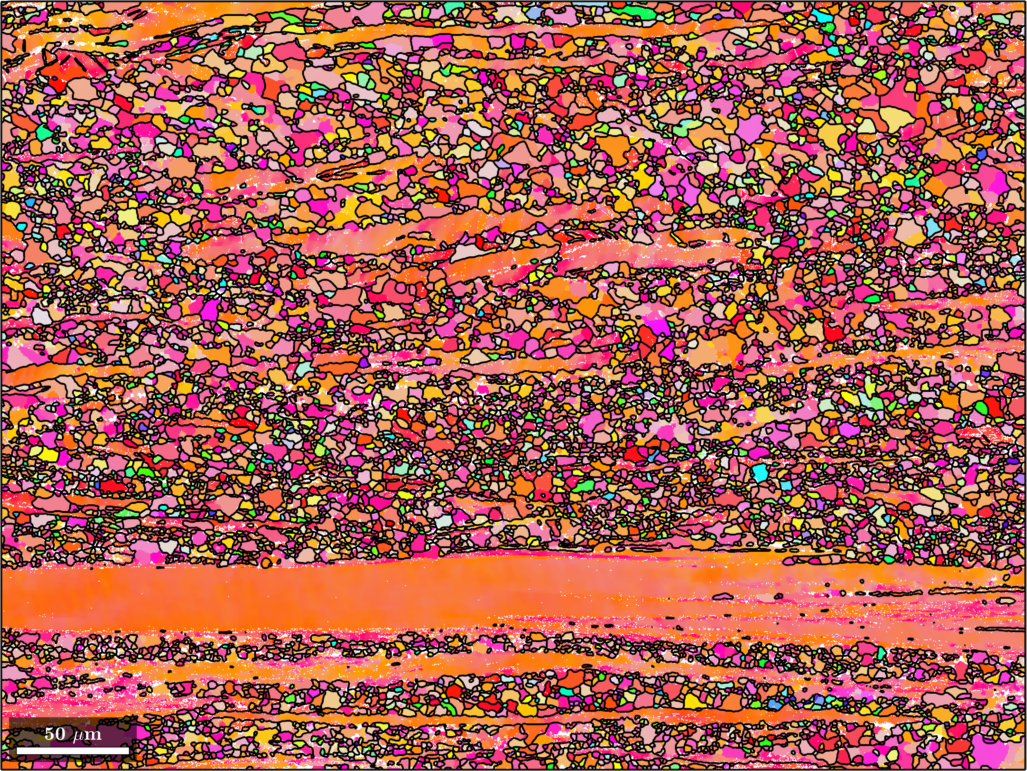}  
  \caption{\label{fig:HCHP8_IPF_EBSD} Parallel plates.}
\end{subfigure}
\caption{EBSD orientation maps for different initial microstructures ($T = 650^{\circ}C ; ~ \dot{\varepsilon} = 1.0 ~ s^{-1}; \varepsilon = 1.0-1.2; ~ dt=100 ~ s$). IPF Y color code.}
\label{fig:LargerOrientationMaps_ImpactInitialMicrostructure}
\end{figure}

Figure \ref{fig:LargerOrientationMaps_ImpactInitialMicrostructure} consists in orientation maps of samples deformed in the following conditions $T=650^{\circ}C; ~ \dot{\varepsilon}=1.0 ~ s^{-1}; ~ \varepsilon = 1.0-1.2$ and held at $650^{\circ}C$ for $100 ~ s$.

\clearpage

\section{Model parameters}\label{app:ModelParameters}

\begin{table}[h!]
\centering
\begin{tabular}{c c c c}\toprule
& Parameter & Symbol & Value \\
\midrule
\multirow{3}{*}{Grain boundary} & Mobility pre-exponential factor & $M_0$  & $1.35 \times 10^{-7}$ $m^4 \cdot J^{-1} \cdot s^{-1} $ \\
& Mobility activation energy & $Q_m$ & $134000 ~ J \cdot mol^{-1}$ \\
& HAGB energy &  $\gamma_{max}$ & $0.22 ~ J \cdot m^{-2}$ \cite{Dunlop2007_Rx} \\
\midrule
\multirow{5}{*}{Mechanical behavior} & Burgers vector & $b$  & $3.23 ~ \AA$  \cite{Dunlop2007_PlaticDef} \\
& Dislocation density behind RX front & $\rho_0$ & $1 \times 10^{13} ~ m^{-2}$ \\
& Shear modulus &  $\mu$ & $42.52-0.022 \times T ~ (K) ~ GPa$ \cite{Moon2006} \\
& Taylor factor & $M_{\textsc{T}}$ & $2.3$ \cite{Dunlop2007_PlaticDef, Shen2013}\\
& Geometrical factor & $\alpha_{\textsc{T}}$ & $0.2$ \cite{Mughrabi2016}\\
\midrule
\multirow{8}{*}{Hardening and recovery} & \multirow{3}{*}{Hardening coefficients} & $K^1_0$  & $1.6 \times 10^{14} ~ m^{-2}$\\
& & $m^h$  & $0.14$\\
& & $Q_h$  & $1.71 \times 10^{5} ~ J \cdot mol^{-1}$\\
& \multirow{3}{*}{Hardening coefficients} & $K^2_0$  & $712$\\
& & $Q_r$  & $3.42 \times 10^{5} ~ J \cdot mol^{-1}$\\
& & $m^r$  & $0.11$\\
& \multirow{2}{*}{Yield stress coefficients} & $\sigma^i_0$  & $-17.8 ~ MPa$\\
& & $\sigma^s_0$  & $2.86 ~ MPa$\\
\midrule
\multirow{4}{*}{Subgrain formation} & Grain reference diameter ($\mu m$) & $D_0$  & $1.0$ \\
& Minimum subgrain disorientation (${\circ}$) & $\theta_0$ & $1.0$ \\
& Fixed coefficient &  $m$ & $2$ \\
& LAGB dislocation set number & $\eta$ & 2\\
\bottomrule
\end{tabular}
\caption{\label{tab:ModelParameters} Definition of model parameters.}
\end{table}

\section{Additional simulation results}\label{app:AdditionalSimulationResults}

\begin{figure}[h!]
\centering
\begin{subfigure}{0.49\textwidth}
  \centering
  \includegraphics[width=1.0\linewidth]{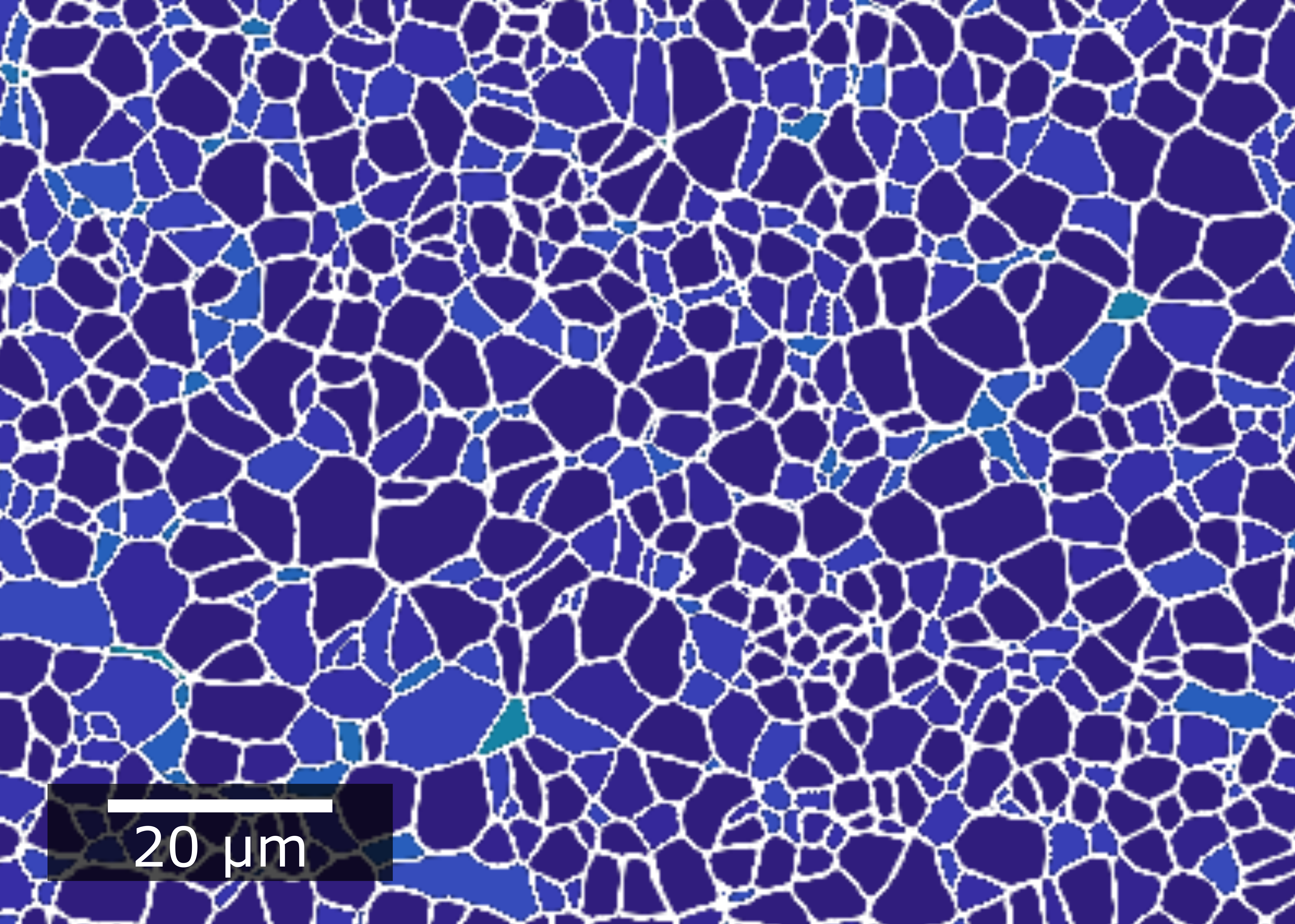}  
  \caption{\label{fig:PerGrain}  $Per ~ grain$.}
\end{subfigure}\\
\begin{subfigure}{0.49\textwidth}
  \centering
  \includegraphics[width=1.0\linewidth]{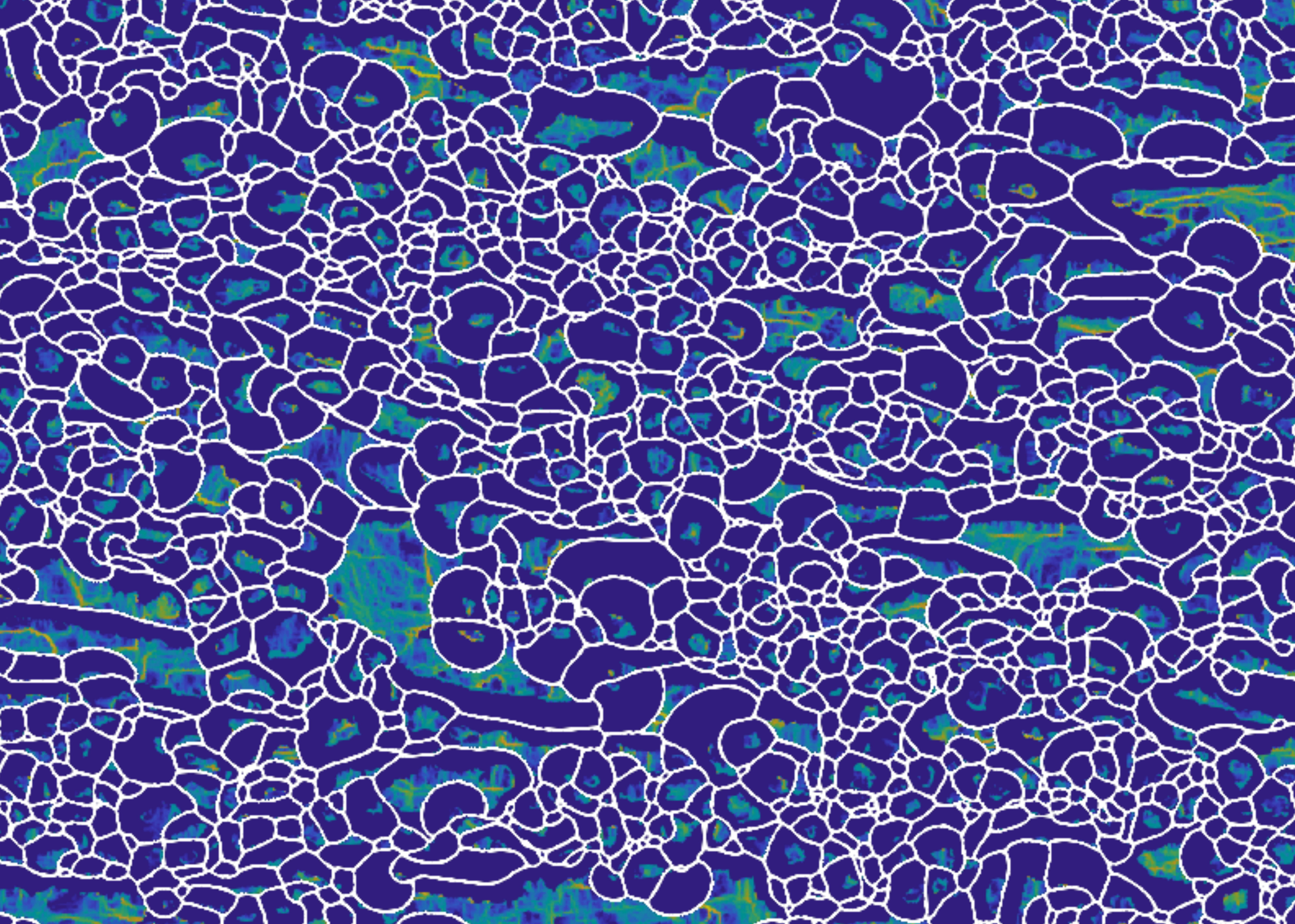} 
  \caption{\label{fig:PerInterface} $Per ~ interface; ~ w_{avg}=100 ~ nm$.}
\end{subfigure}
\begin{subfigure}{0.49\textwidth}
  \centering
  \includegraphics[width=1.0\linewidth]{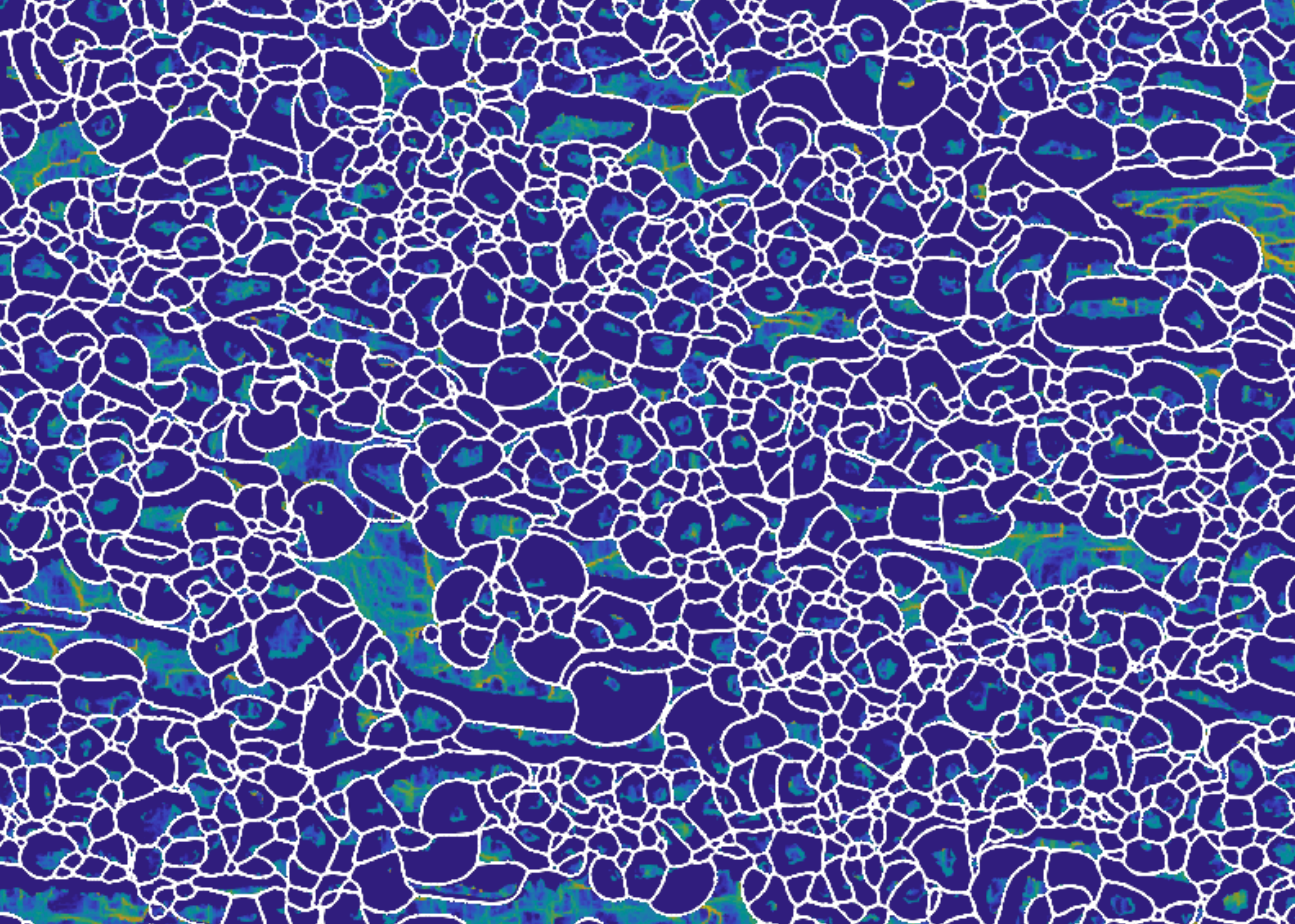}  
  \caption{\label{fig:PerInterfaceLargerThickness} $Per ~ interface; ~ w_{avg}=200 ~ nm$.}
\end{subfigure}\\
\begin{subfigure}{0.49\textwidth}
  \centering
  \includegraphics[width=1.0\linewidth]{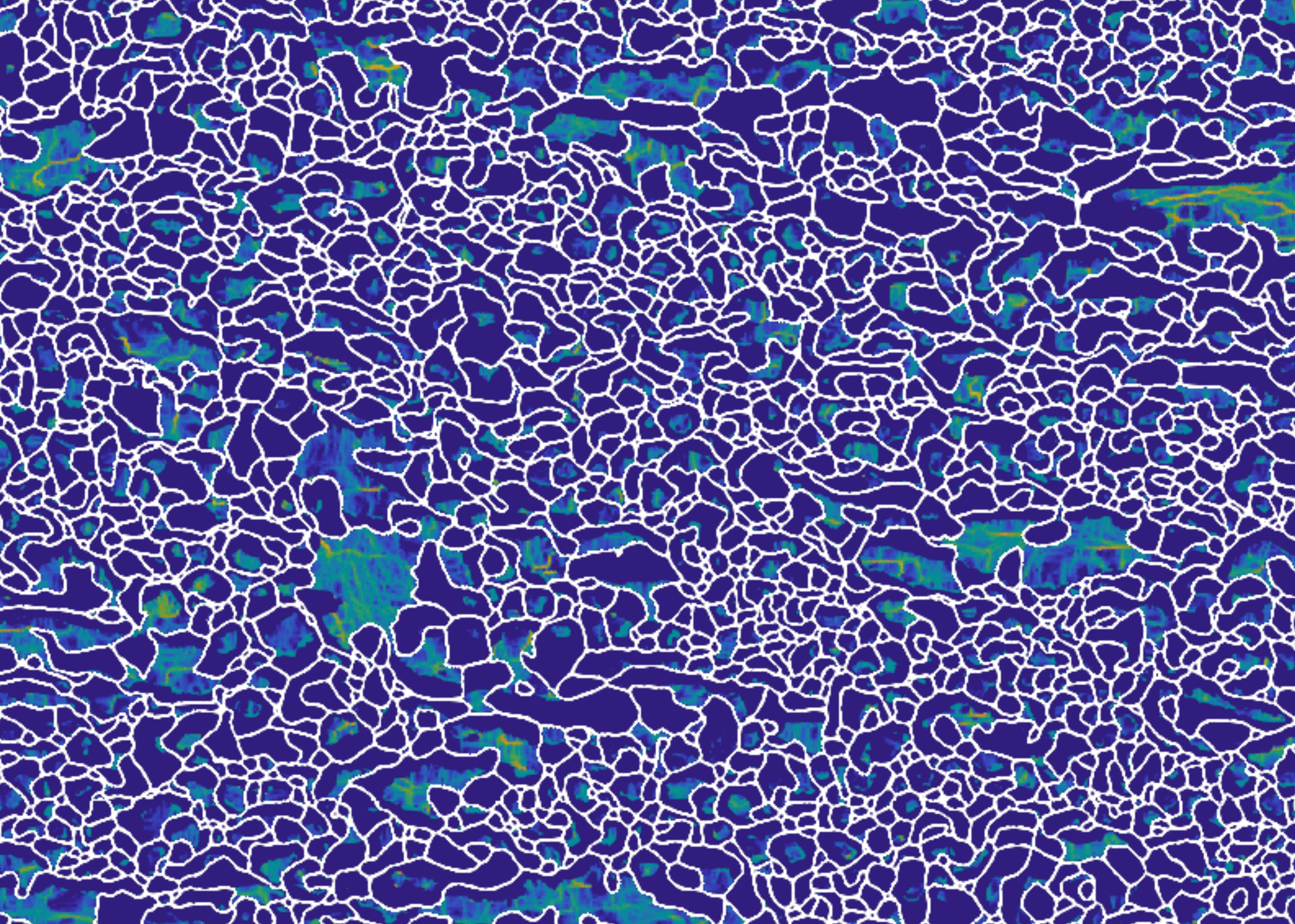}  
  \caption{\label{fig:FullHetero} $Local$.}
\end{subfigure}
\begin{subfigure}{0.49\textwidth}
  \centering
  \includegraphics[width=1.0\linewidth]{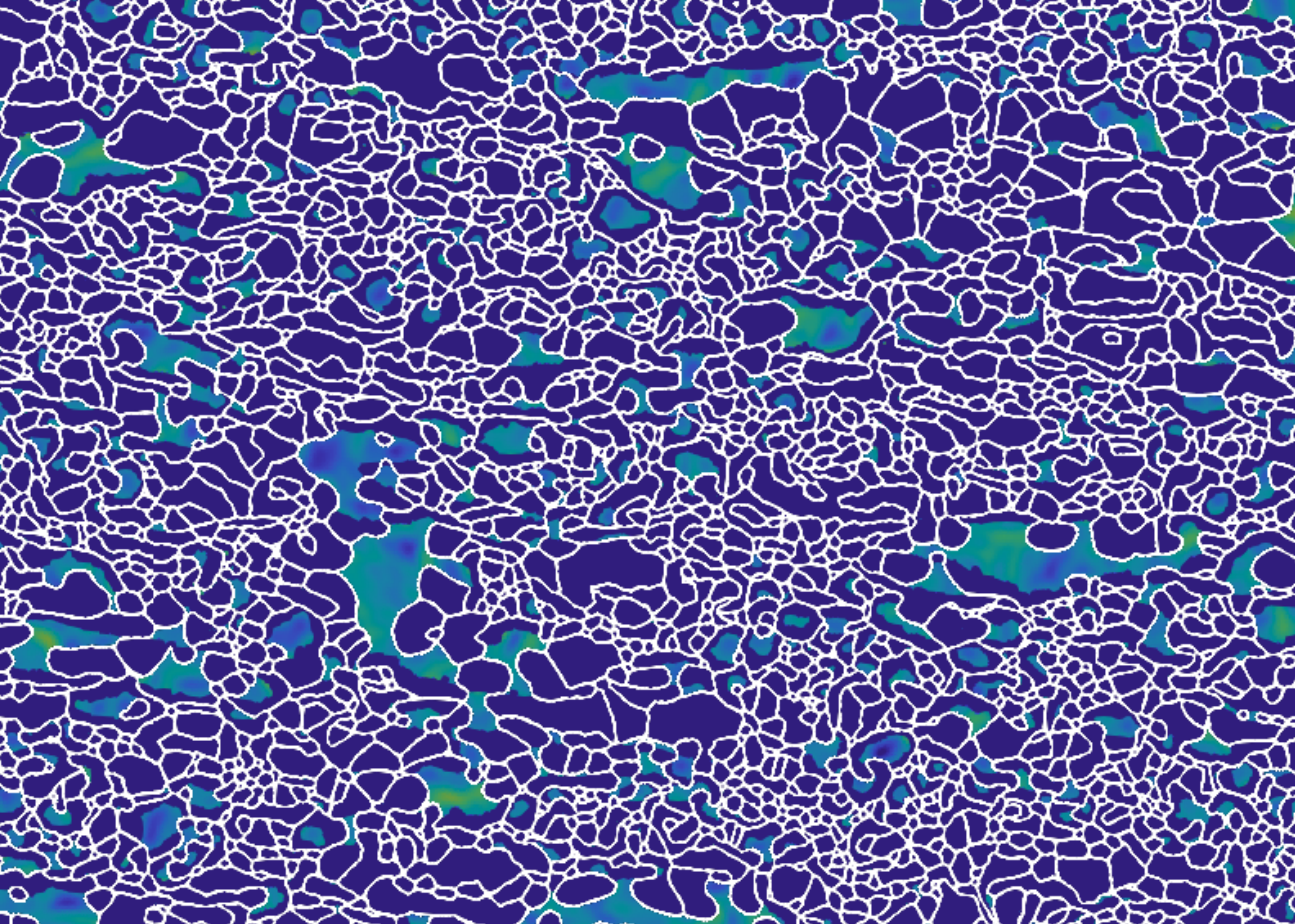}  
  \caption{\label{fig:FullHeteroGaussian} $Local; ~ Gaussian ~ filtered$.}
\end{subfigure}
\caption{GND density maps from simulation results considering intragranular heterogeneous GND density fields. \textbf{BW} initial microstructure. ($T = 650^{\circ}C ; ~ \dot{\varepsilon} = 1.0 ~ s^{-1}; \varepsilon = 1.2; ~ dt=100 ~ s$).}
\label{fig:PDRXhetGND_GNDmaps}
\end{figure}

\begin{figure}[H]
    \centering
    \includegraphics[width=0.5\linewidth]{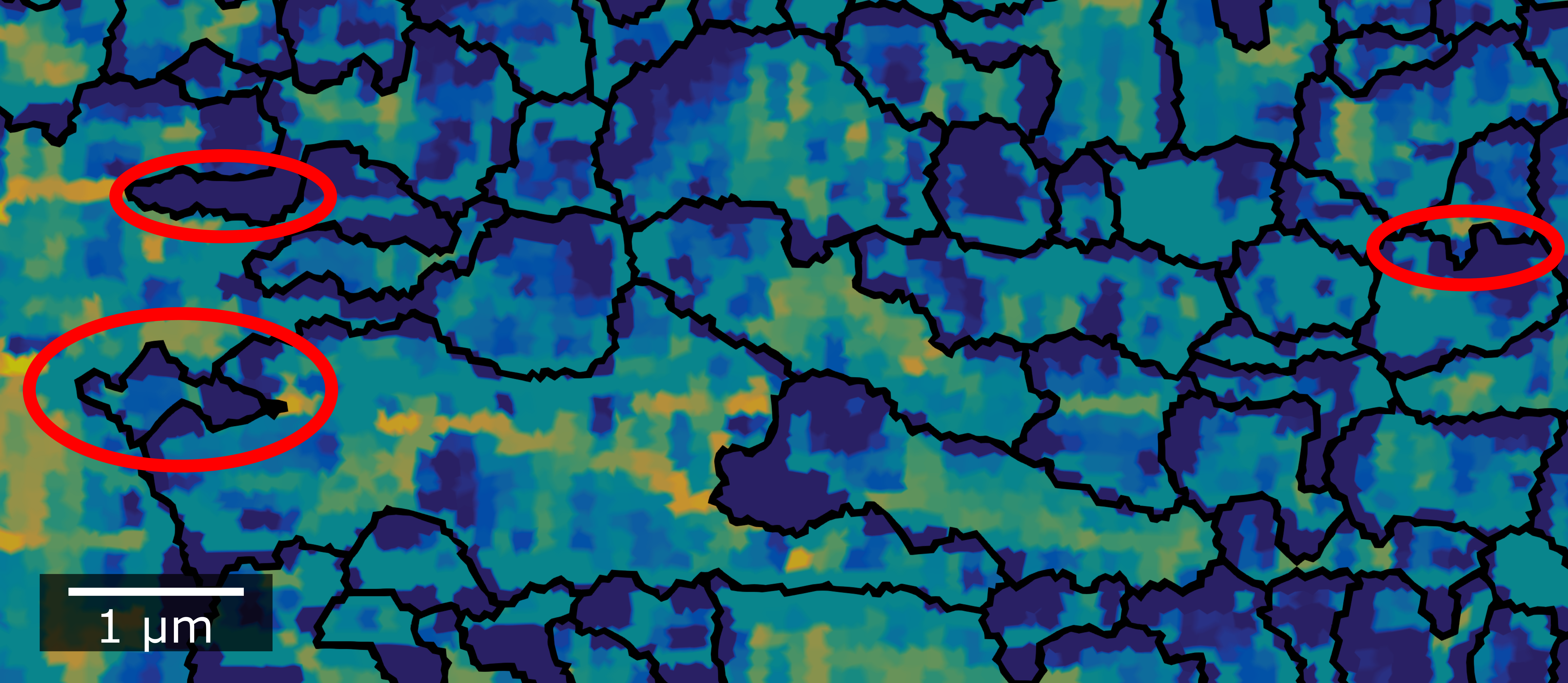}
    \caption{GND map after $5 ~ s$ of heat treatment at $650^{\circ}C$, predicted using local formulation with no pre-processing operation. Some very tortuous grain boundaries are circled in red.}
    \label{fig:TortuosityHetGND}
\end{figure}

\clearpage

\bibliographystyle{unsrt}
\bibliography{template}  

\end{document}